\documentclass[traditabstract]{aa}
\usepackage{graphicx}
\usepackage{enumerate}
\usepackage{epsfig}
\usepackage{color}
\usepackage{txfonts}
\usepackage{natbib}
\bibpunct{(}{)}{;}{a}{}{,} % to follow the A&A style
\begin{document}

\title{Mapping a stellar disk into a boxy bulge:\\The outside-in part of the Milky Way bulge formation}

\titlerunning{Mapping a stellar disk into a boxy bulge}

%\author{P. Di Matteo\inst{1},  A. Gomez\inst{1},  C. Babusiaux\inst{1}, M. Haywood\inst{1}, F. Combes\inst{2},  B. Semelin\inst{2}}%,  O. Snaith\inst{1}}
\author{P. Di Matteo\inst{1}, M. Haywood\inst{1}, A. G$\acute{\textrm{o}}$mez\inst{1},  L. van Damme\inst{1},  F. Combes\inst{2},  A. Hall$\acute{\textrm{e}}$\inst{1}, B. Semelin\inst{2, 3},  M.~D. Lehnert\inst{4}, D.~Katz\inst{1}}

\authorrunning{Di Matteo et al.}

\institute{GEPI, Observatoire de Paris, CNRS, Universit\'e
  Paris Diderot, 5 place Jules Janssen, 92190 Meudon, France\\
\email{paola.dimatteo@obspm.fr}
\and
LERMA, Observatoire de Paris, CNRS, 61 Av. de l$'$Observatoire, 75014 Paris, France
\and Universit$\rm \acute{e}$ Pierre et Marie Curie, 4 place Jussieu, 75005 Paris, France 
\and Institut dÕAstrophysique de Paris, 98 bis Bd Arago, 75014 Paris, France
}

\date{Accepted, Received}

\abstract{By means of idealized, dissipationless N-body simulations
which follow the formation and subsequent buckling of a stellar bar, we
study the characteristics of boxy/peanut-shaped bulges and compare them
with the properties of the stellar populations in the Milky Way bulge.
The main  results of our modeling, valid for the general family of
boxy/peanut shaped bulges, are the following: \emph{(i)} because of
the spatial redistribution  in the disk initiated at the epoch of bar
formation, stars from the innermost
regions to the outer Lindblad resonance of the stellar bar are
mapped into a boxy bulge; \emph{(ii)} the contribution of stars to the
local bulge density depends on their birth radius:  stars born in the
innermost disk tend to dominate the innermost regions of the boxy bulge,
while stars originating closer to the OLR are preferentially found in the outer
regions of the boxy/peanut structure; \emph{(iii)} stellar birth radii
are imprinted in the bulge kinematics, the larger the birth radii of
stars ending up in the bulge, the greater their rotational support and
the higher their line-of-sight velocity dispersions (but note that this
last trend depends on the bar viewing angle); \emph{(iv)} the higher the
classical bulge-over-disk ratio, the larger its fractional contribution
of stars at large vertical distance from the galaxy mid-plane. Comparing
these results with the properties of the stellar populations of the
Milky Way's bulge recently revealed by the ARGOS survey, we conclude
that: \emph{(I)} the two most metal-rich populations of the MW bulge,
labeled A and B in the ARGOS survey, originate in the disk, with the
population of A having formed on average closer to the Galaxy center
than the population of component B; \emph{(II)} a massive (B/D$\sim$0.25)
classical spheroid can be excluded for the Milky Way, thus confirming
previous findings that the Milky Way bulge is composed of populations
that mostly have a disk origin.  On the basis of their chemical and
kinematic characteristics, the results of our modeling suggests that the
populations A, B and C, as defined by the ARGOS survey, can be associated,
respectively, with the inner thin disk, to the young thick and to the
old thick disk, following the nomenclature recently suggested for stars
in the solar neighborhood by Haywood et al. (2013).}


\maketitle

%%
%% no 1,7,8,14,15,21,22,28,29,35
\section{Introduction}

Boxy and peanut shaped bulges are present in about half of edge-on disk
galaxies \citep{lut00}. The closest example of a boxy bulge can be found
in our Galaxy \citep{okuda77, maihara78,
weiland94, dwek95}.  Even if some studies \citep{binney85, whit88} have
proposed that these structures can be formed during accretion events,
their high frequency and relation to the fraction of barred galaxies
in disks \citep{eskridge00, menendez07, marinova07, aguerri09} suggest
that a more common mechanism may be responsible for shaping the central
regions of galaxies, giving them their boxy- or peanut-shaped morphology.

A number of numerical investigations \citep{combes81, pfenniger91,
athanassoula05, martinez06} have indeed shown that boxy bulges can be
manifestations of secular processes that occur in disk galaxies such
as thick stellar bars seen edge on. During their evolution, stellar
bars can indeed go through one (or multiple) buckling phase(s), which
are the consequences of vertical instabilities, and depending on the
bar viewing angle, the resulting thick structure can appear boxy, if
observed mainly along the bar major axis, or peanut-shaped, if observed
mainly along the bar minor axis. It is also possible that a combination
of these mechanisms, satellite accretion and bar instability, may be
responsible for some of the observed bulge morphologies \citep{mihos95}.

Observations suggest that boxy bulges do not represent a homogeneous
class of objects. Studies of galaxies indeed demonstrate that
boxy bulges display a range of properties in their kinematics and stellar populations
\citep{williams11}. Some of them show a constant rotation with height above
the plane, while some others do not,  thus indicating that cylindrical
rotation does not necessarily characterize these structures;  some of them
show negative vertical metallicity gradients, that can be accompanied
by positive [$\alpha$/Fe] gradients, while some are more homogenous,
indicating that different stellar populations can dominate these
structures at different vertical distances from the galaxy midplane.
The complexity of the observed characteristics may be also due to
the concomitant presence of a classical\footnote{In the following,
by classical bulge we mean a spheroidal component, not formed by disk
instabilities, but rather through mergers or some dissipative collapse
at early phases of the galaxy formation.} bulge in the inner regions of
some galaxy disks.  It is not simple to identify the presence of such
components in boxy or peanut-shaped structures by the characteristics
of their stellar populations and/or kinematics. The existence of a
vertical metallicity gradient, for example, does not necessarily imply the
presence of a classical bulge \citep{bekki11, martinez13}; the detection
of cylindrical rotation does not necessarily imply that the boxy bulge
is the result of pure bar instabilities \citep{saha12, saha13}.

The question of understanding how much of a classical bulge is present
in structures otherwise mostly shaped by secular evolution processes is
fundamental not only in interpreting observations of galaxies, but also
for understanding the formation and evolution of the central regions
of the Milky Way. Indeed, over the last two decades, a number of studies
have elucidated the complexity of the boxy, peanut-shaped structure
at the center of our Galaxy \citep{mcwilliam94, zoccali06, lecureur07,
zoccali08, babusiaux10, shen10, ness12, ness13a, ness13b, gonzalez13,
bensby13}, but no consensus has yet been reached on how to interpret
these important results.

A number of studies have shown evidence of the presence of a metal
poor, $\alpha$-enriched component, whose kinematic properties are
significantly different from that of the metal-rich, nearly solar
[$\alpha$/Fe] bulge component \citep{babusiaux10, hill11, ness13b}.
\citet{babusiaux10}, for example, have pointed out the presence of
two distinct populations along the bulge minor axis, with distinct
kinematic properties: a metal-poor population ([Fe/H]$\sim$-0.3 dex)
whose radial velocity dispersion is constant with latitude, and a metal
rich population ([Fe/H]$>$~0.1 dex) whose radial velocity dispersion
decreases substantially with the distance from the Galactic mid-plane. The
contribution of these two populations changes with latitude, the metal
rich component disappearing when moving away from the plane, where the
metal poor population is becoming dominant.  By comparing the data with
the N-body model of \citet{fux99}, \citet{babusiaux10} interpreted
these two populations as being the signature in the inner Galactic
disk of the simultaneous presence of a classical metal-poor bulge and a
metal-rich population with bar-like kinematics.  However, the presence of a classical
bulge in the Milky Way disk and even its role in explaining the characteristics
of the observed metal poor, $\alpha$-enhanced population is debated.
 \citet{shen10}, for example, 
have questioned the existence of any classical bulge in the Milky Way,
ruling out the possibility that our Galaxy has a classical bulge with
a mass greater than $\sim$15$\%$ of the stellar disk mass.

In addition, recent studies have pointed out the similarities between
the metal poor populations of the galactic bulge and the thick disk
population at the solar vicinity \citep{melendez08, ryde10, alves10,
bensby10, gonzalez11, ness13a}. The ARGOS survey \citep{freeman13},
in particular, is currently mapping the Galactic bulge over a large
extent of latitudes and longitudes, contributing significantly to our
understanding of how the bulge populations differ in their spatial
redistribution, chemical properties and kinematics \citep{ness12,
ness13a, ness13b}. The results of this survey mainly suggest the
existence of at least three primary components in the Milky Way bulge. Two
components (defined respectively as component A and B in their paper),
with [Fe/H]$>$$-$0.5 dex, are part of the boxy/peanut shaped bulge,
with component B ([Fe/H]$\sim$-0.25 dex) being dynamically hotter than
component A ([Fe/H]$\sim$0.1 dex), rotating 20$\%$ faster than A, and
being more prominent at high latitudes; component C ([Fe/H]$<$-0.5 dex)
has the highest radial velocity dispersions in the fields analyzed,
nearly constant both in latitude and longitude, and has been explained
by Ness and collaborators as part of the inner thick disk.

Is it possible to explain the complexity and richness of these data with a
``simple'' scenario for the formation of the Milky Way bulge? How many of
the observed characteristics, for example, are due to a complex accretion
history for the Galaxy and how much is due to secular evolution of the
disk, with the possible contribution of an older and more metal-poor
thick-disk component and/or a classical spheroid?

In this paper, we try to answer to some of these questions by showing
that at least part of the characteristics observed among the stellar
populations of the Galactic bulge, such as the angular momentum support,
radial velocity dispersions, and dependence of the fractional contribution
of different stellar populations on latitude, can be explained as the
simple result of the mapping of a stellar disk onto a boxy/peanut-shaped
bulge.  We will show that  \emph{a large portion of the stellar disk,
from the innermost regions to the outer Lindblad resonance of the
bar,}  is involved in the formation of a boxy/peanut structure, as a
result of the radial migration initiated before the buckling instability
of the formation of the bar.

In particular, we will show that the two populations contributing to the
boxy structure (component A and B in Ness et al papers) have kinematic
and spatial characteristics compatible with a origin in different
regions of the disk. Specifically, we show that, on average, component B
formed from stars with initial radii larger than the stars that comprise
component A. On the basis of the observed characteristics, the spatial
distribution, chemistry and kinematics, we propose that component B is
mostly made of the young MW thick disk \citep[stellar ages between 8 and
10 Gyr, as observed at the solar neighborhood;][]{haywood13}, while A is
mostly made of stars that originated in the inner thin disk. The presence
of a small (B/D=0.1), classical bulge is not excluded and we propose a
possible signature of its presence should be searched for in observations
of the bulge. We will discuss the possibility that the old thick disk
\citep[ages greater than 10 Gyr;][]{haywood13} is the main contributor
to component C, which is not part of the boxy/peanut structure.

The paper is organized as follows: after describing the N-body models
used for the analysis (Sect.~\ref{method}), we will present the main
results from an analysis of these simulations in Sects.~\ref{results}
and ~\ref{discussion}; then in Sect.~\ref{conclusions}, we state the
main conclusions of our work.

\begin{figure}
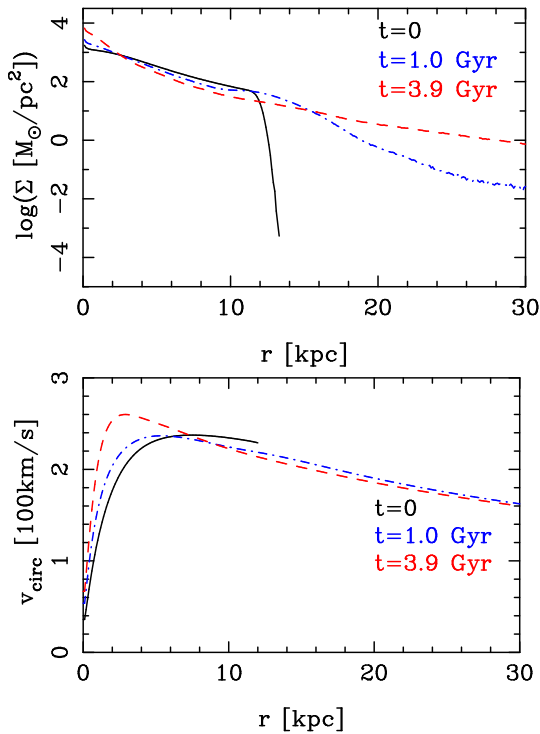

\centering
\includegraphics[width=4.8cm,angle=270]{pface_initfin2_BD0p00.ps}
\includegraphics[width=4.8cm,angle=270]{pvcirc_initfin_BD0p00.ps}
\caption{\emph{(Top panel):} Surface density profiles of the
initial modeled bulgeless galaxy seen face-on (black curve), after
1.0 Gyr of evolution (blue curve) and after 3.9 Gyr of evolution (red
curve). \emph{(Bottom panel):} The initial circular velocity of disk
stars (black curve), after 1.0 Gyr of evolution (blue curve) and 3.9
Gyr of evolution (red  curve).}
\label{density}
\end{figure}

\begin{figure}
\centering
\includegraphics[width=4.8cm,angle=270]{Avstime_gS0BD0p00.ps}
\includegraphics[width=3.2cm,angle=270]{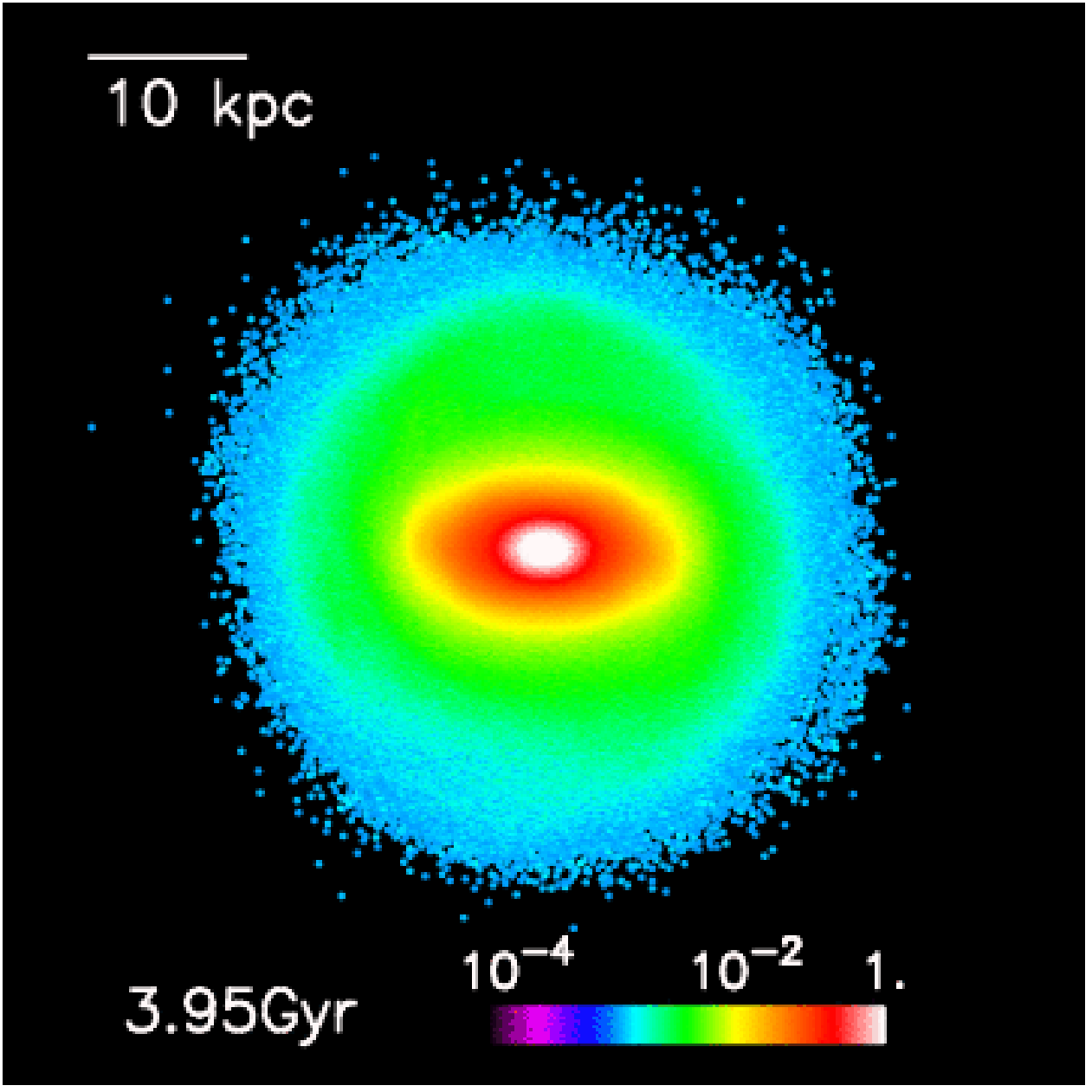}
\includegraphics[width=3.2cm,angle=270]{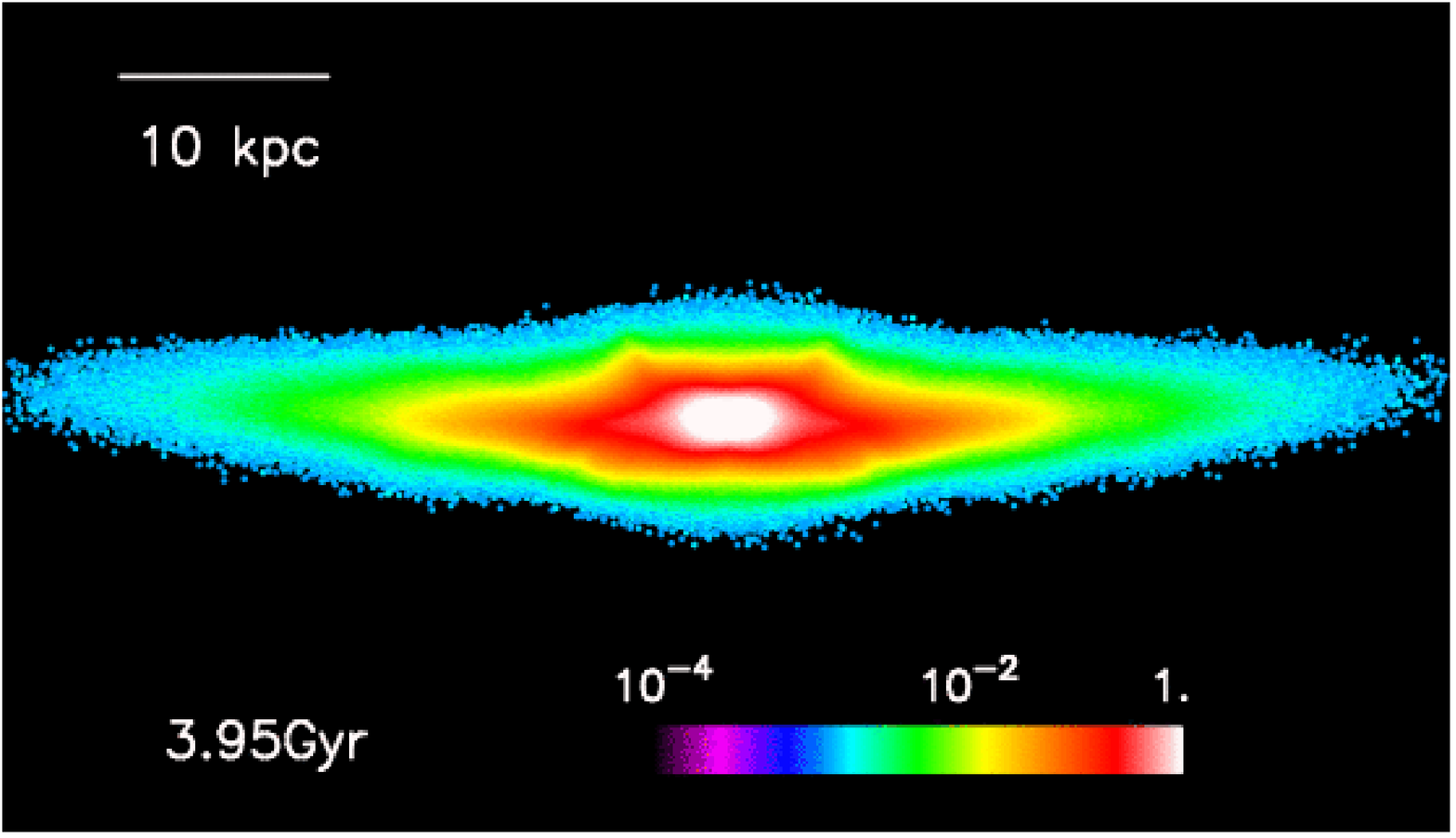}
\caption{\emph{(Top panel:)} Evolution of the $A_1$ (black line),
$A_2$ (red line), $A_4$ (blue line) asymmetries versus time for the
model with B/D=0. The $A_i$ values are normalized to the $m=0$ value,
$A_0$. \emph{(Bottom panels:)} Face-on and edge-on views of the stellar
component of the bulgeless disk galaxy at the end of the simulation. }
\label{asym}
\end{figure}

\begin{figure*}
\centering
\includegraphics[width=3.cm,angle=270]{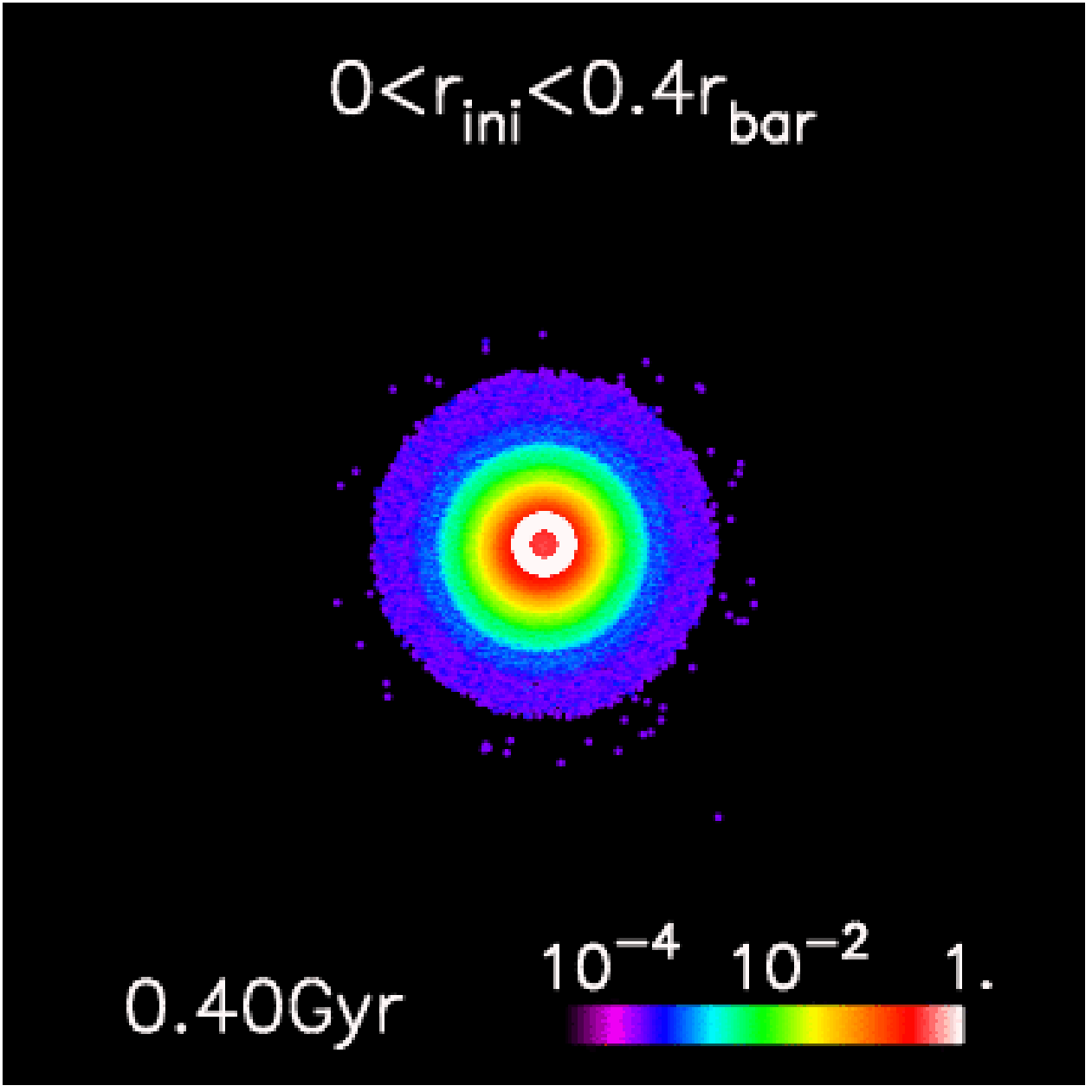}
\includegraphics[width=3.cm,angle=270]{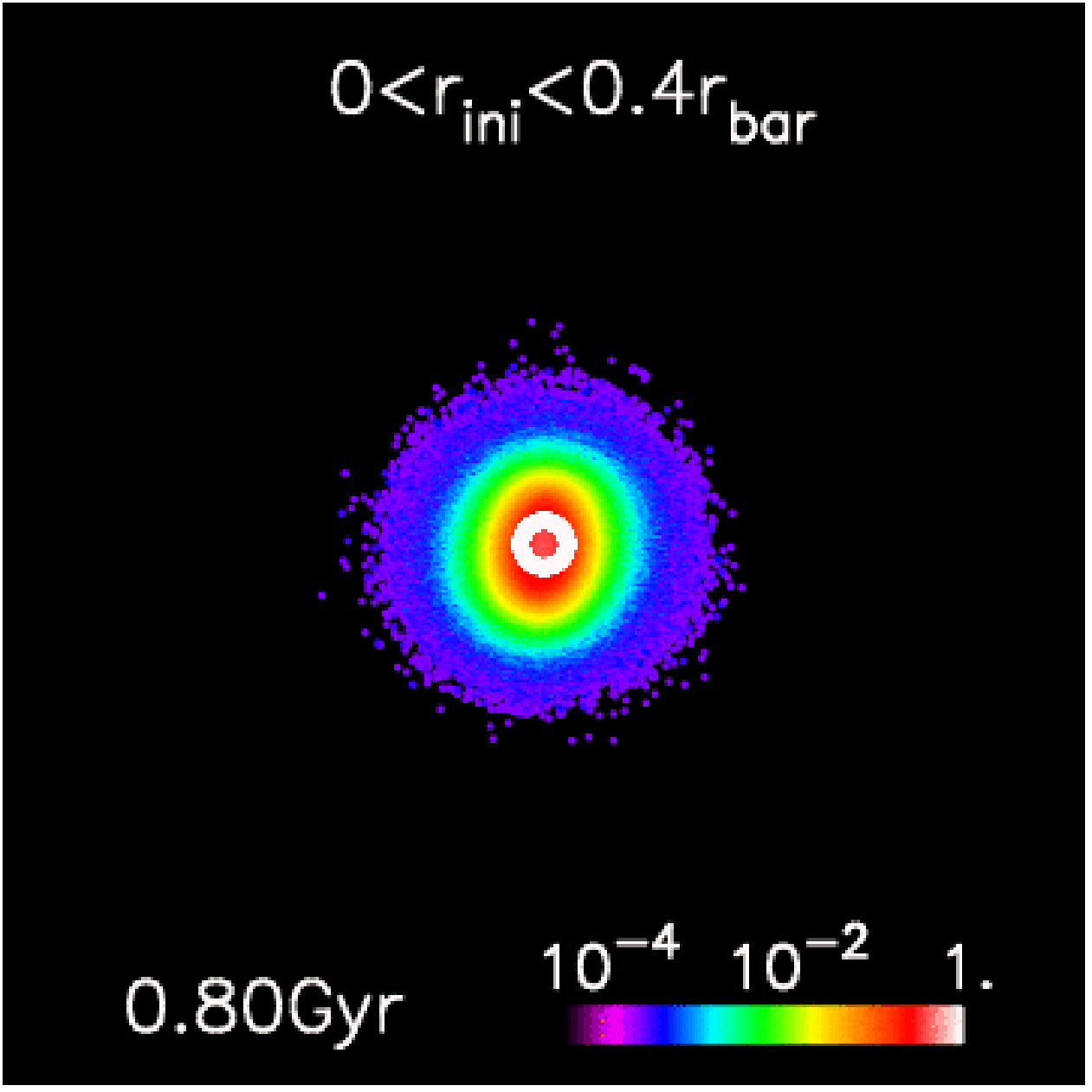}
\includegraphics[width=3.cm,angle=270]{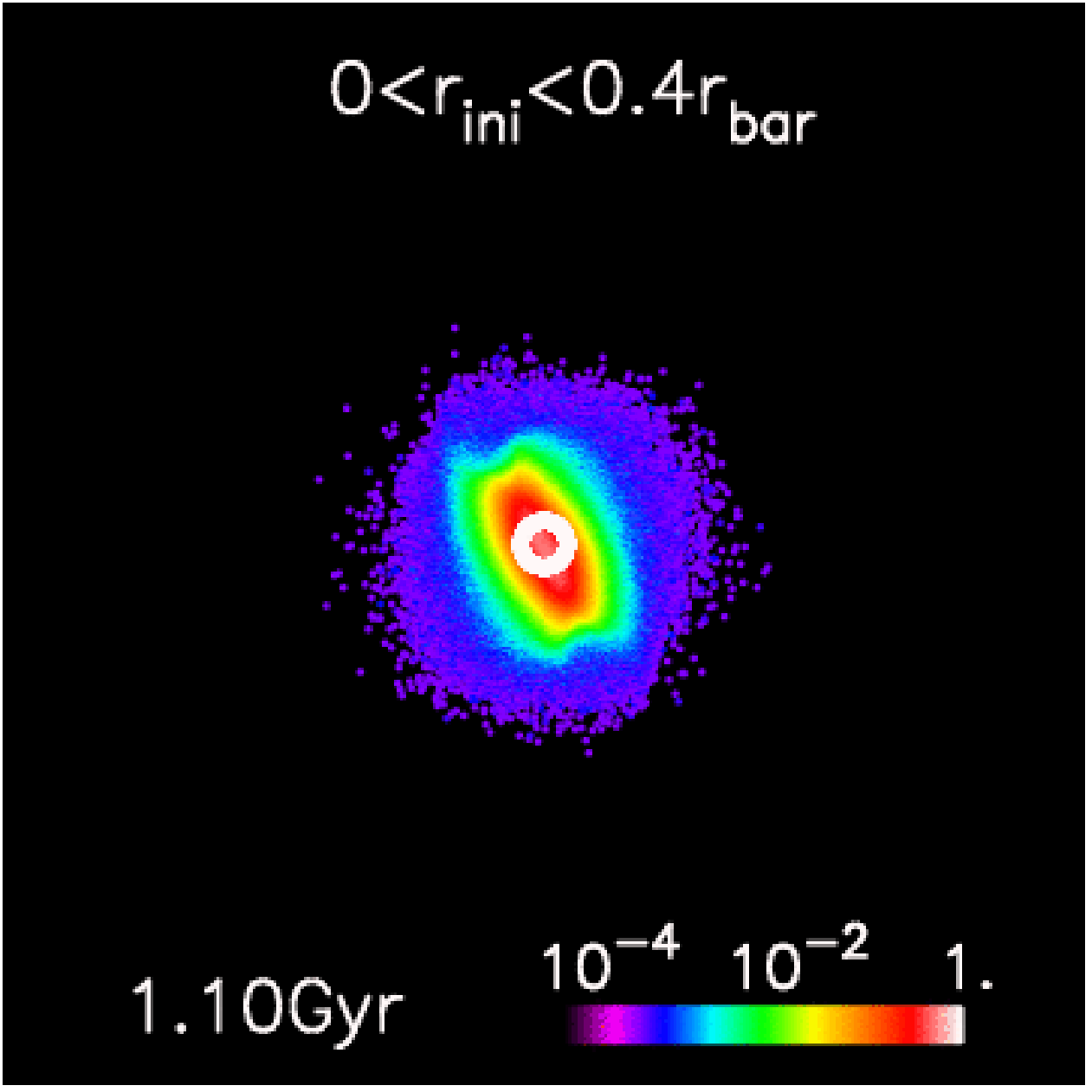}
\includegraphics[width=3.cm,angle=270]{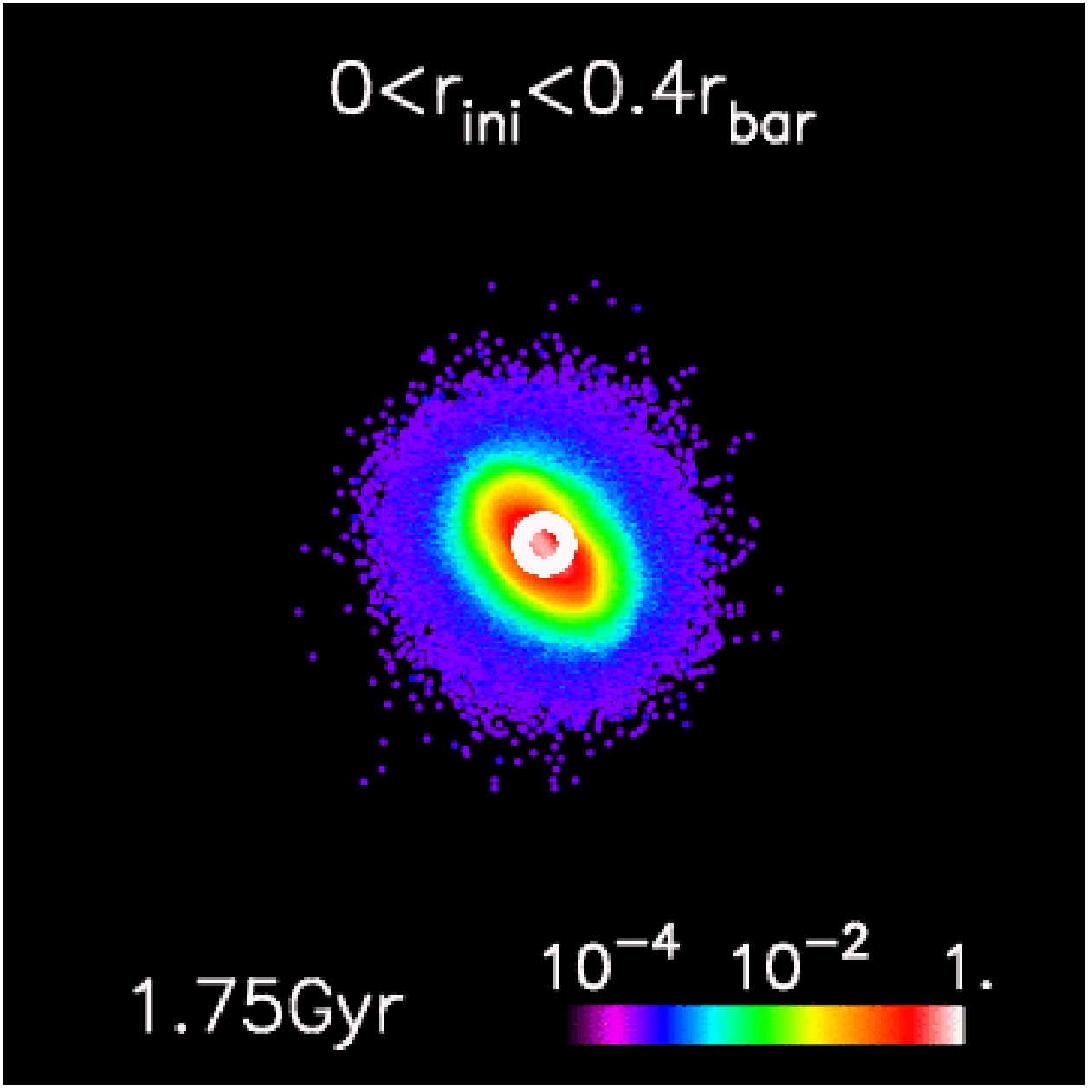}
\includegraphics[width=3.cm,angle=270]{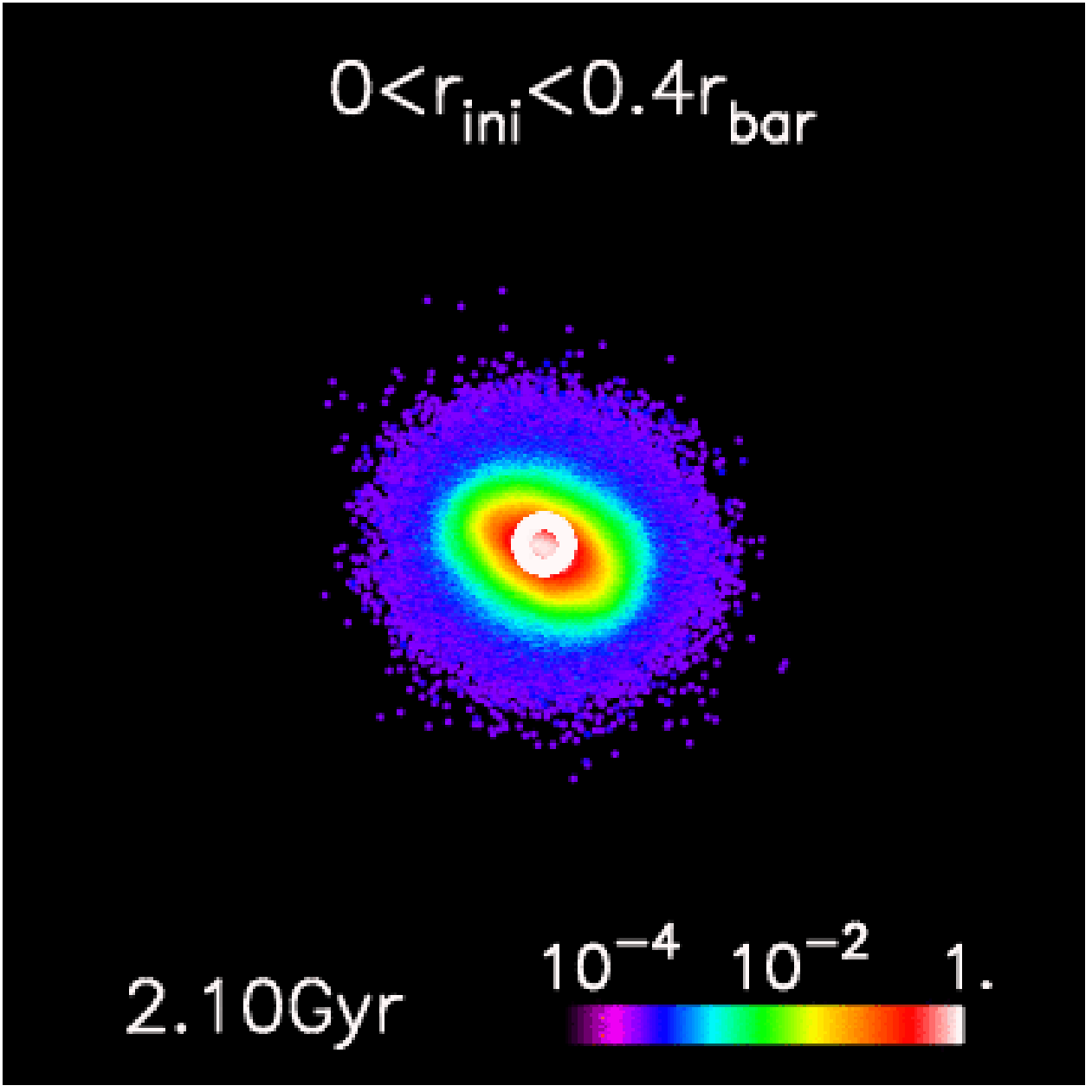}
\includegraphics[width=3.cm,angle=270]{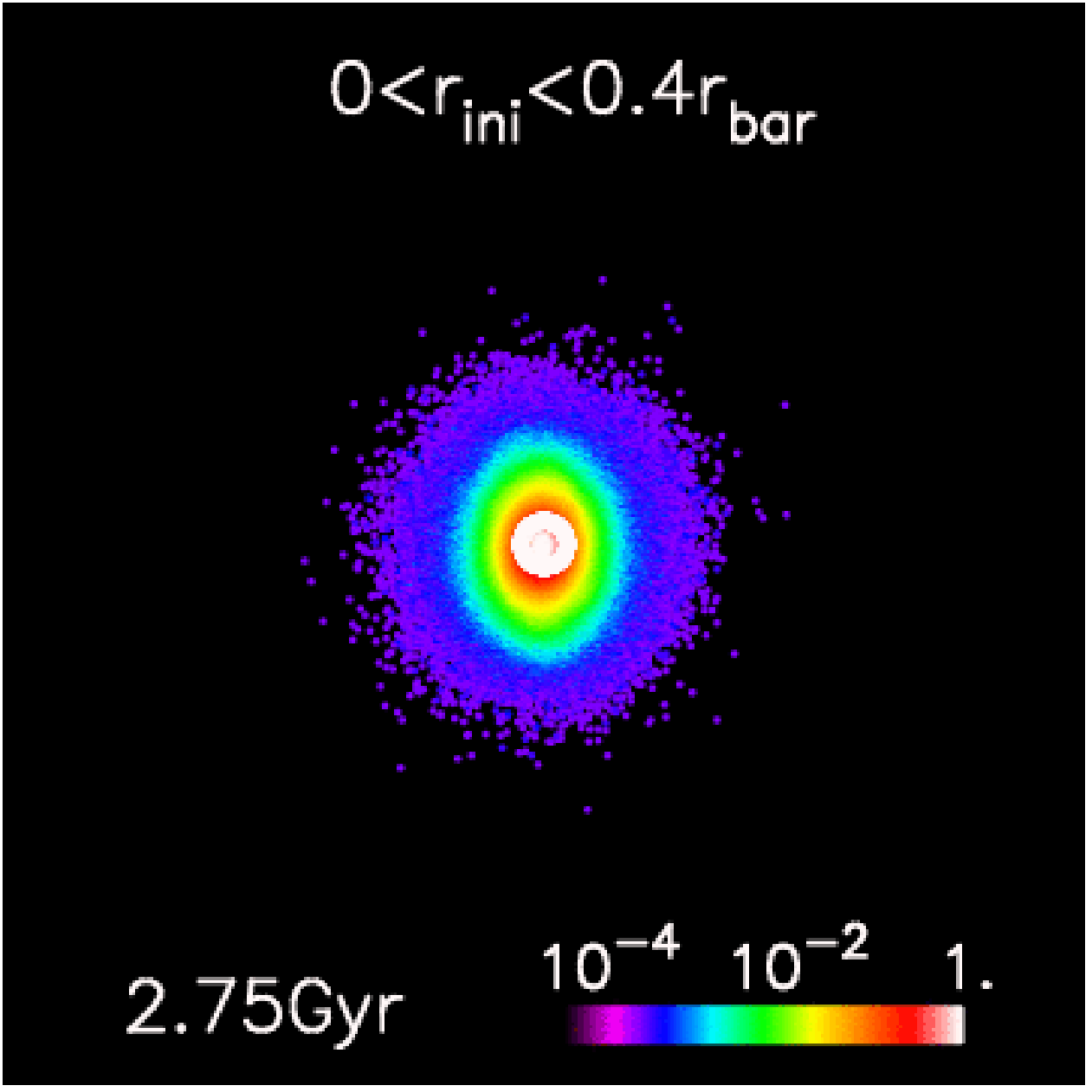}

\includegraphics[width=3.cm,angle=270]{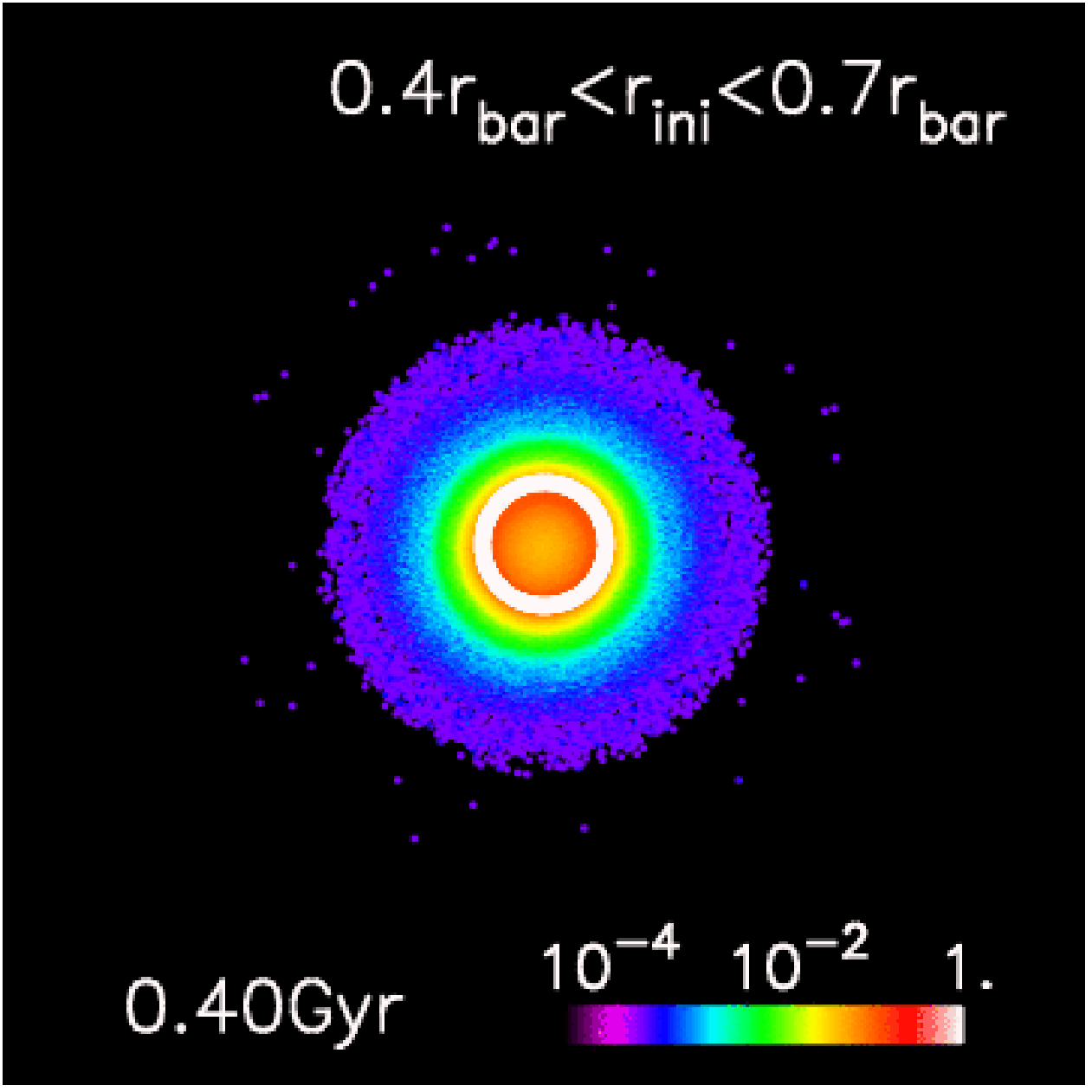}
\includegraphics[width=3.cm,angle=270]{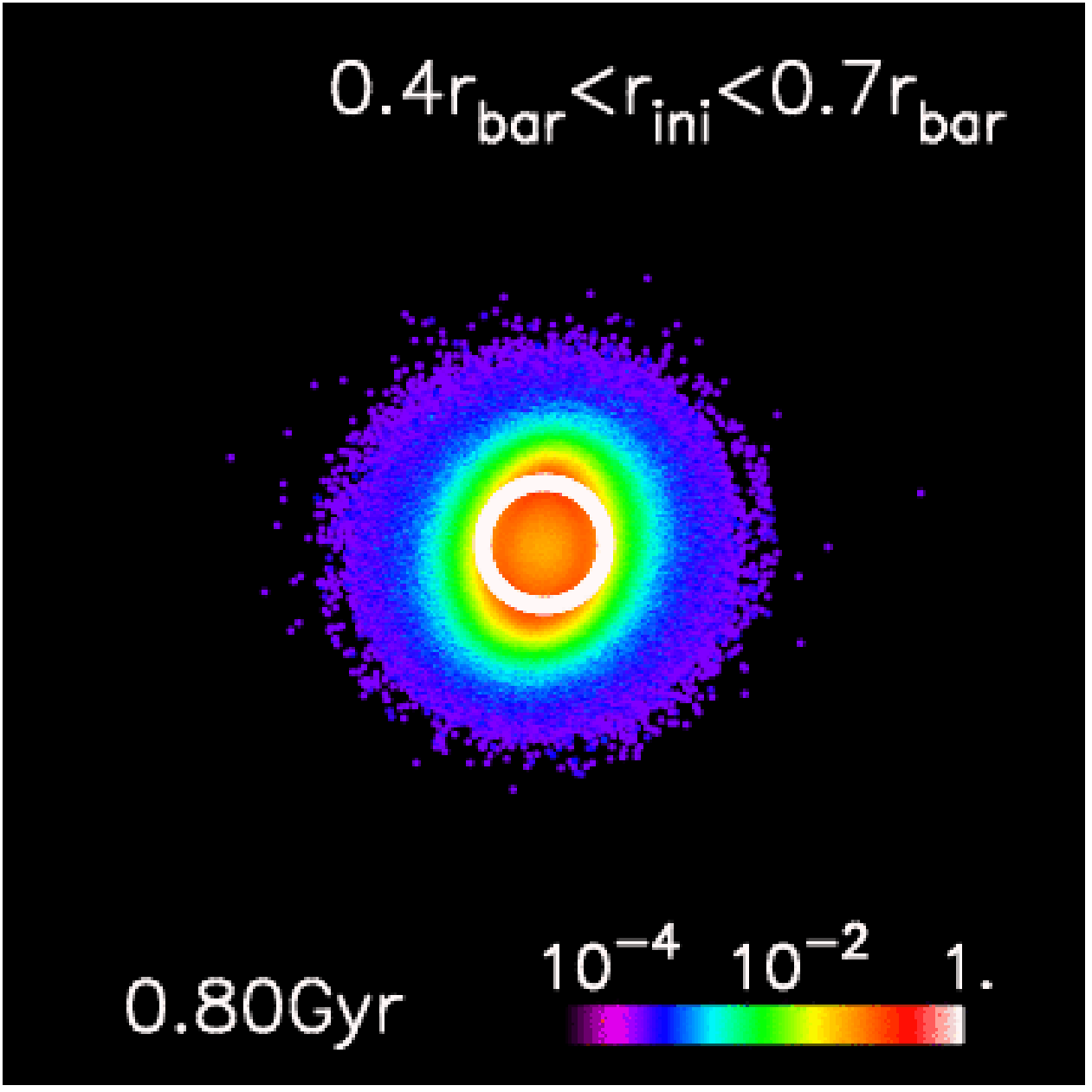}
\includegraphics[width=3.cm,angle=270]{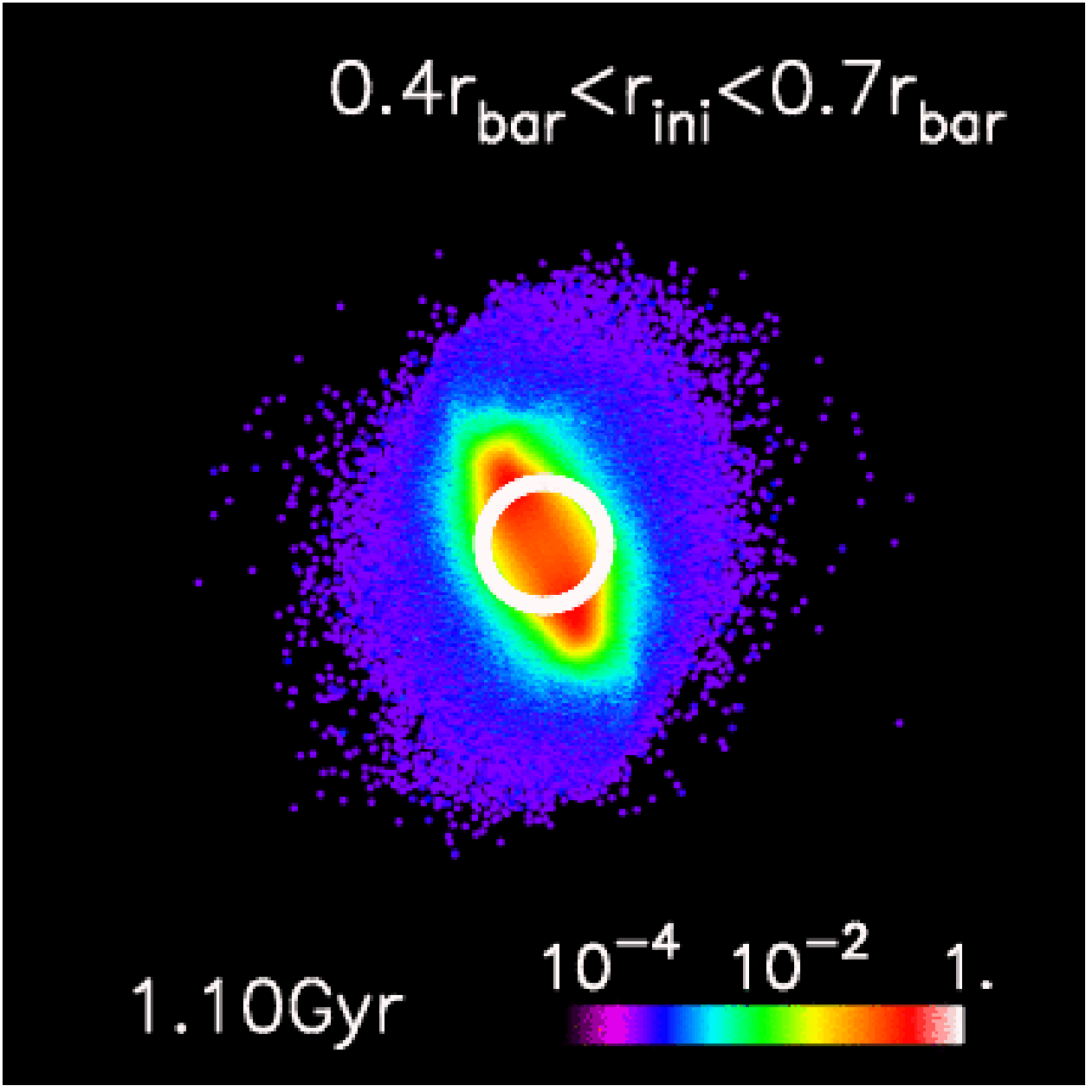}
\includegraphics[width=3.cm,angle=270]{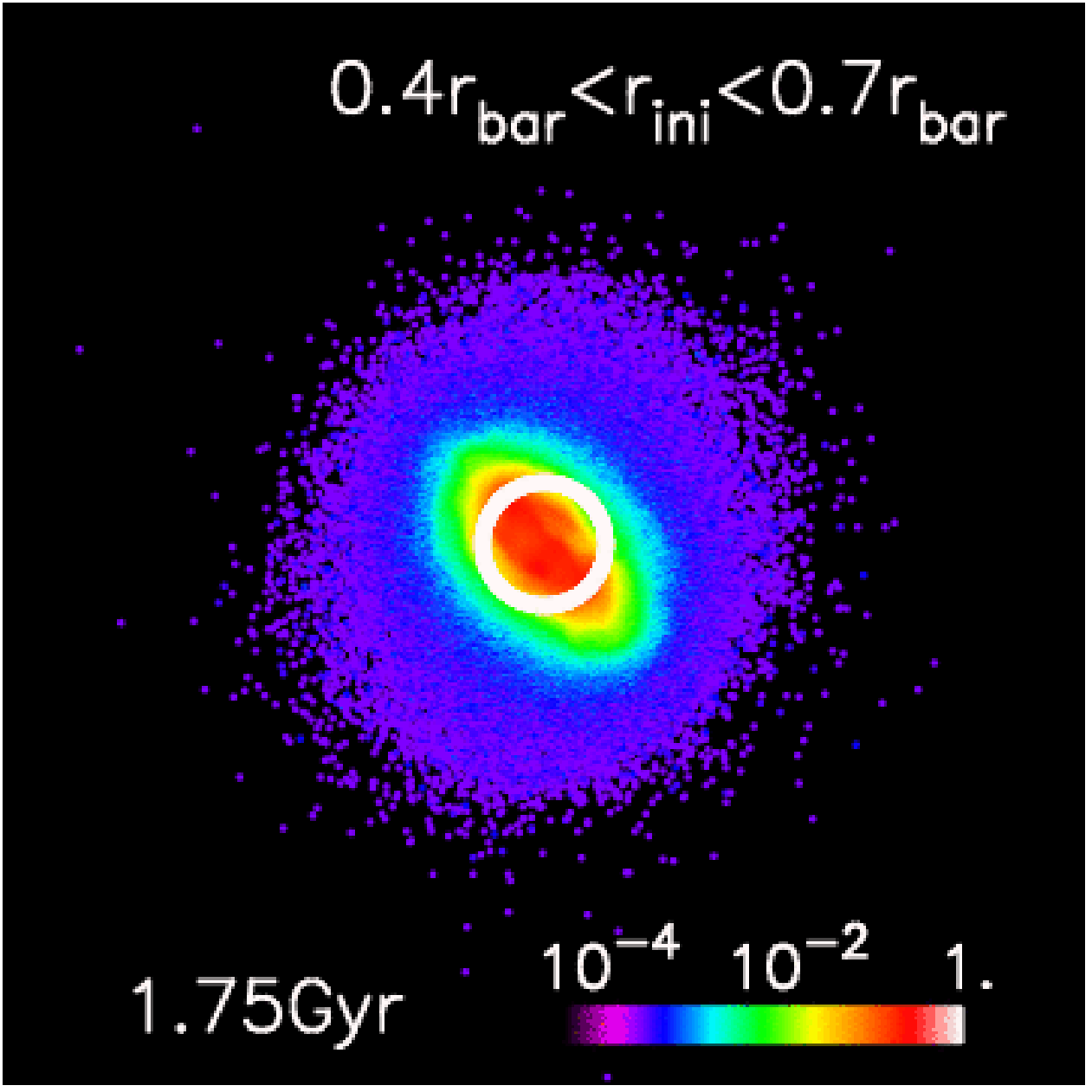}
\includegraphics[width=3.cm,angle=270]{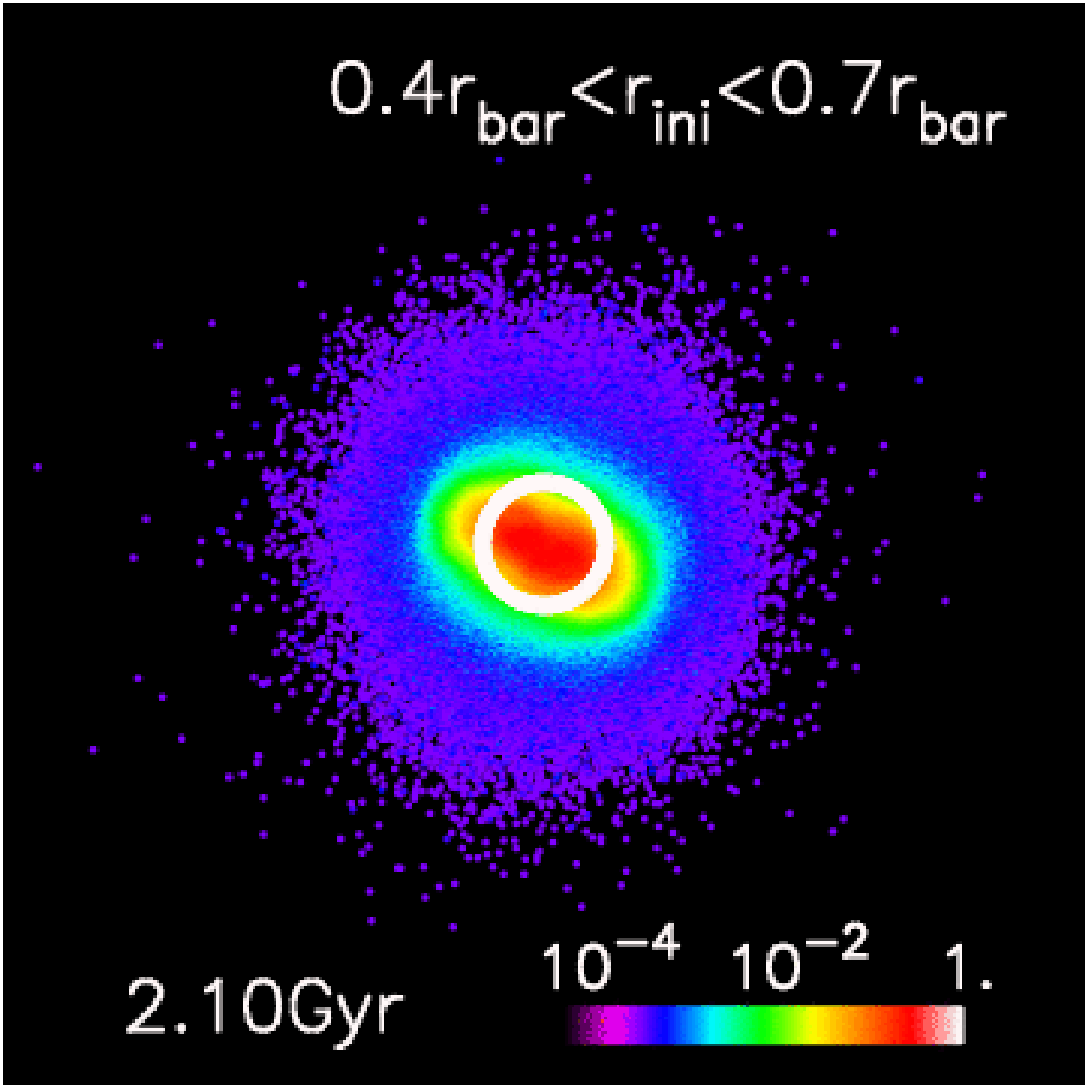}
\includegraphics[width=3.cm,angle=270]{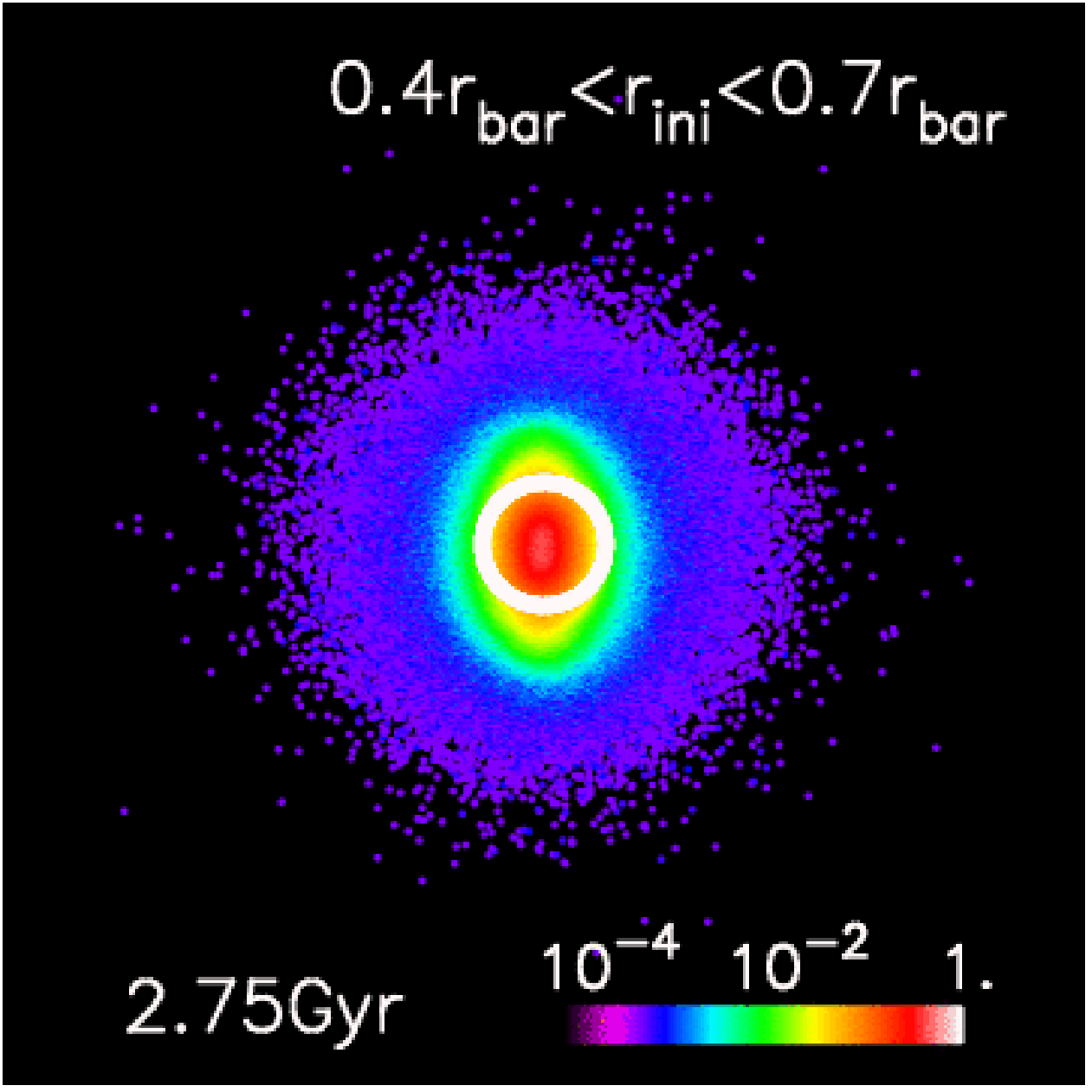}

\includegraphics[width=3.cm,angle=270]{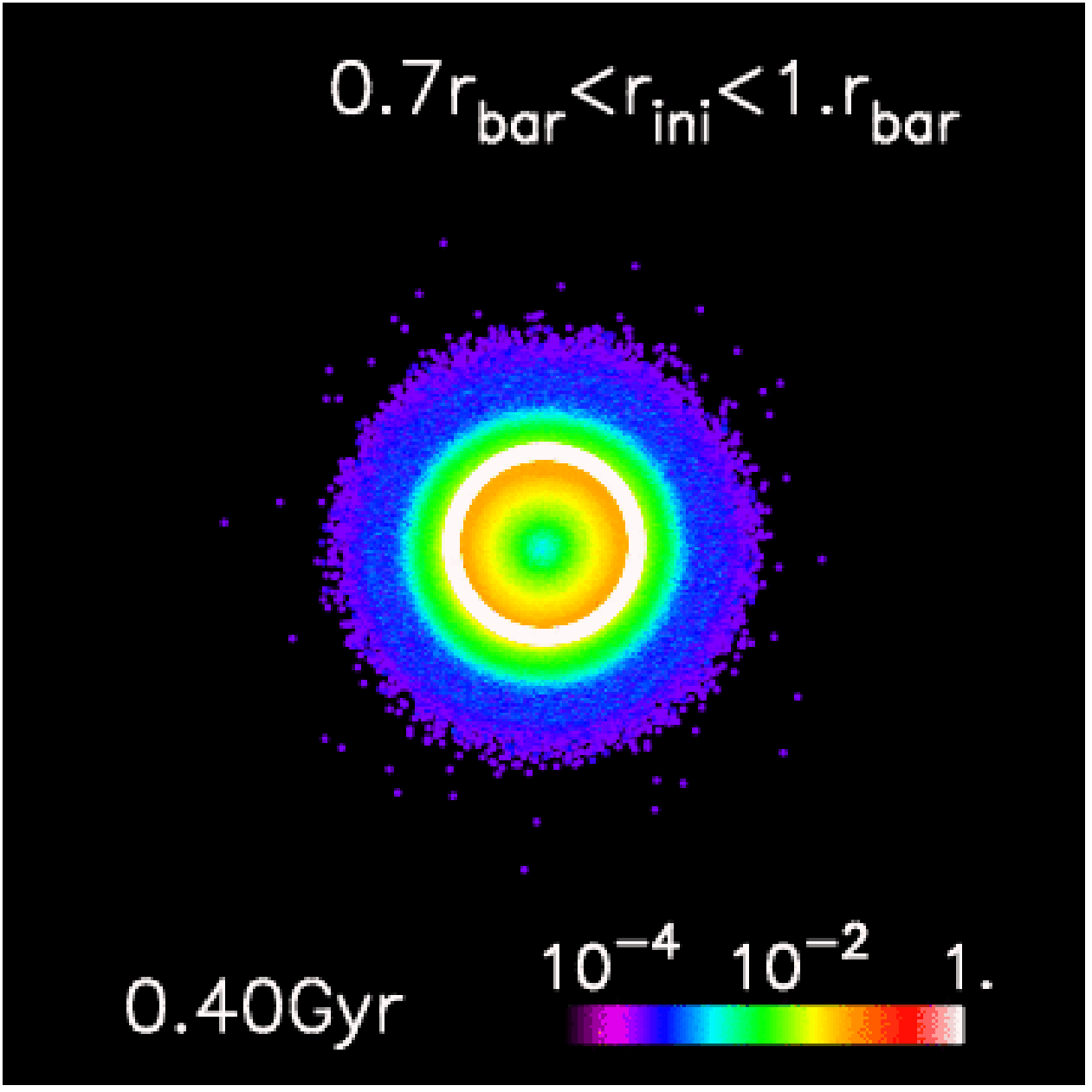}
\includegraphics[width=3.cm,angle=270]{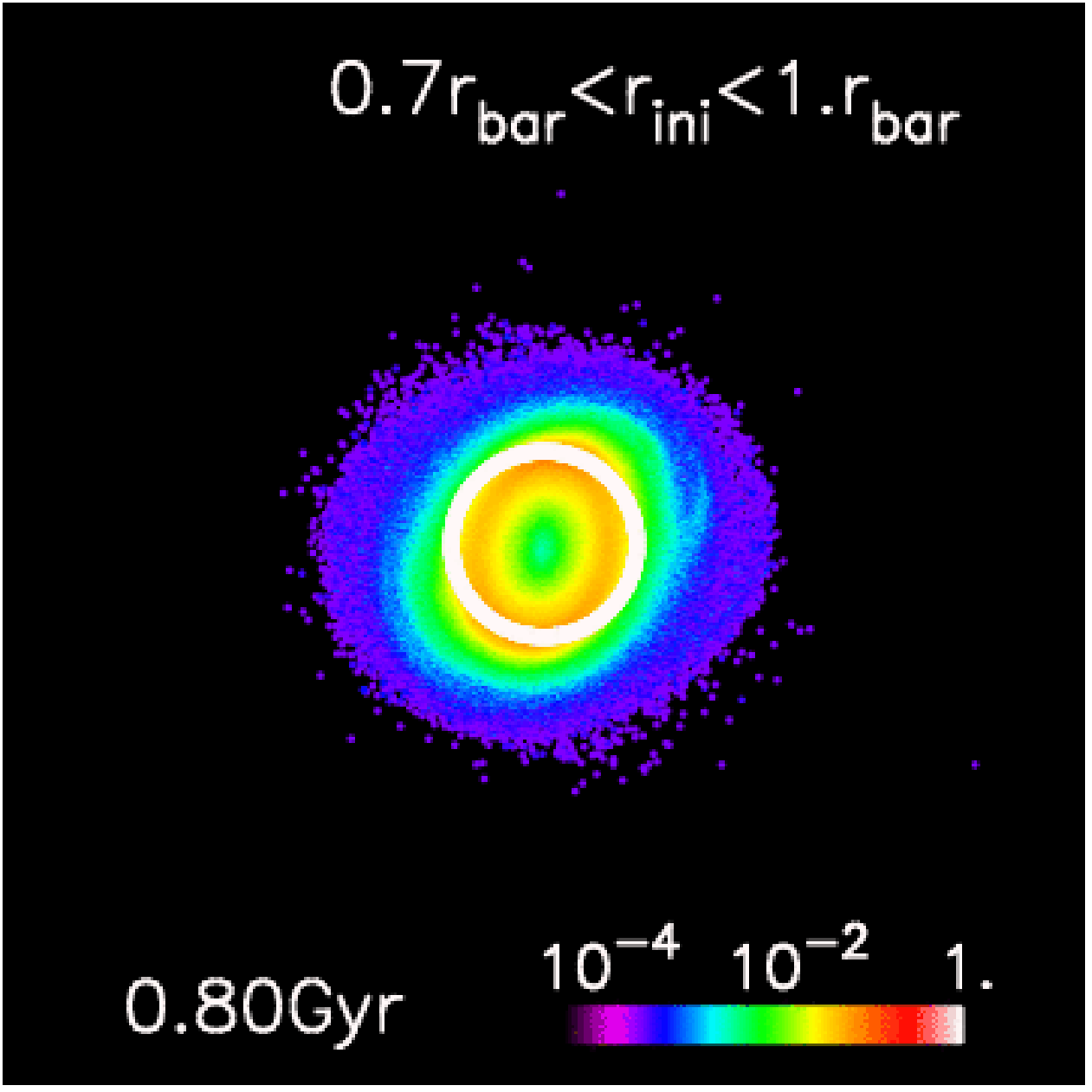}
\includegraphics[width=3.cm,angle=270]{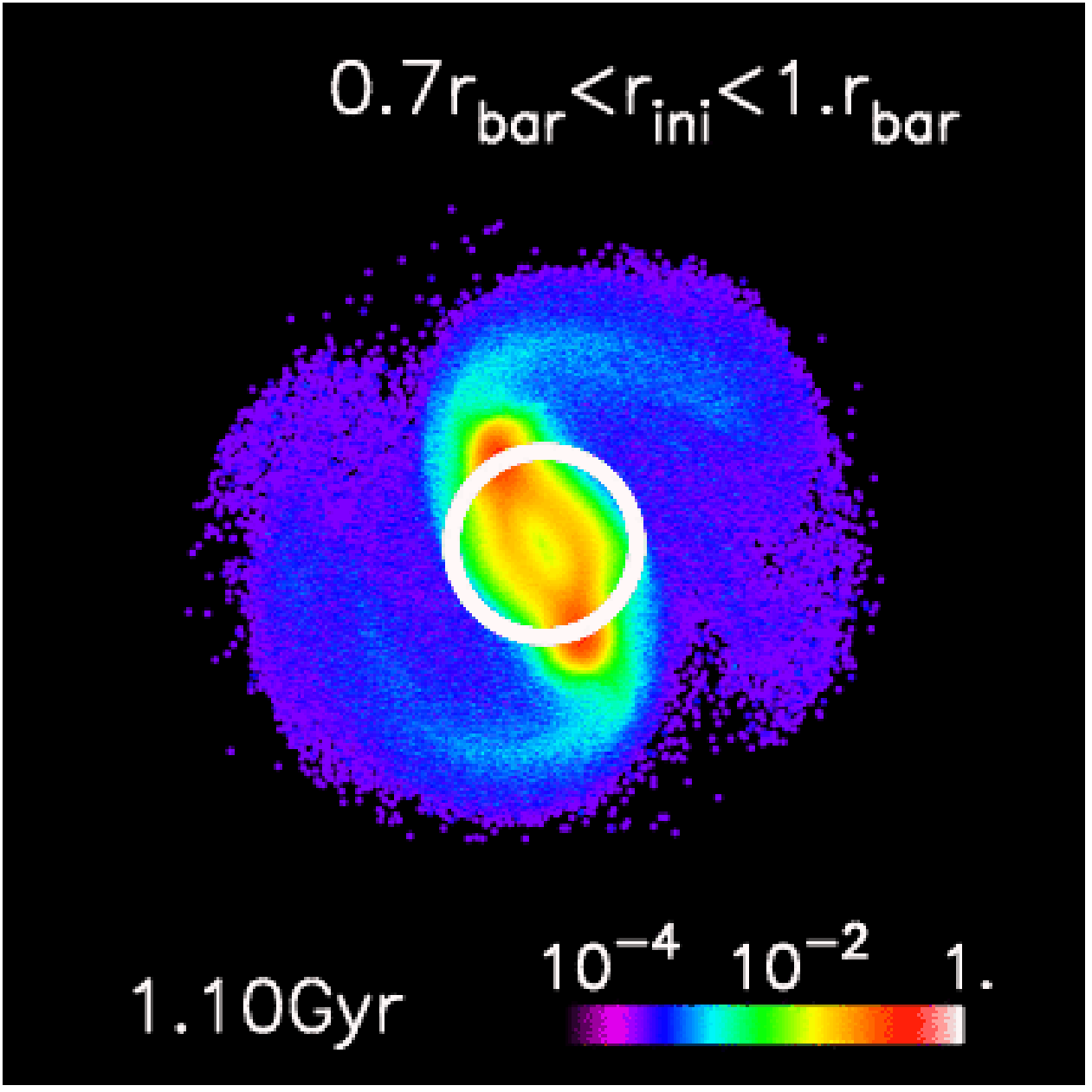}
\includegraphics[width=3.cm,angle=270]{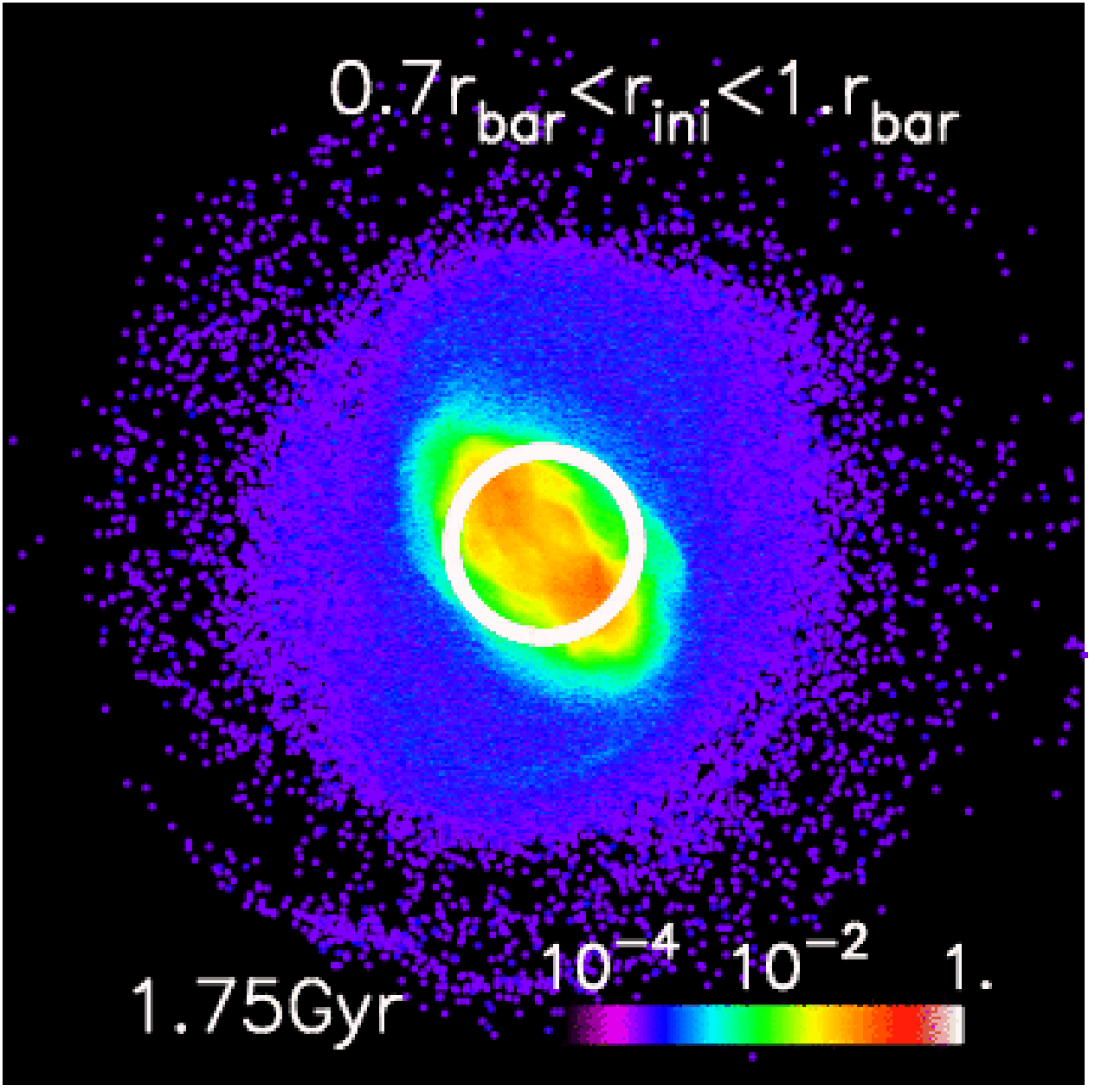}
\includegraphics[width=3.cm,angle=270]{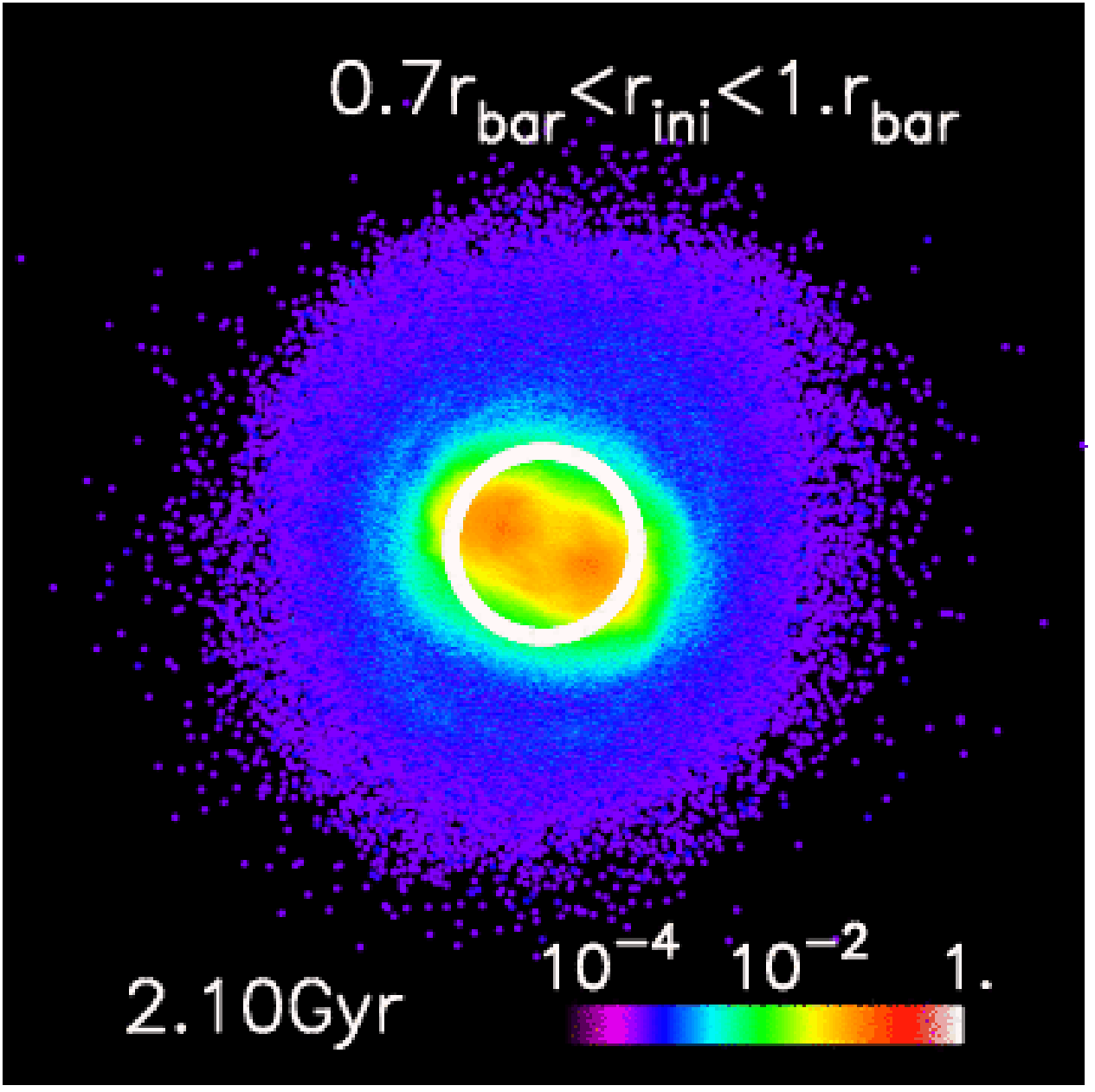}
\includegraphics[width=3.cm,angle=270]{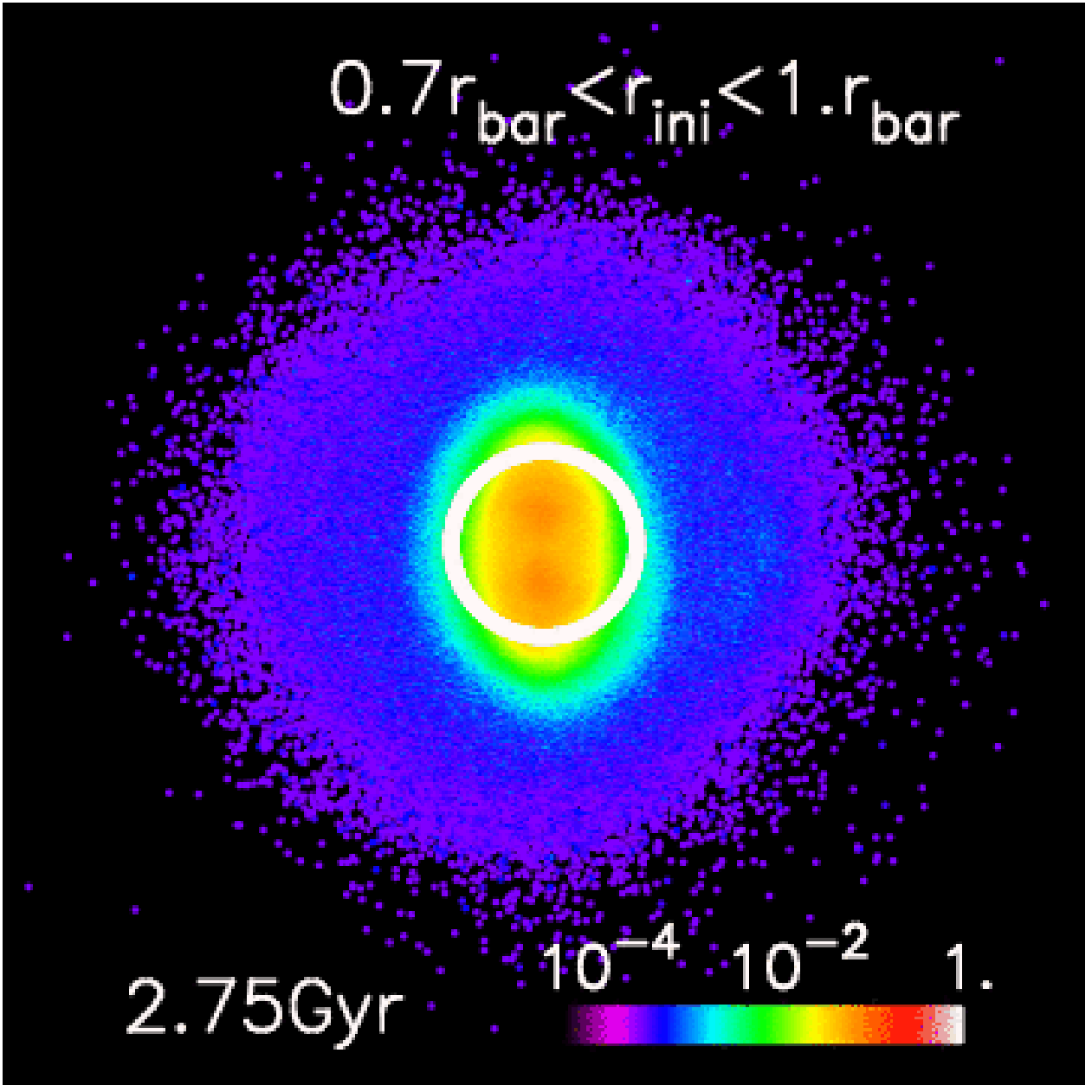}

\includegraphics[width=3.cm,angle=270]{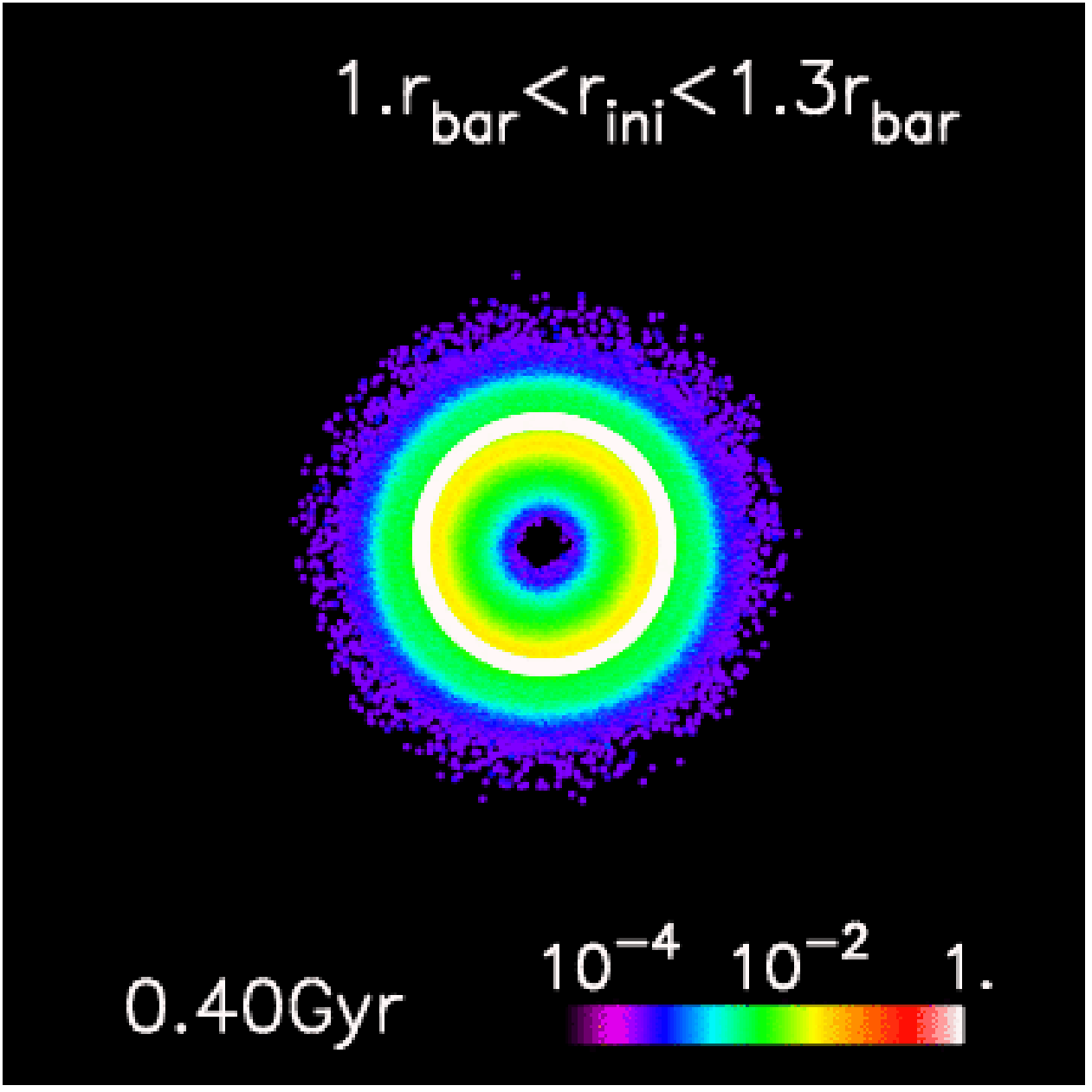}
\includegraphics[width=3.cm,angle=270]{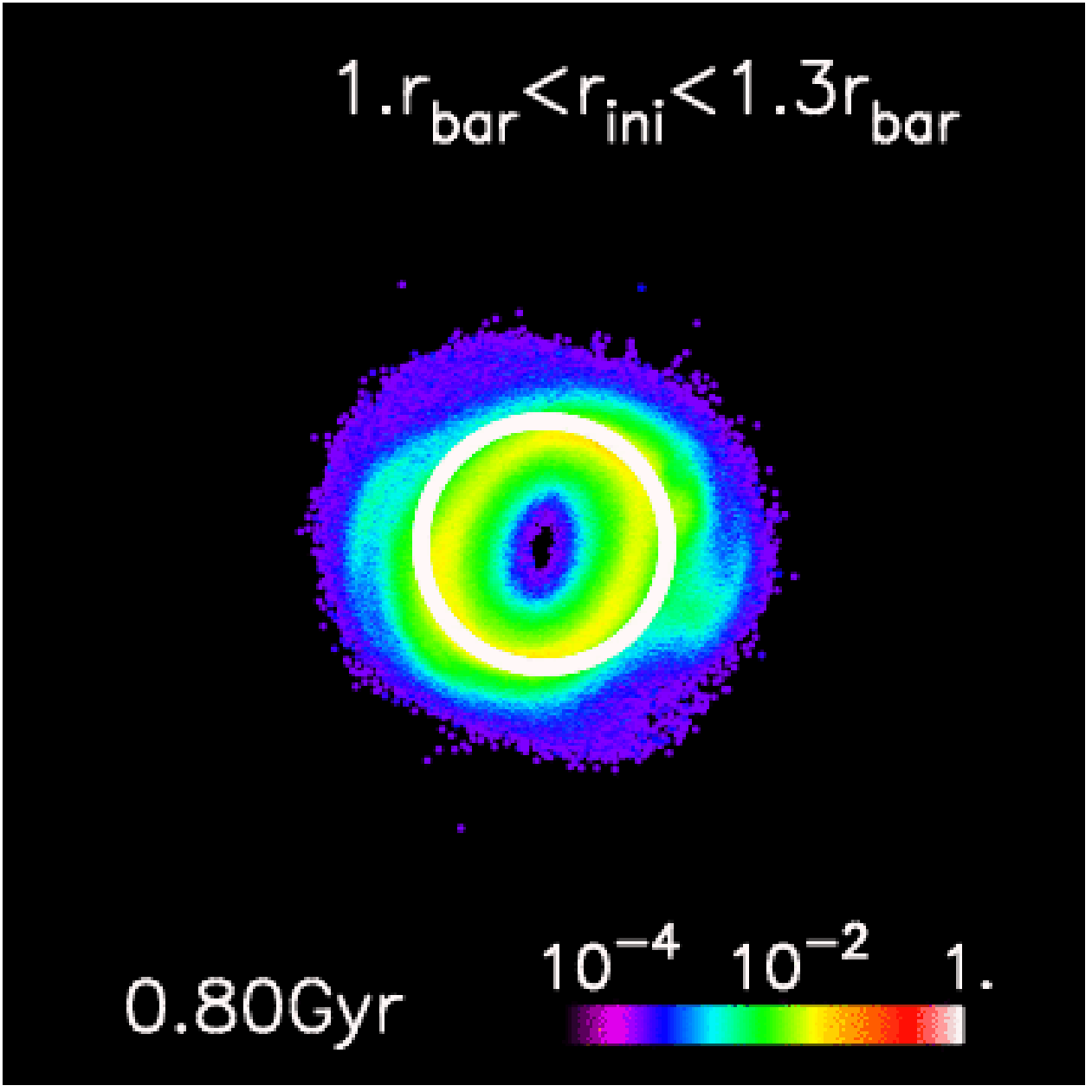}
\includegraphics[width=3.cm,angle=270]{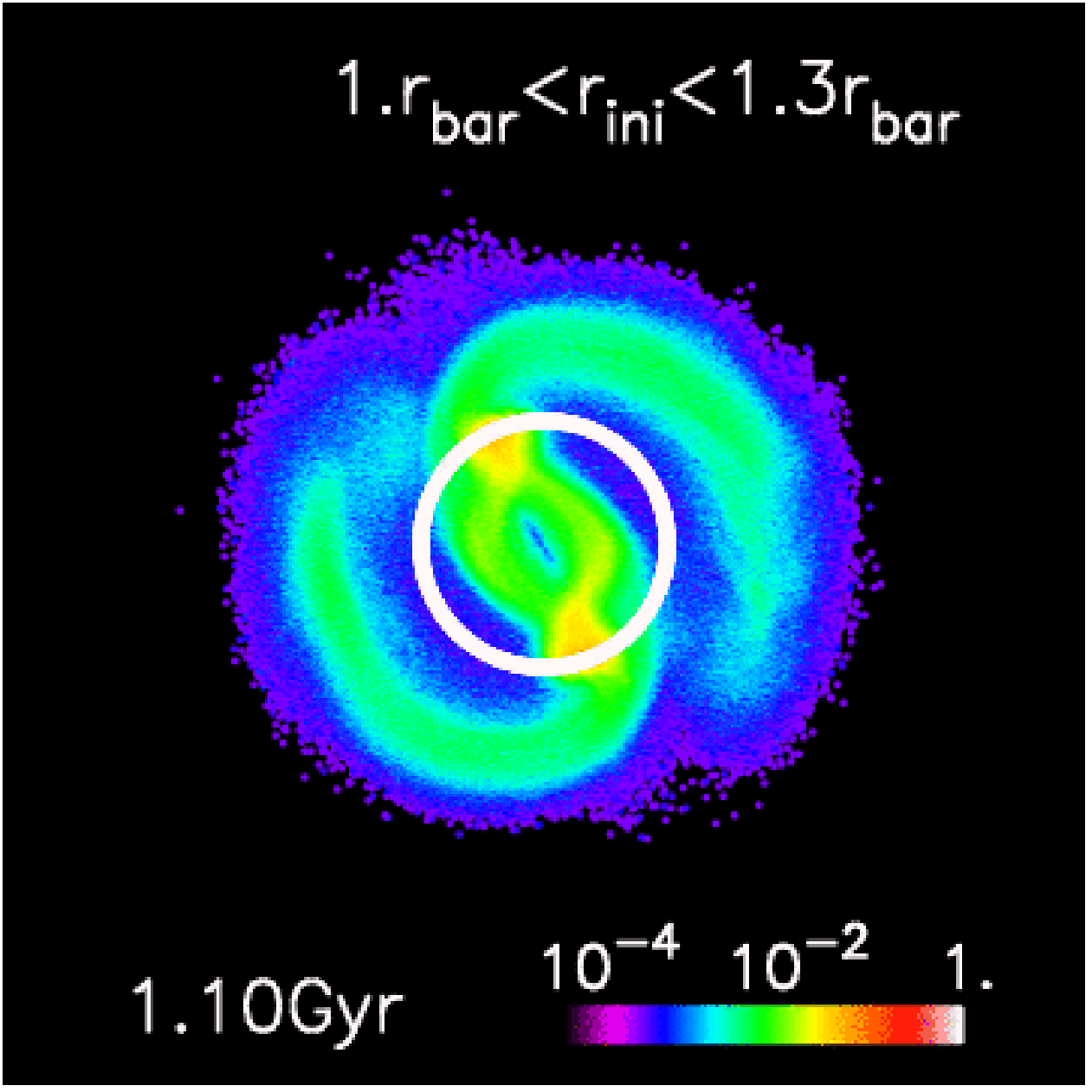}
\includegraphics[width=3.cm,angle=270]{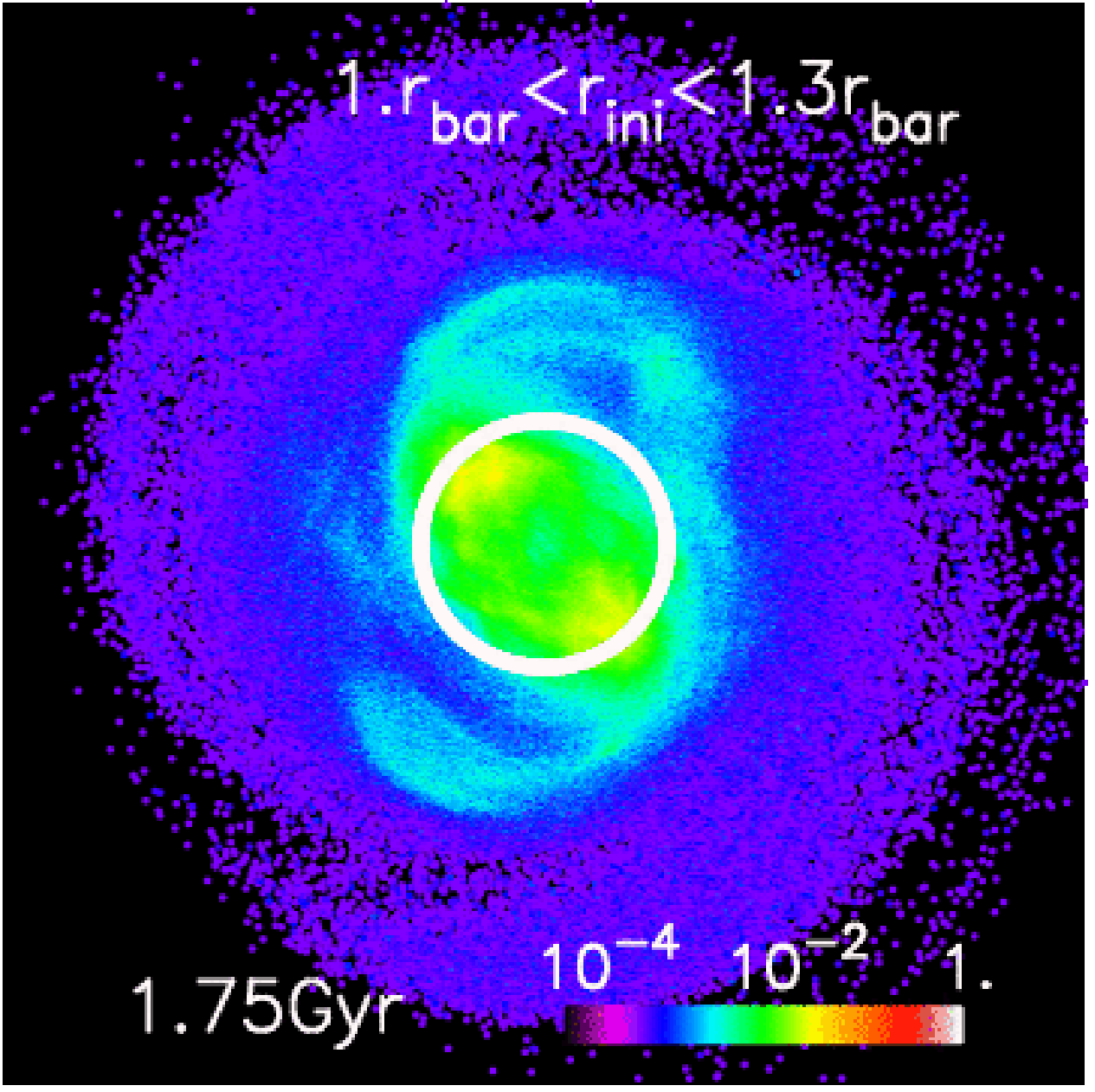}
\includegraphics[width=3.cm,angle=270]{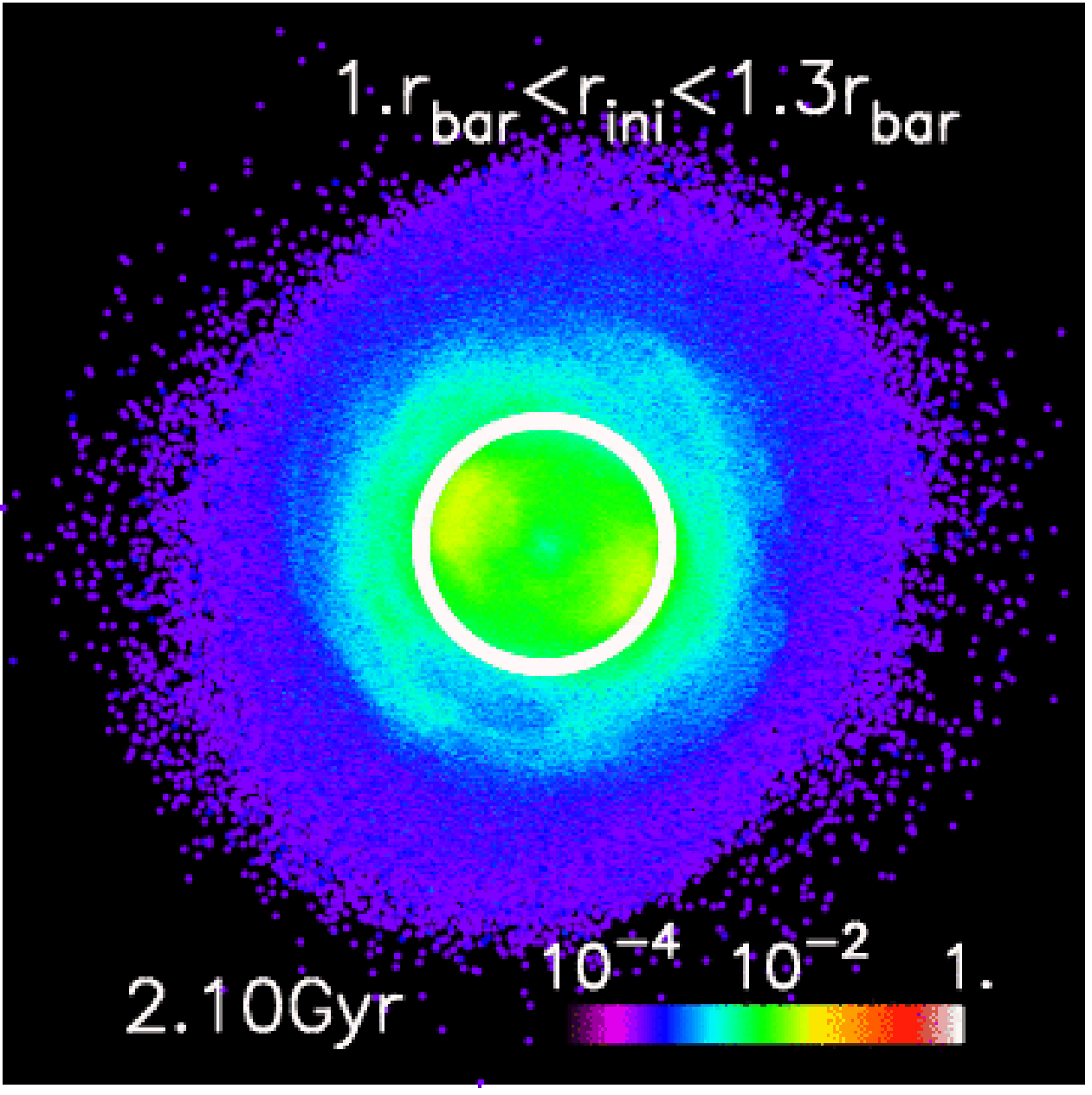}
\includegraphics[width=3.cm,angle=270]{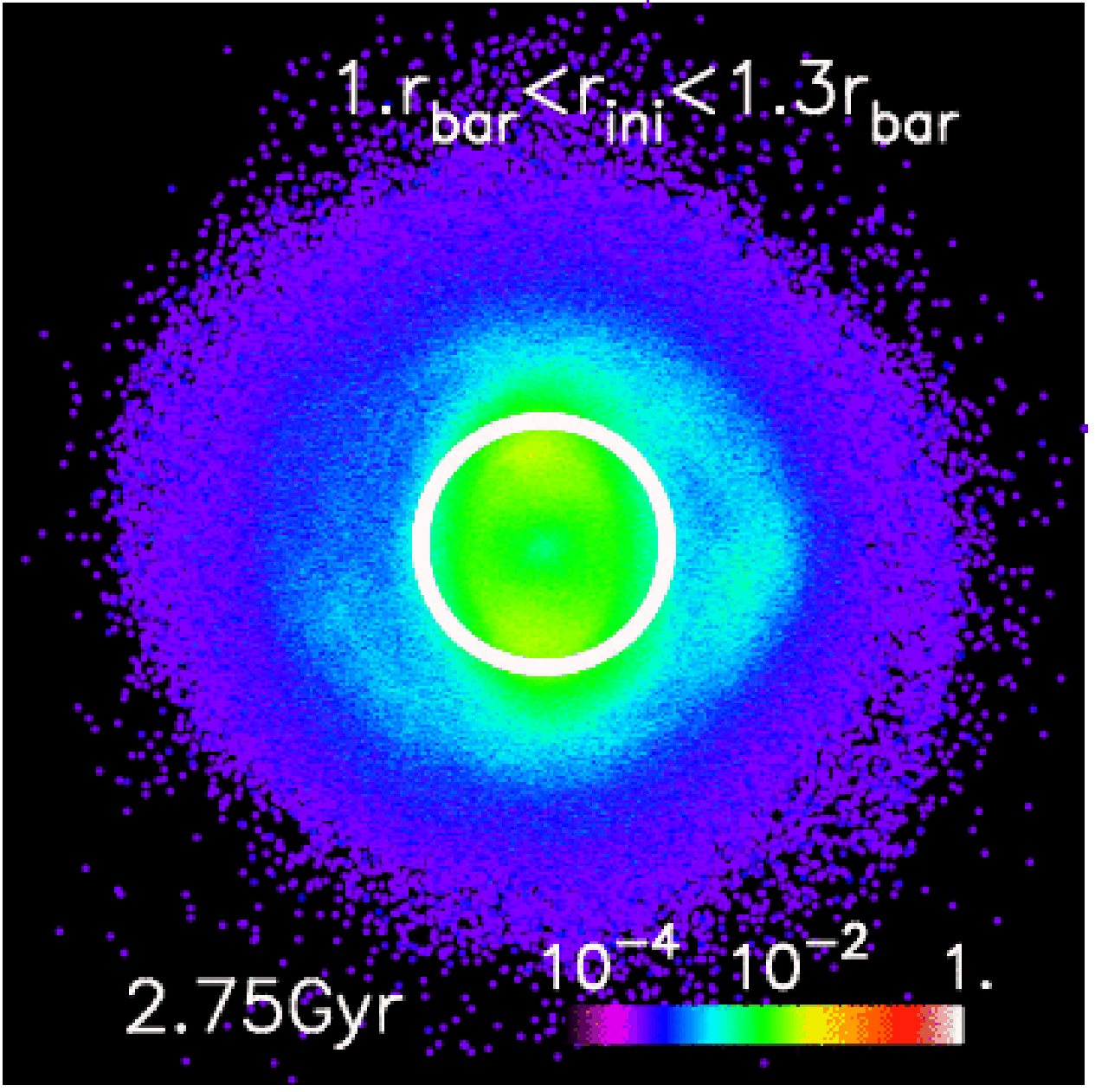}

\includegraphics[width=3.cm,angle=270]{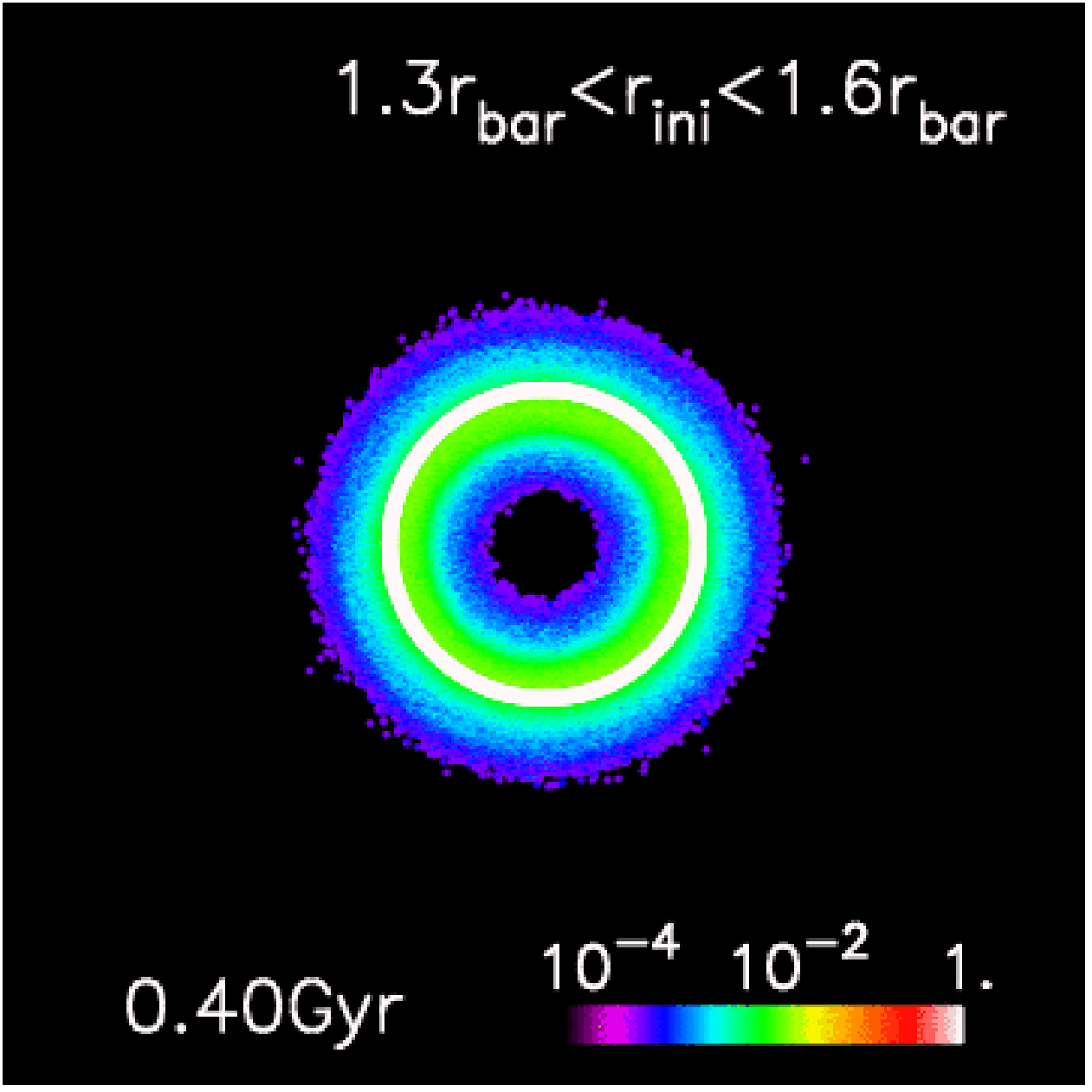}
\includegraphics[width=3.cm,angle=270]{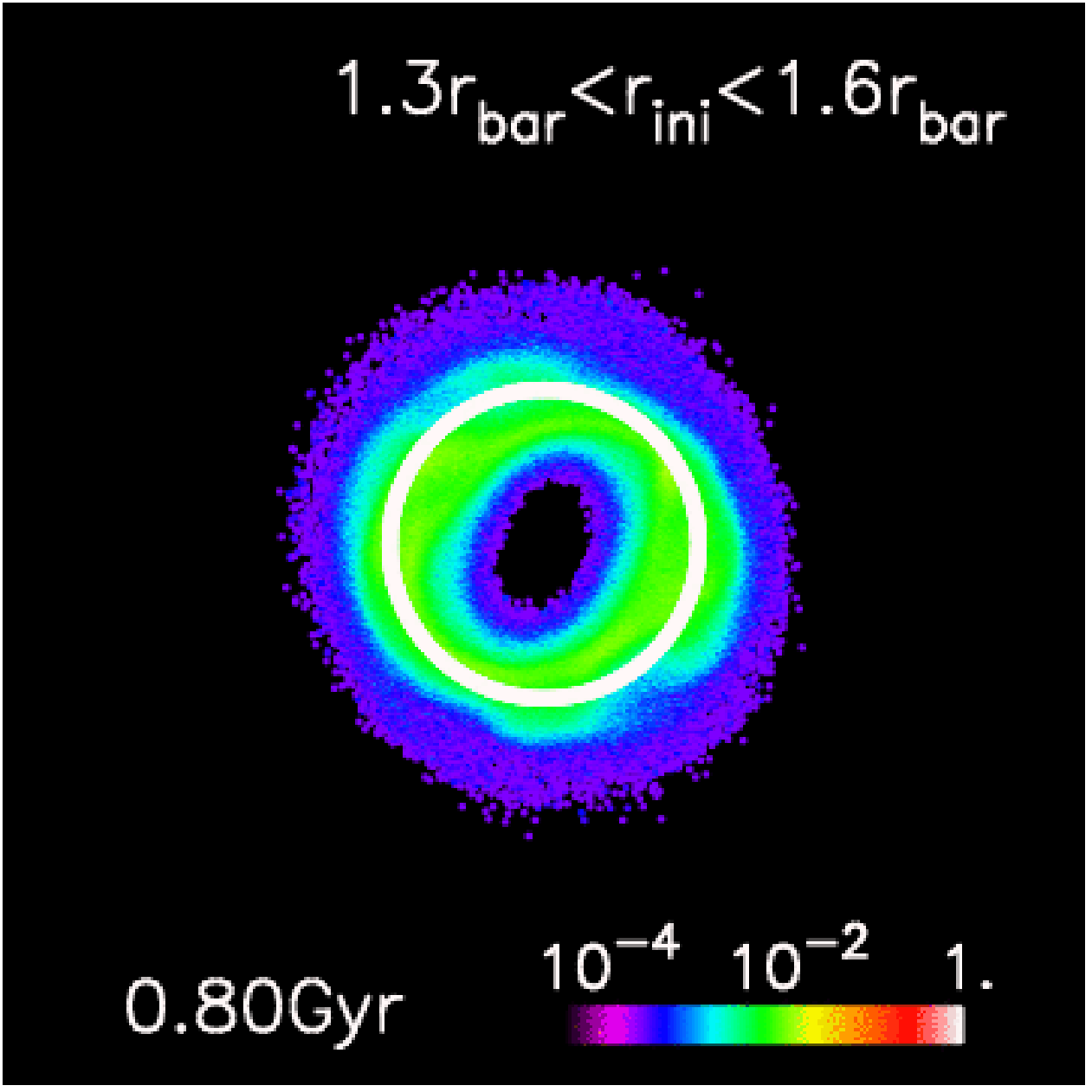}
\includegraphics[width=3.cm,angle=270]{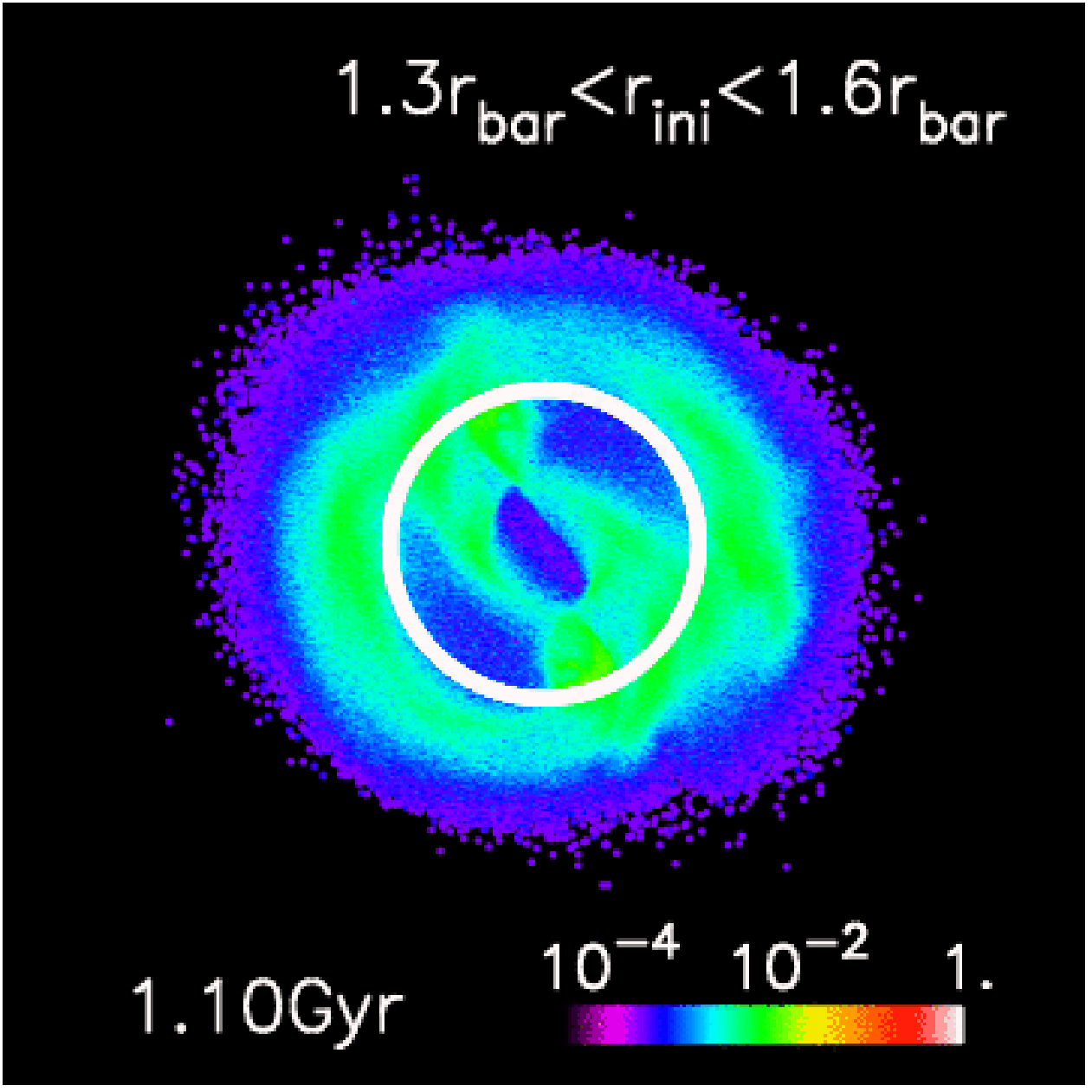}
\includegraphics[width=3.cm,angle=270]{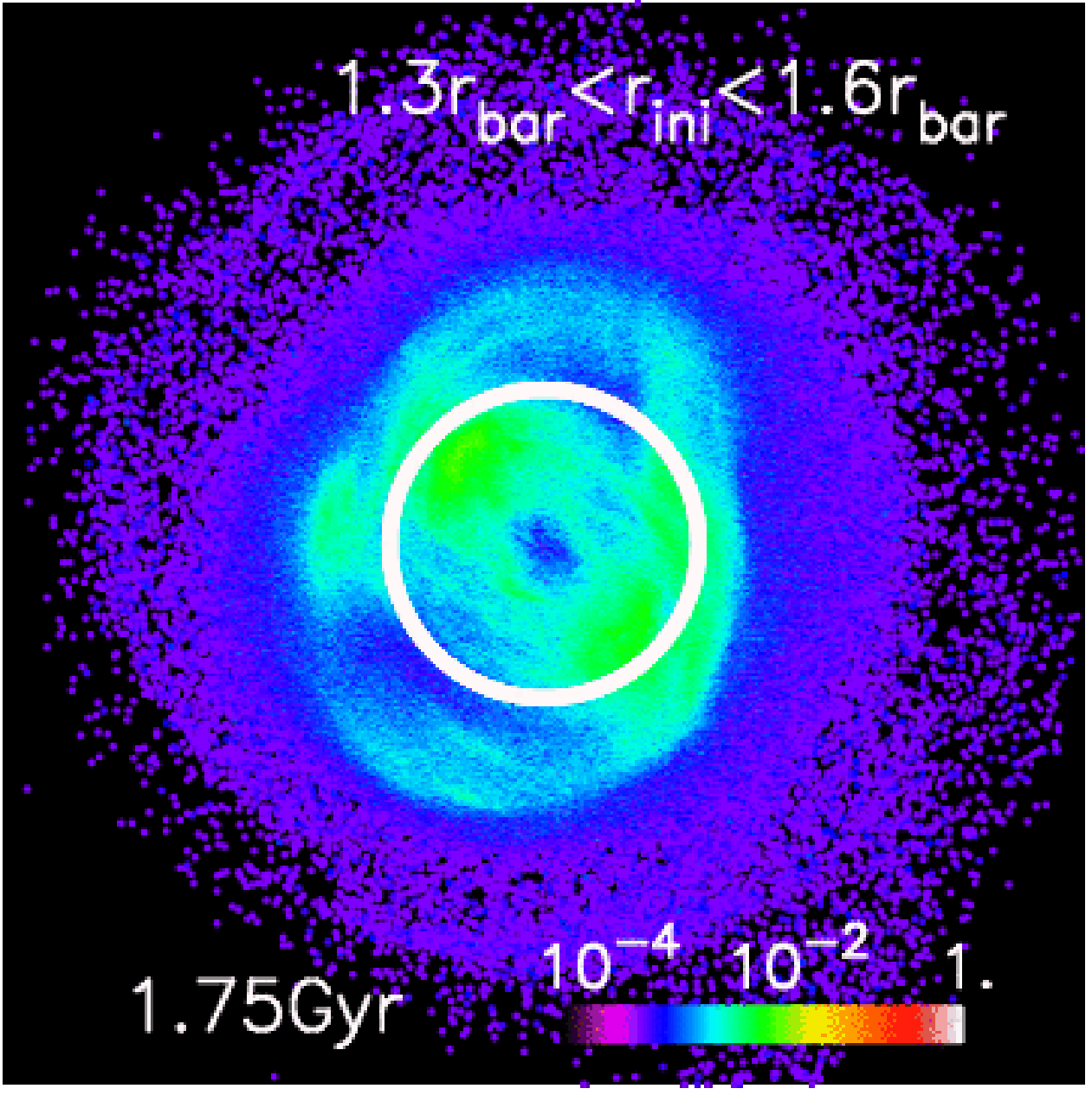}
\includegraphics[width=3.cm,angle=270]{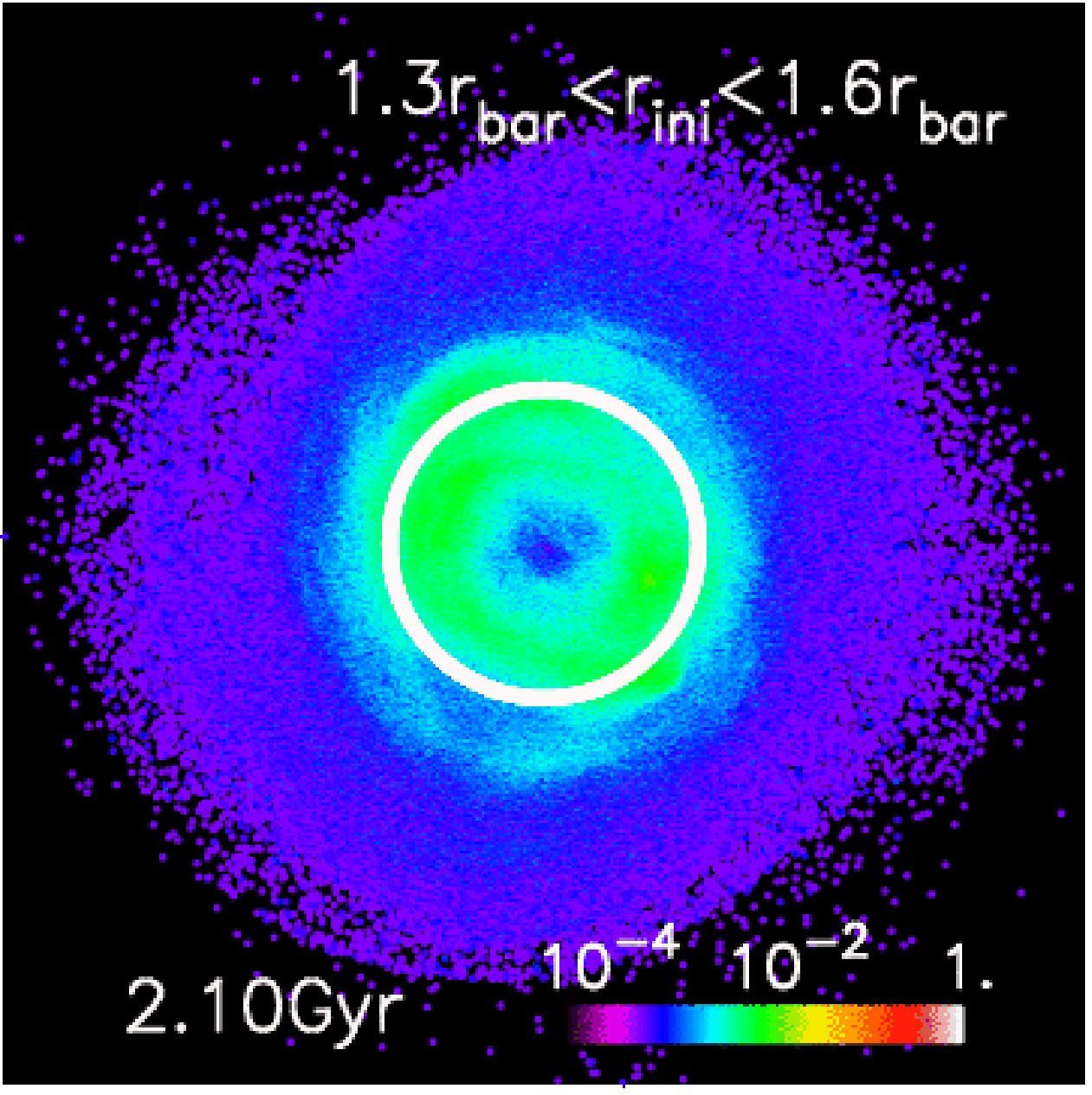}
\includegraphics[width=3.cm,angle=270]{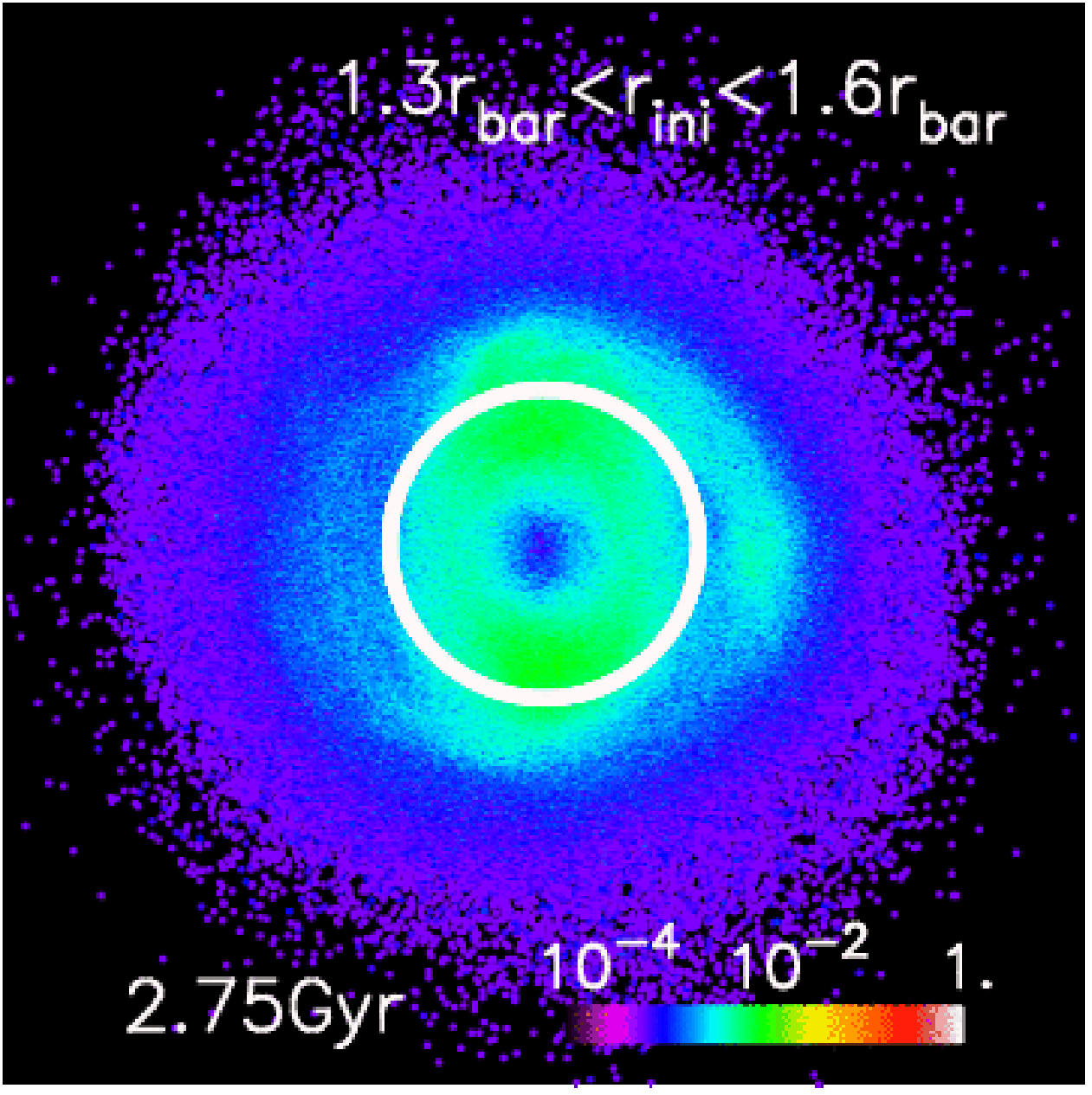}

\includegraphics[width=3.cm,angle=270]{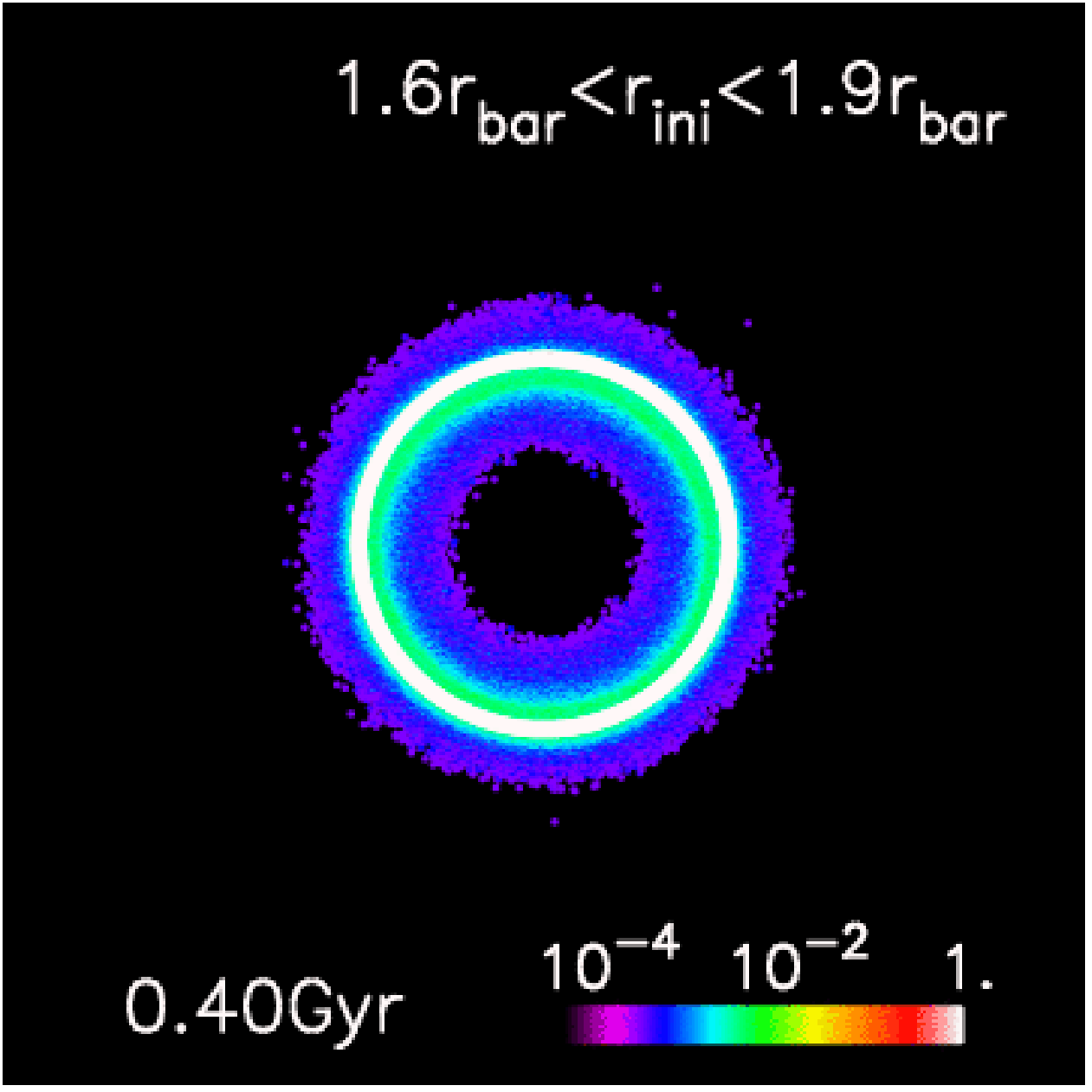}
\includegraphics[width=3.cm,angle=270]{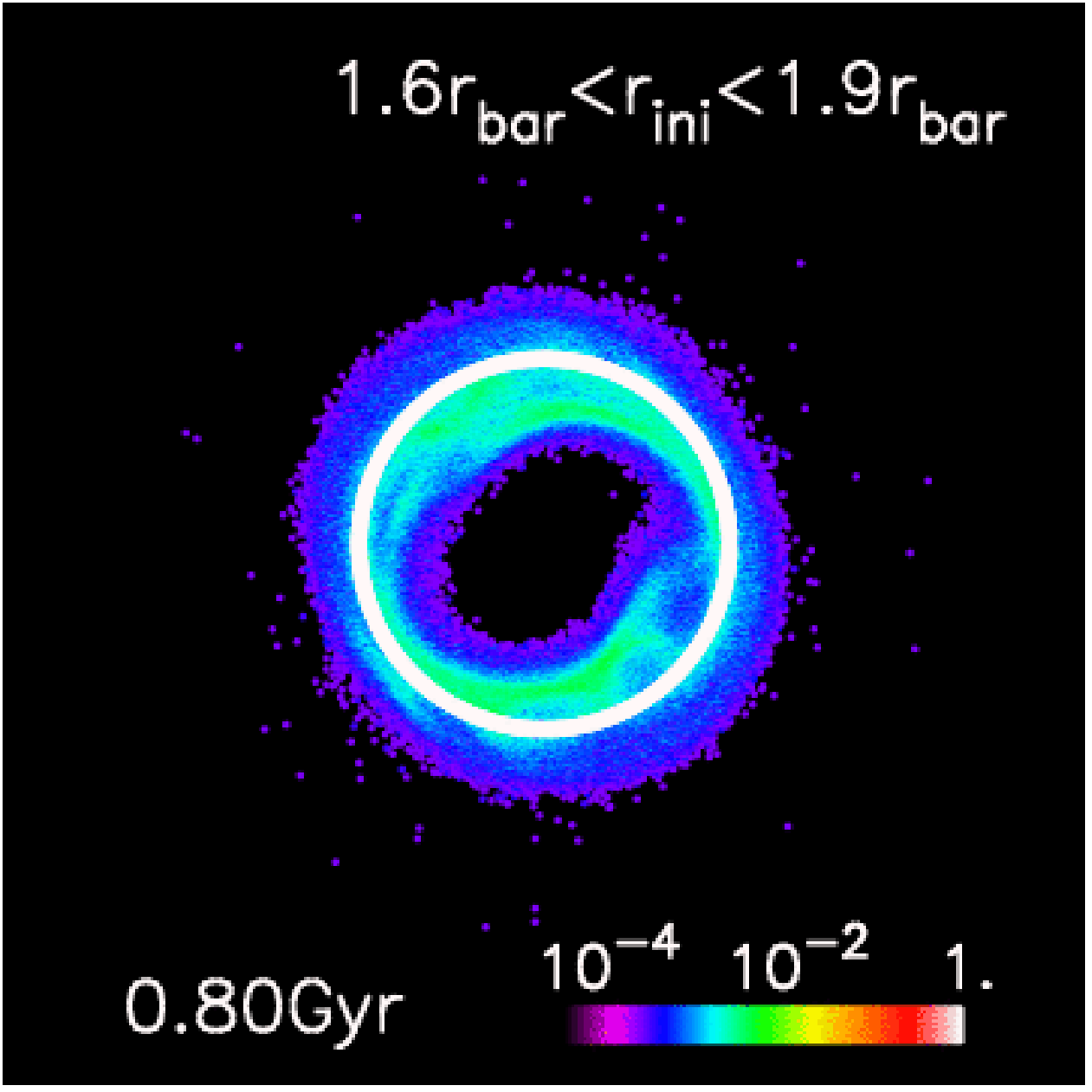}
\includegraphics[width=3.cm,angle=270]{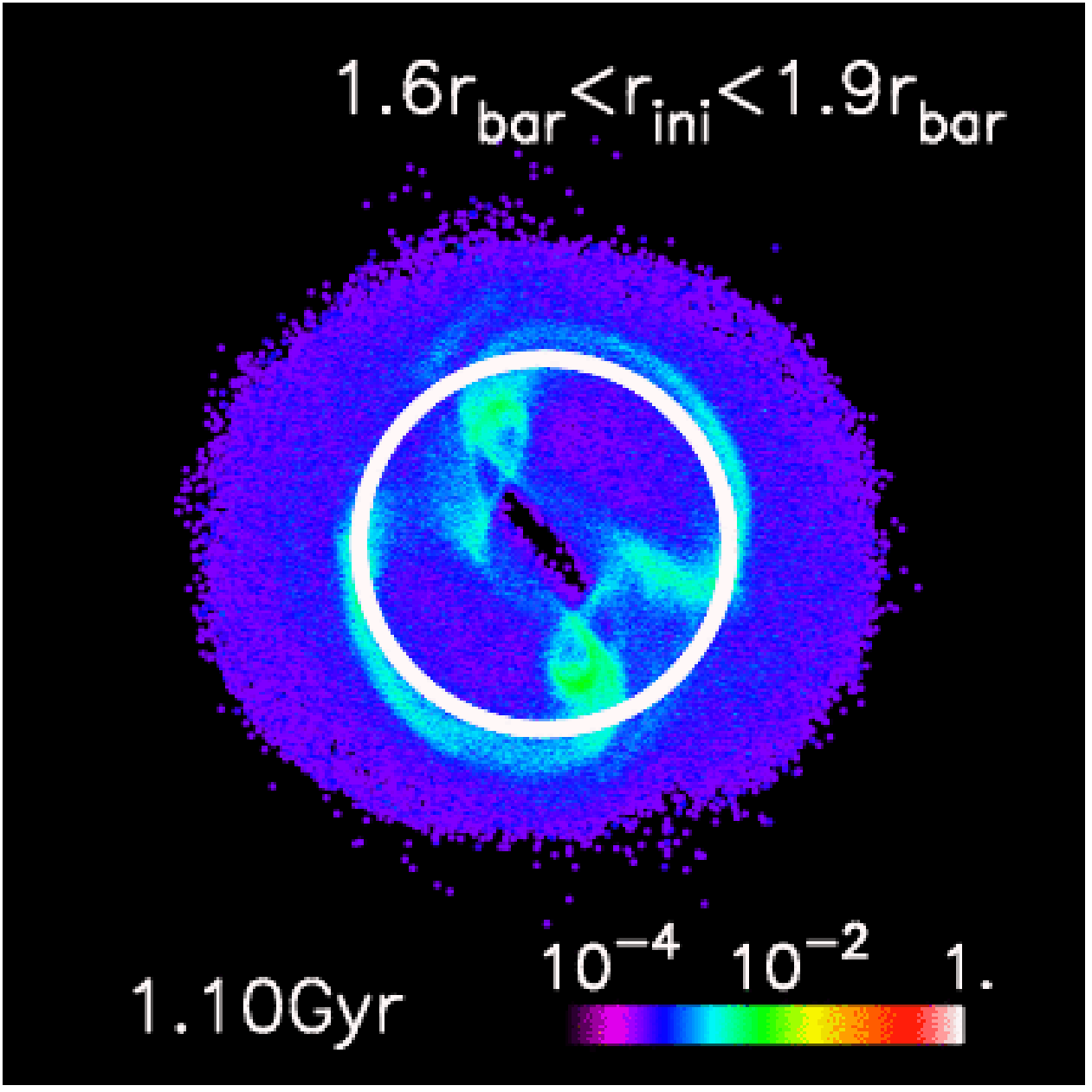}
\includegraphics[width=3.cm,angle=270]{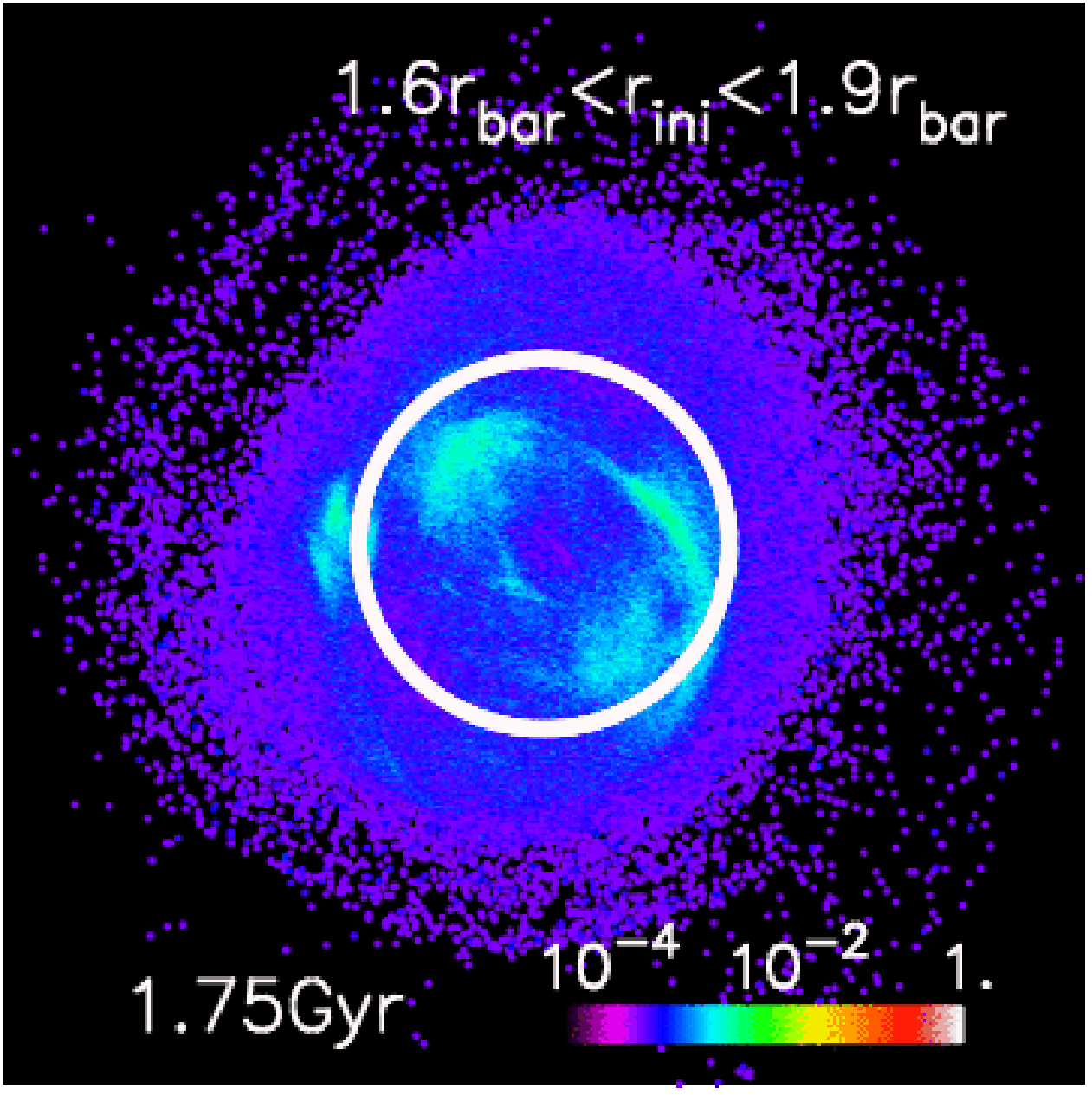}
\includegraphics[width=3.cm,angle=270]{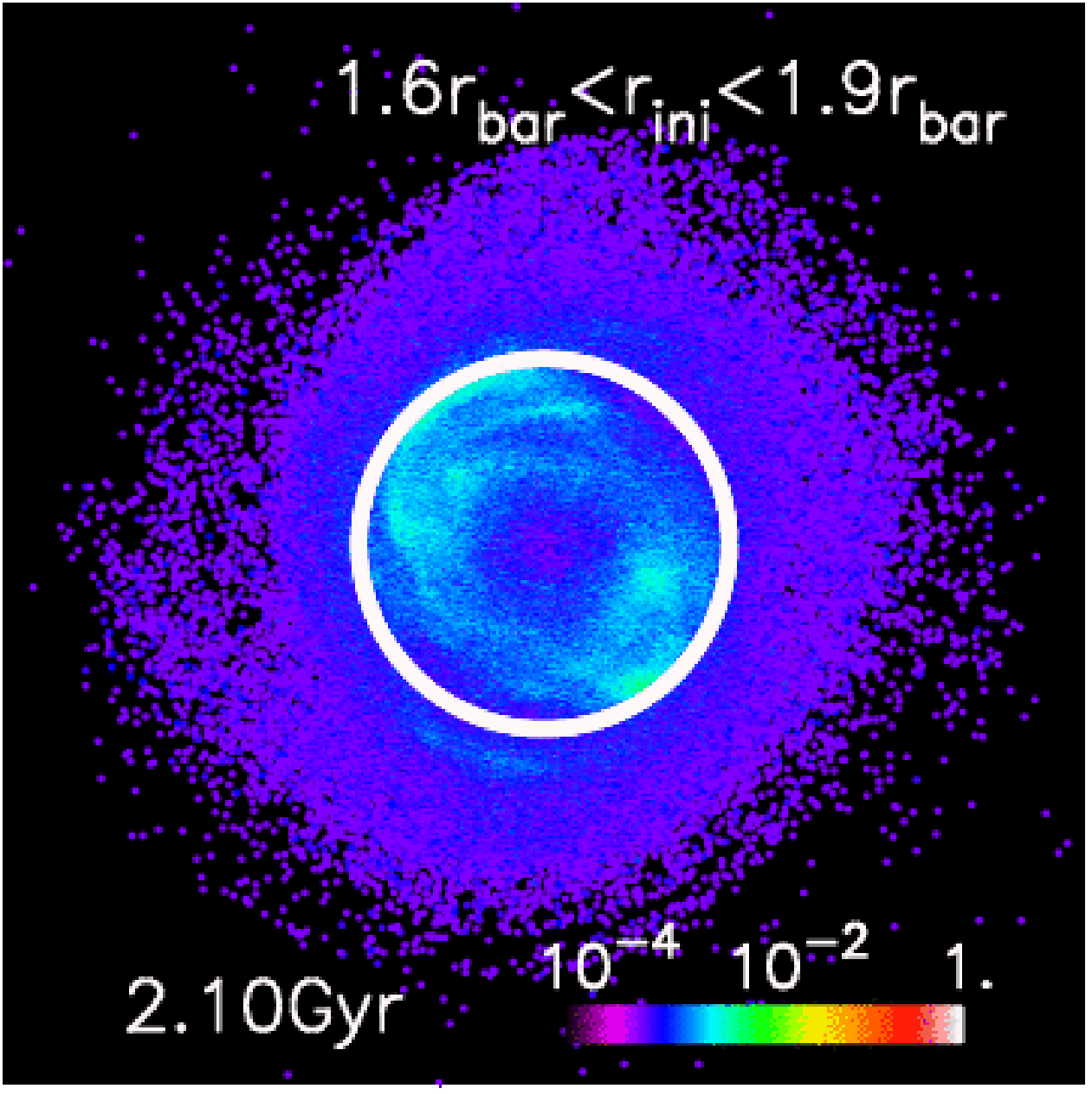}
\includegraphics[width=3.cm,angle=270]{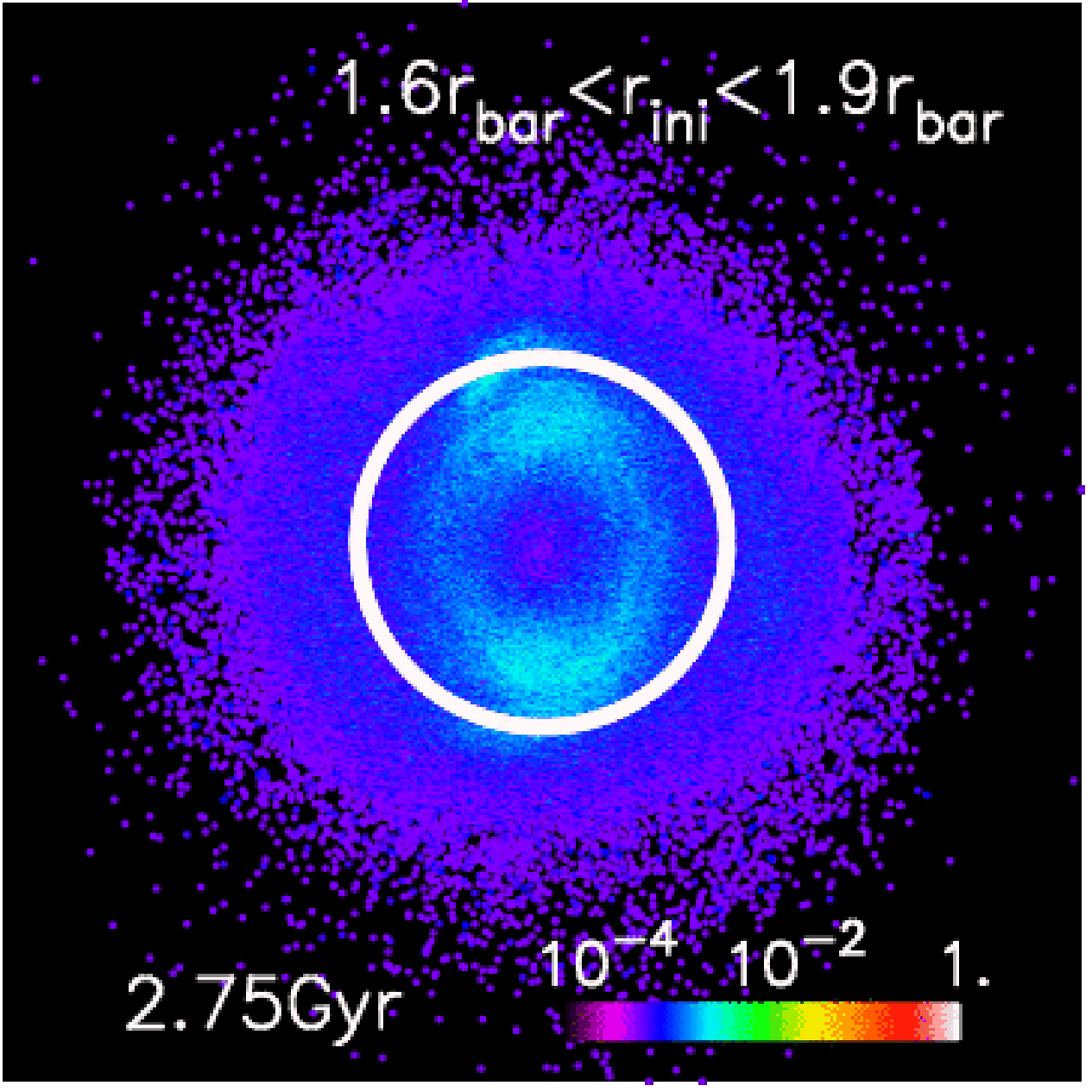}

\caption{\emph{(From top to bottom:)} Face-on density distribution of
stars of various birth radii:  $\rm{r_{ini}} \le 0.4r_{bar}$;
$0.4r_{bar} \le \rm{r_{ini}} \le 0.7r_{bar}$; $0.7r_{bar} \le \rm{r_{ini}}
\le r_{bar}$; $r_{bar}\le \rm{r_{ini}} \le 1.3r_{bar}$; $1.3r_{bar}
\le \rm{r_{ini}} \le 1.6r_{bar}$; $1.6r_{bar} \le \rm{r_{ini}} \le
1.9r_{bar}$. Different columns correspond to different times, as
indicated. In each panel, the average initial radius is indicated by a
white circle.}
\label{redistrib}
\end{figure*}

\section{Initial conditions and numerical method}\label{method}

The set of three simulations analyzed in this paper are the same
already presented in \citet{dimatteo13}: they consist of an isolated
disk, with a varying bulge-to-disk ratio (B/D=0., 0.1 and 0.25,
respectively), and containing no gas.  The dark halo and the optional
bulge are modeled as a Plummer sphere  \citep{BT87}. The dark halo has
a mass  $M_{\rm H}=1.02\times10^{11}M_{\odot}$ and a characteristic
radius  $r_{\rm H}$=10~kpc. The bulge, when present, does not
rotate initially (but see Sect.~\ref{bulge} for its final rotational
content), has a mass $M_{\rm B}=9\times10^9M_{\odot}$ and characteristic
radius $r_{\rm B}$=1.3~kpc, for the case with B/D=0.1, and  $M_{\rm
B}=2.2\times10^{10}M_{\odot}$ and $r_{\rm B}$=2~kpc, for the case with
B/D=0.25. The stellar disk follows a Miyamoto-Nagai density profile
\citep{BT87}, with mass $M_*=9\times10^{10}M_{\odot}$  and vertical
and radial scale lengths given by $h_*$=0.5~kpc and $a_*$=4~kpc,
respectively. The initial disk size is 13 kpc (that is,
initial stellar positions are generated between r=0 and r=13 kpc; see
Appendix~\ref{app1} for models employing an initially more extended
stellar disk), and the Toomre parameter is set equal to Q=1.8.
The galaxy is represented by $N_{\rm tot}=30 720 000$ particles
redistributed among dark matter ($N_{\rm H}=10 240 000$) and stars
($N_{\rm stars}=20 480 000$).  To initialize particle velocities, we
adopted the method described in  \citet{hern93}. The amplitude of
the initial fluctuations around the equilibrium virial ratio $Q_{vir}$
in the resulting models is $\Delta Q_{vir}/Q_{vir}\sim$1\% over the
first Gyr of evolution.

To model galaxy evolution, we employed a Tree-SPH code, in which
gravitational forces are calculated using a hierarchical tree method
(Barnes \& Hut 1986). The code has been presented in \citet{sem02}
and we refer the reader to this paper for a full description. For the
dissipationless simulations analyzed in this paper, the SPH part of the
code has been switched off, and gravitational forces are calculated
using a tolerance parameter $\theta=0.7$ and include terms up to the
quadrupole order in the multiple expansion. A Plummer potential is
used to soften gravity at scales smaller than $\epsilon=50~{\rm pc}$.
With this spatial resolution, it is possible to resolve the vertical
structure of thin disks, and follow small scale inhomogeneities.

The equations of motion are integrated over 4 Gyr, using a leapfrog
algorithm with a fixed time step of $\Delta t=0.25$ $\times $  $10^5~{\rm
yr}$. Some of the main characteristics of the simulated models
and their evolution are shown in Fig.~\ref{density}.

\section{Results}\label{results}

\begin{figure}
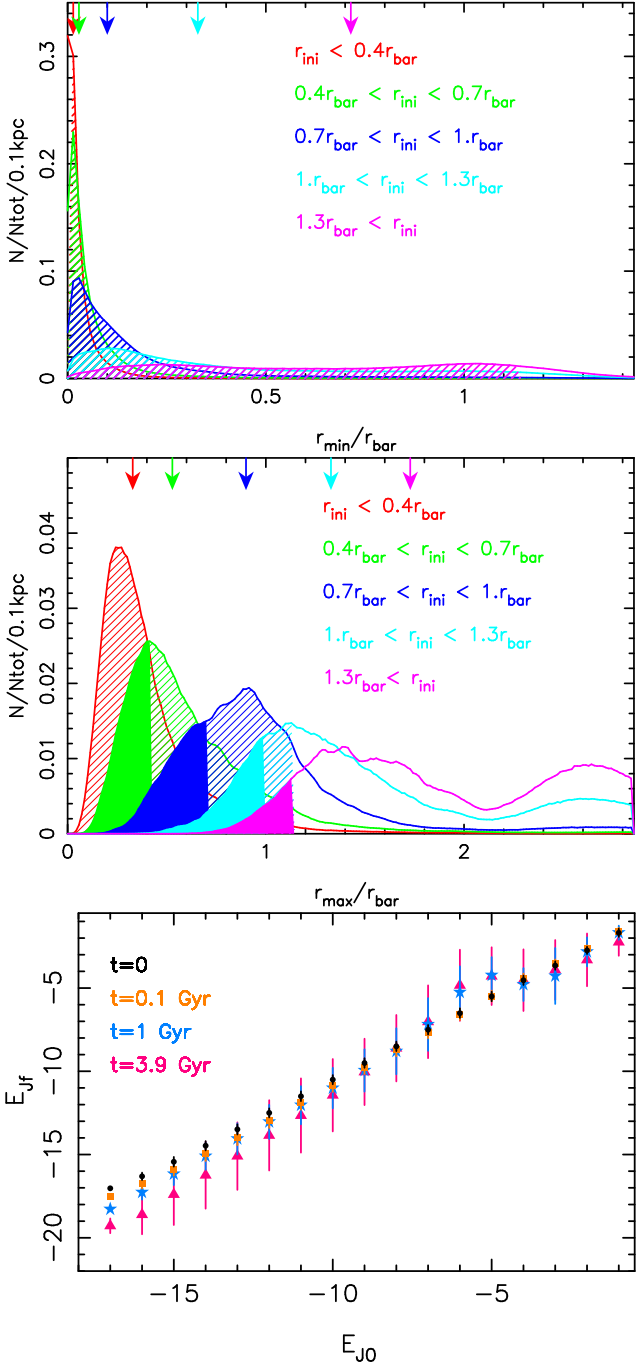

\centering
\vspace{1.5cm}
\includegraphics[width=6.cm,angle=270]{prminreg2_gS0_q1p8_BD0p00.ini.ps}
\includegraphics[width=6.cm,angle=270]{prmaxreg2_gS0_q1p8_BD0p00.ini.ps}
\includegraphics[width=6.cm,angle=270]{pJac.ps}
\caption{\emph{(Top panel: )} Distribution of the orbital pericenter
radii of stars of different provenance in the disk, as indicated in
the legend.  The dashed area represents stars whose pericenter radii are
inside the bar region. The arrows on the top axis indicate the medians of
the five distributions. \emph{(Middle panel: )} Same as the top panel,
but for the distribution of orbital apocenter radii. The dashed area
represents stars whose apocenter radii are inside the bar region, the
full area stars whose apocenter radii are also smaller then their birth
radii. Distances are in units of the initial bar scale length,
$r_{bar}$. \emph{(Bottom panel: )} Temporal evolution of Jacoby
energy in a rotating frame with pattern speed $\Omega=25$km/s/kpc. For
each bin in the initial Jacoby energy $E_{J0}$, the corresponding value
$E_{Jf}$ at time $t$ is plotted, together with the standard deviation of
the distribution. Different colors correspond to different times during
the evolution of the disk: initial time (black);  axisymmetric stellar
distribution (orange); epoch of strong thin stellar bar before buckling
(blue); final configuration, after the formation of the boxy bulge
(red). }
\label{rmax}
\end{figure}

\begin{table}
\caption{Fraction of stars in the bar region: For five different regions of provenance in the disk (first column), the percentage of stars with pericenter radii (second column) and apocenter radii (third column) inside the bar are shown.  Also, the percentage of stars inside the bar  whose apocenters are  smaller than their birth radii is given (fourth column).}              % title of Table
\label{table1}      % is used to refer this table in the text
\centering                                      % used for centering table
\begin{tabular}{c c c c}          % centered columns (4 columns)
\hline\hline                        % inserts double horizontal lines
& \% $r_{min} < r_{bar}$ & \% $r_{max} < r_{bar}$ &  \% $r_{max} < r_{bar}$ \\
& & & \& $r_{max} < r_{ini}$  \\    % table heading
\hline                                   % inserts single horizontal line
$r_{ini}\le 0.4r_{bar}$& 100 &    98 &  -- \\
$0.4r_{bar}< r_{ini} \le 0.7r_{bar}$ & 100 & 90 & 32 \\
$0.7r_{bar}< r_{ini} \le r_{bar}$ & 99 & 75 & 27 \\
$r_{bar}< r_{ini} \le 1.3r_{bar}$ & 93 & 30 & 18 \\
$1.3r_{bar}< r_{ini}$ & 85 &  8 &  8 \\
 %   1 & 50 & $-837$ & 970 \\      % inserting body of the table

\hline                                             %inserts single line
\end{tabular}
\end{table}

Several studies have pointed out the possibility that stars observed today
at a given location of a galaxy disk may be born in very different regions
from those where they are currently observed \citep{sel02,deb06, ros08,
ros08b, min10, min11,  bru11, min11, min12}. All these investigations have
explored the impact this redistribution has on the observed properties
of stellar disks, in particular on their external regions. None has yet
investigated in detail the effect of this redistribution on the properties
of the bar and bulge regions of disk galaxies. In the following sections,
we show that this impact is significant and cannot be neglected when
interpreting the observed properties of boxy/peanut shaped structures
in the Milky Way and other galaxies.

\begin{figure}
\centering
\includegraphics[width=3.cm,angle=270]{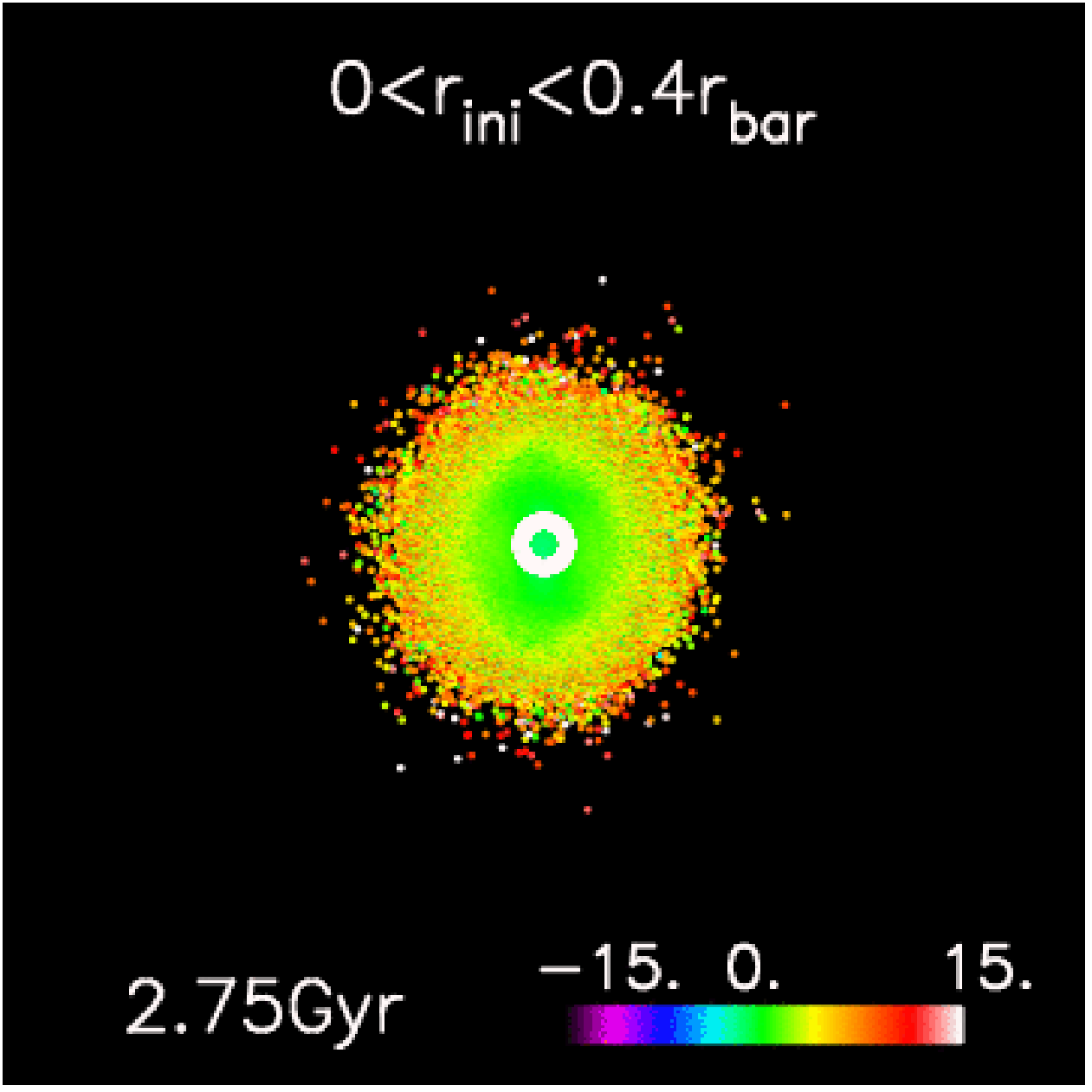}
\hspace{-0.2cm}
\includegraphics[width=3.cm,angle=270]{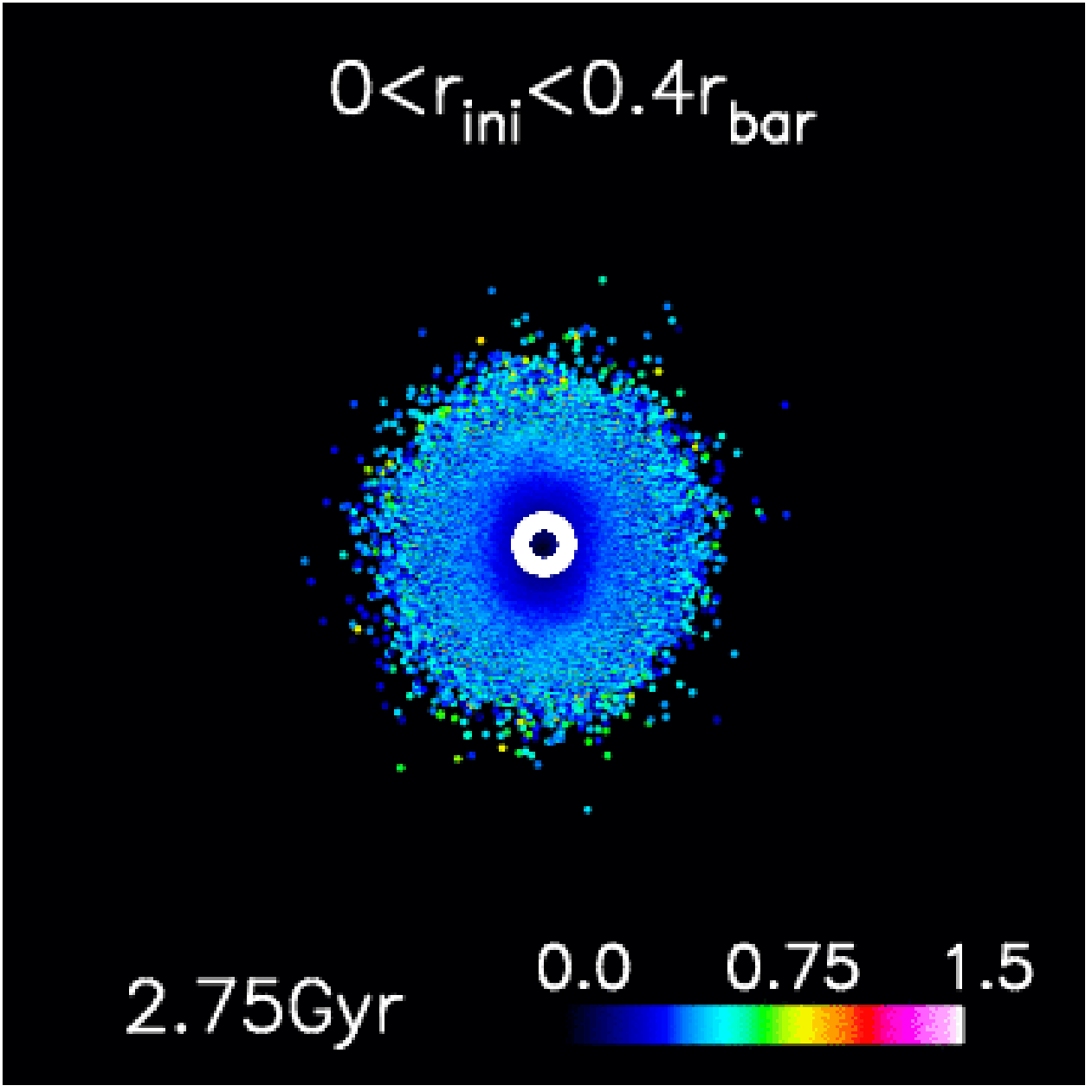}
\hspace{-0.2cm}
\includegraphics[width=3.cm,angle=270]{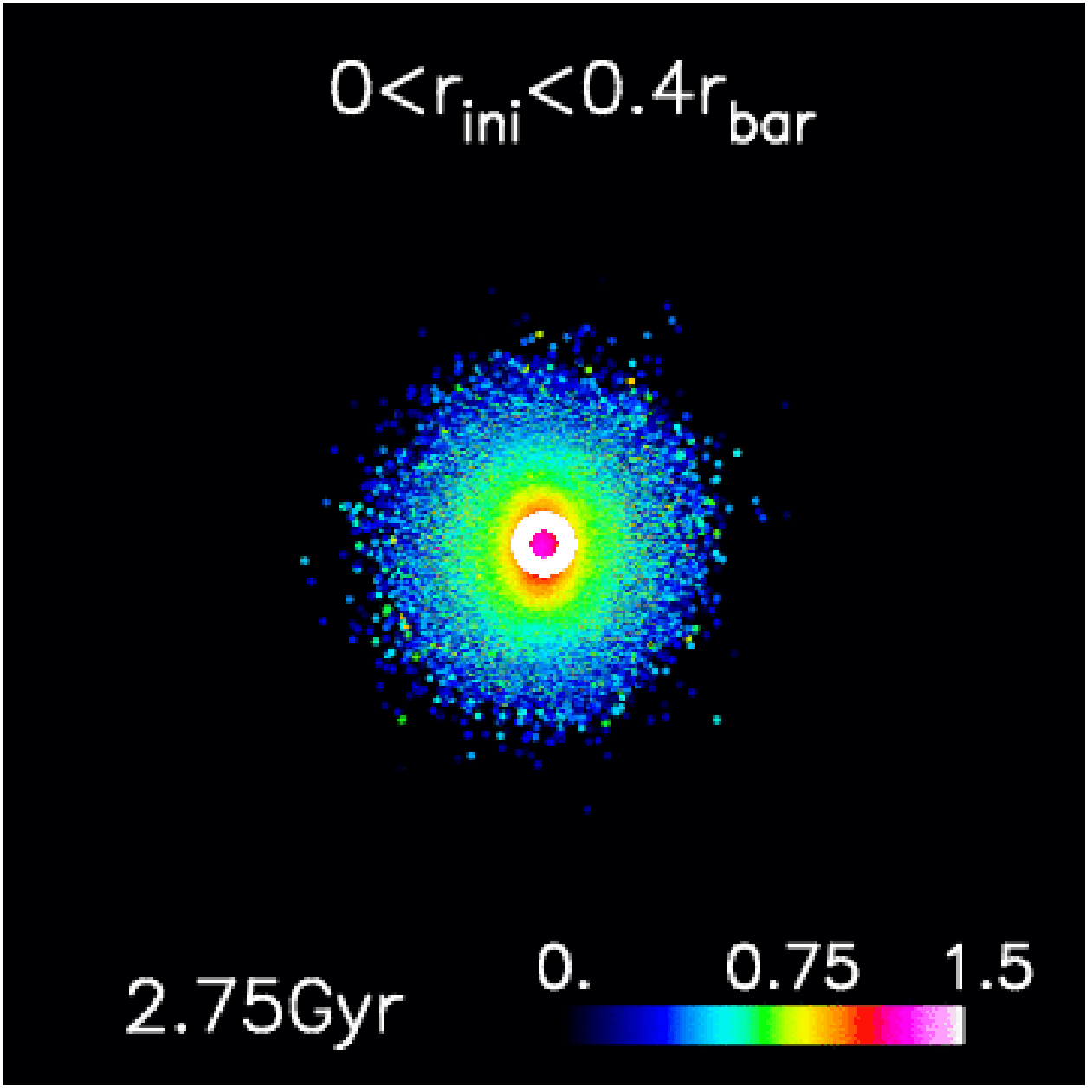}

\includegraphics[width=3.cm,angle=270]{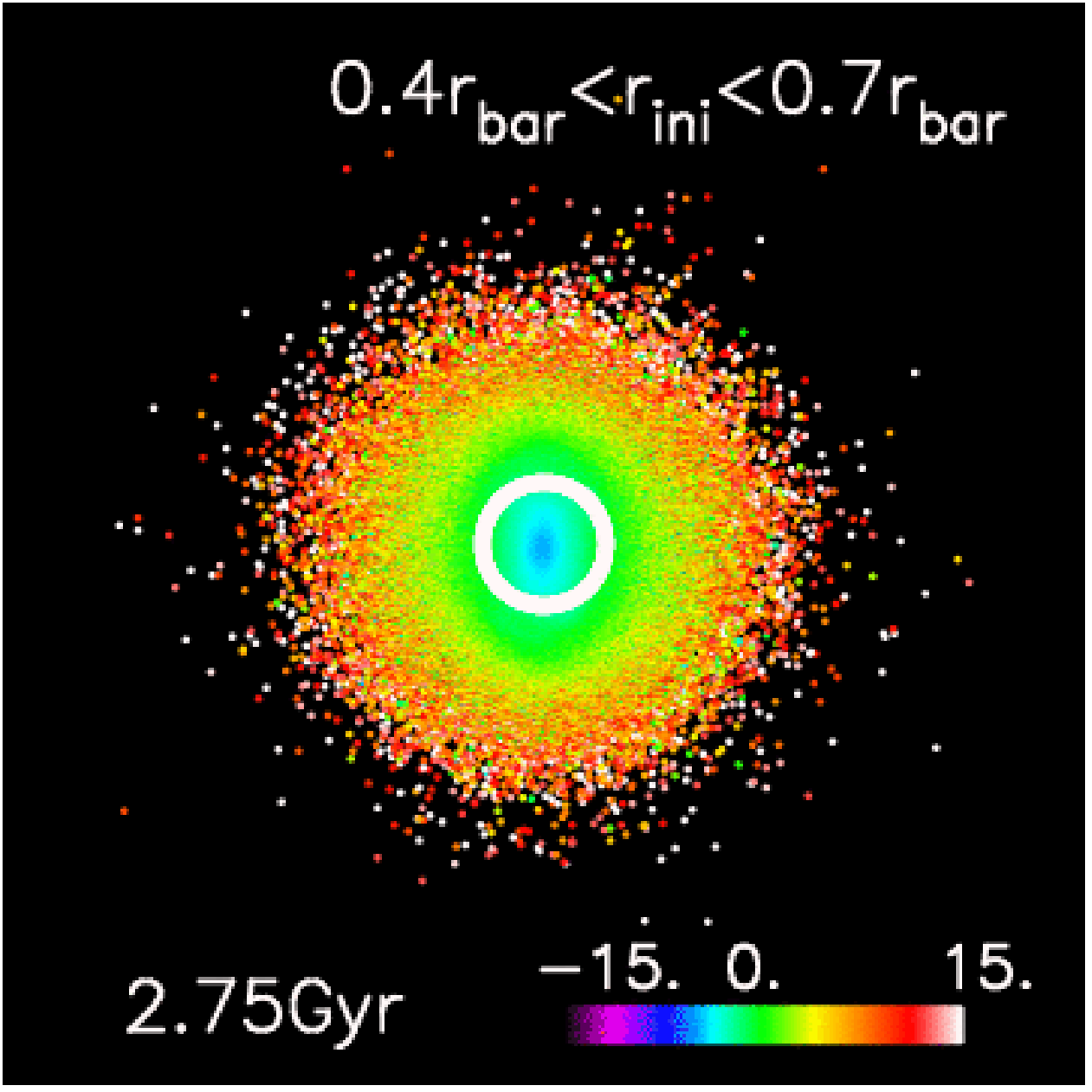}
\hspace{-0.2cm}
\includegraphics[width=3.cm,angle=270]{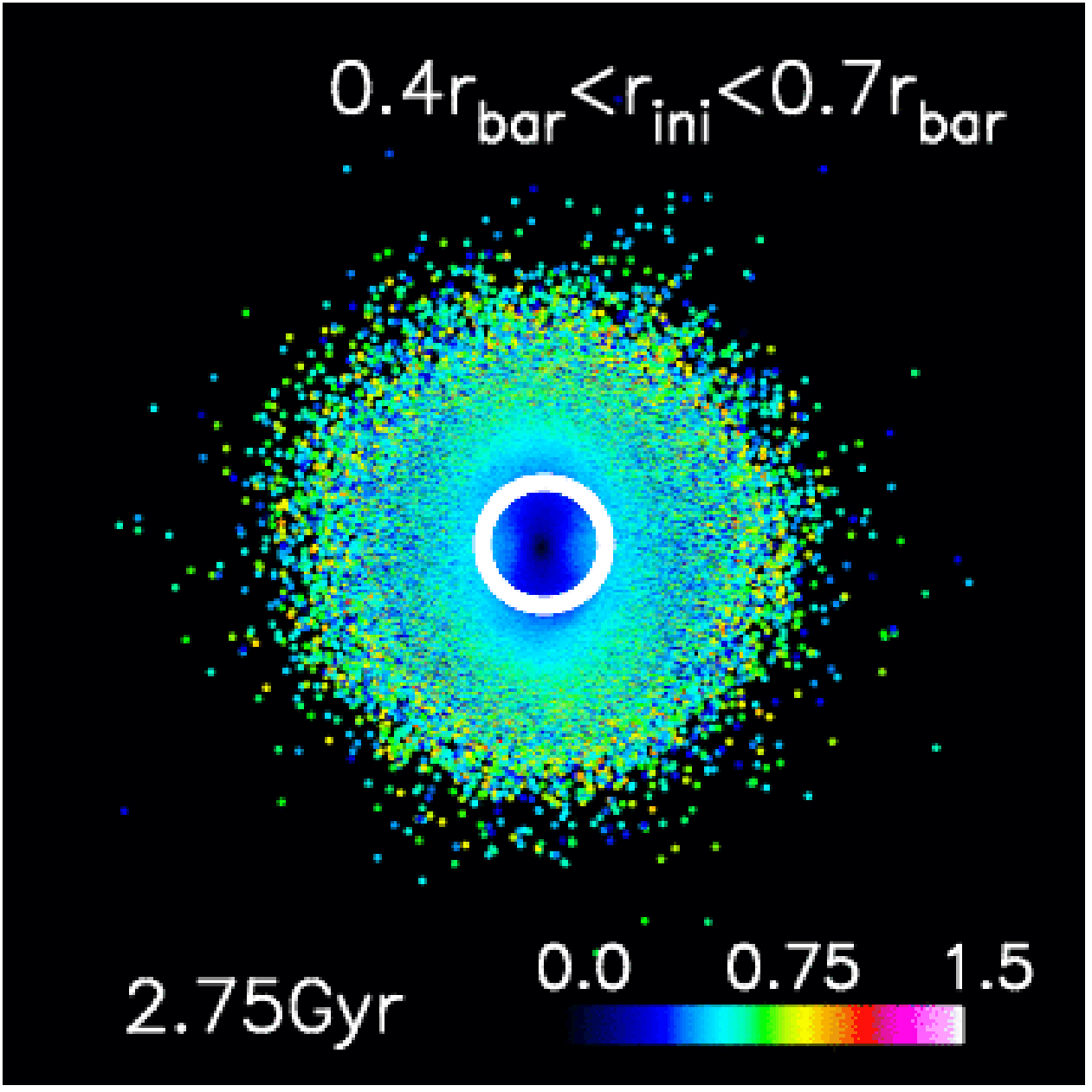}
\hspace{-0.2cm}
\includegraphics[width=3.cm,angle=270]{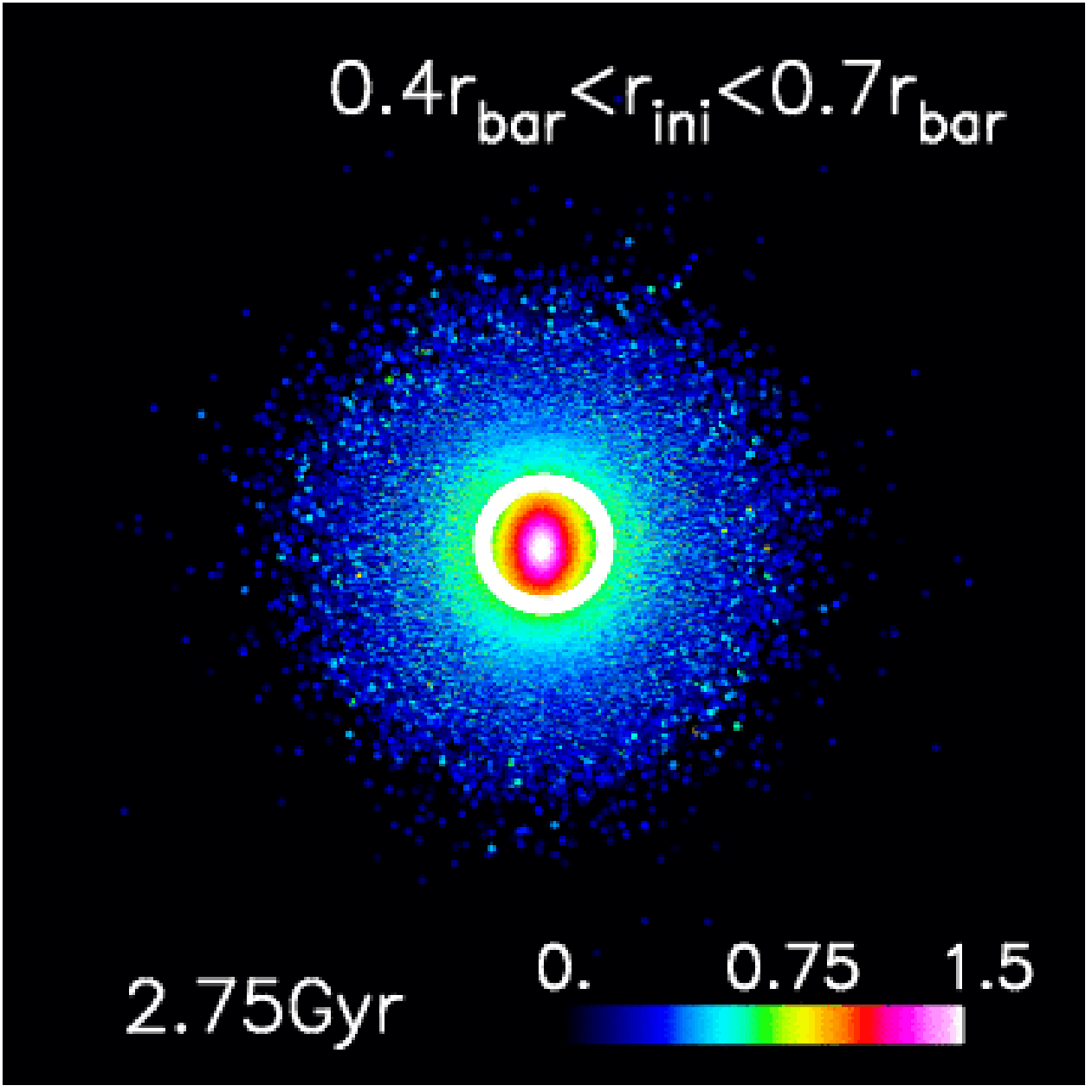}

\includegraphics[width=3.cm,angle=270]{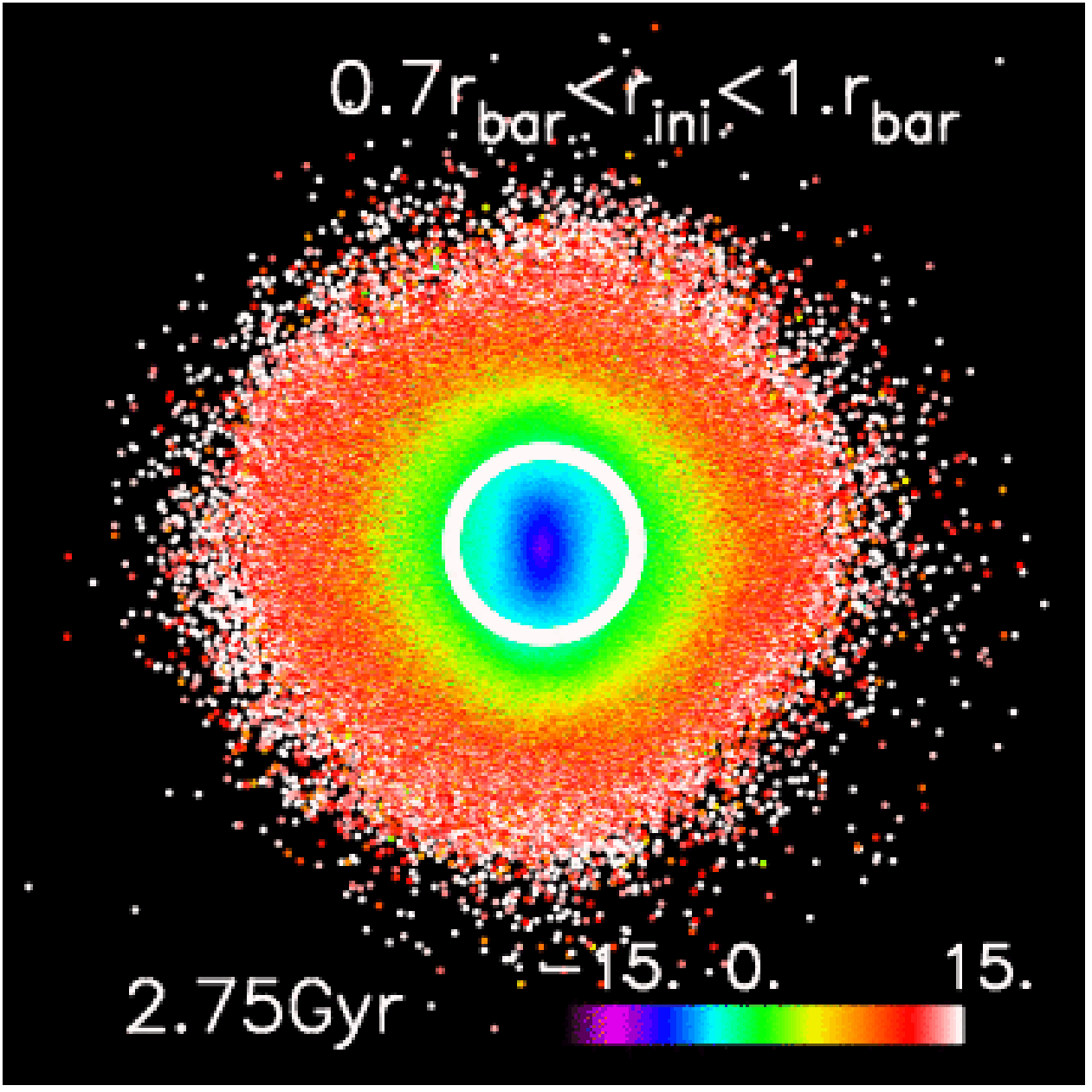}
\hspace{-0.2cm}
\includegraphics[width=3.cm,angle=270]{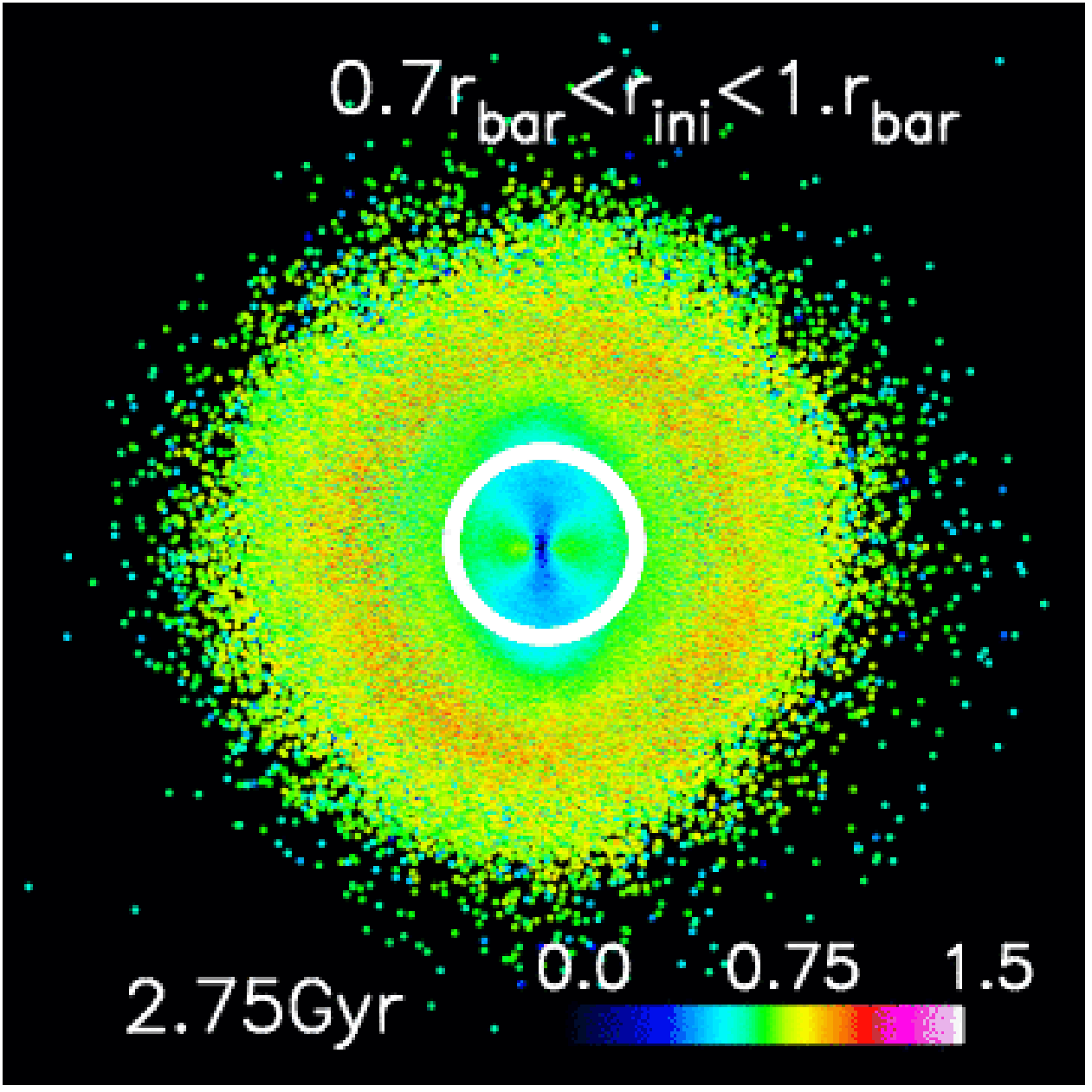}
\hspace{-0.2cm}
\includegraphics[width=3.cm,angle=270]{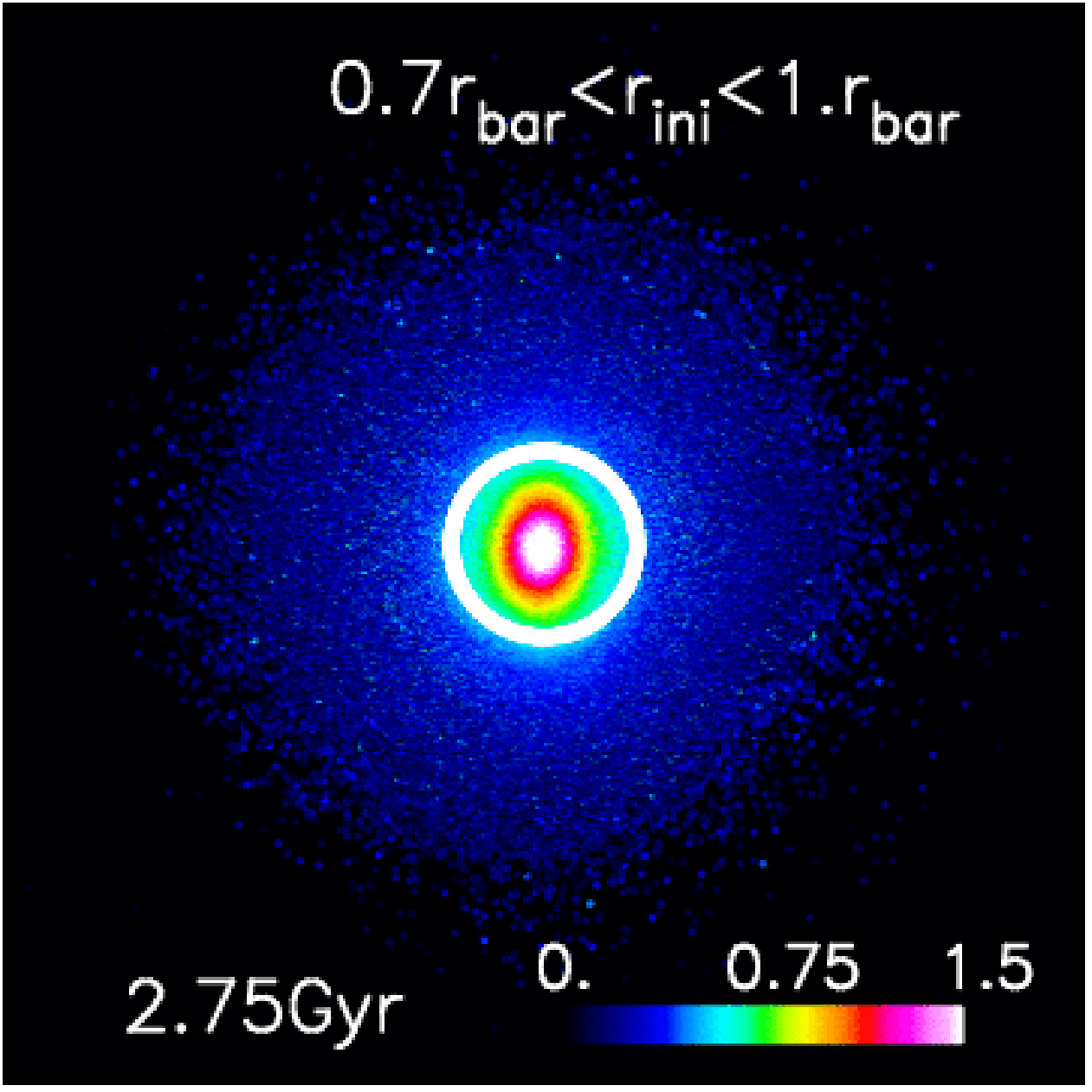}

\includegraphics[width=3.cm,angle=270]{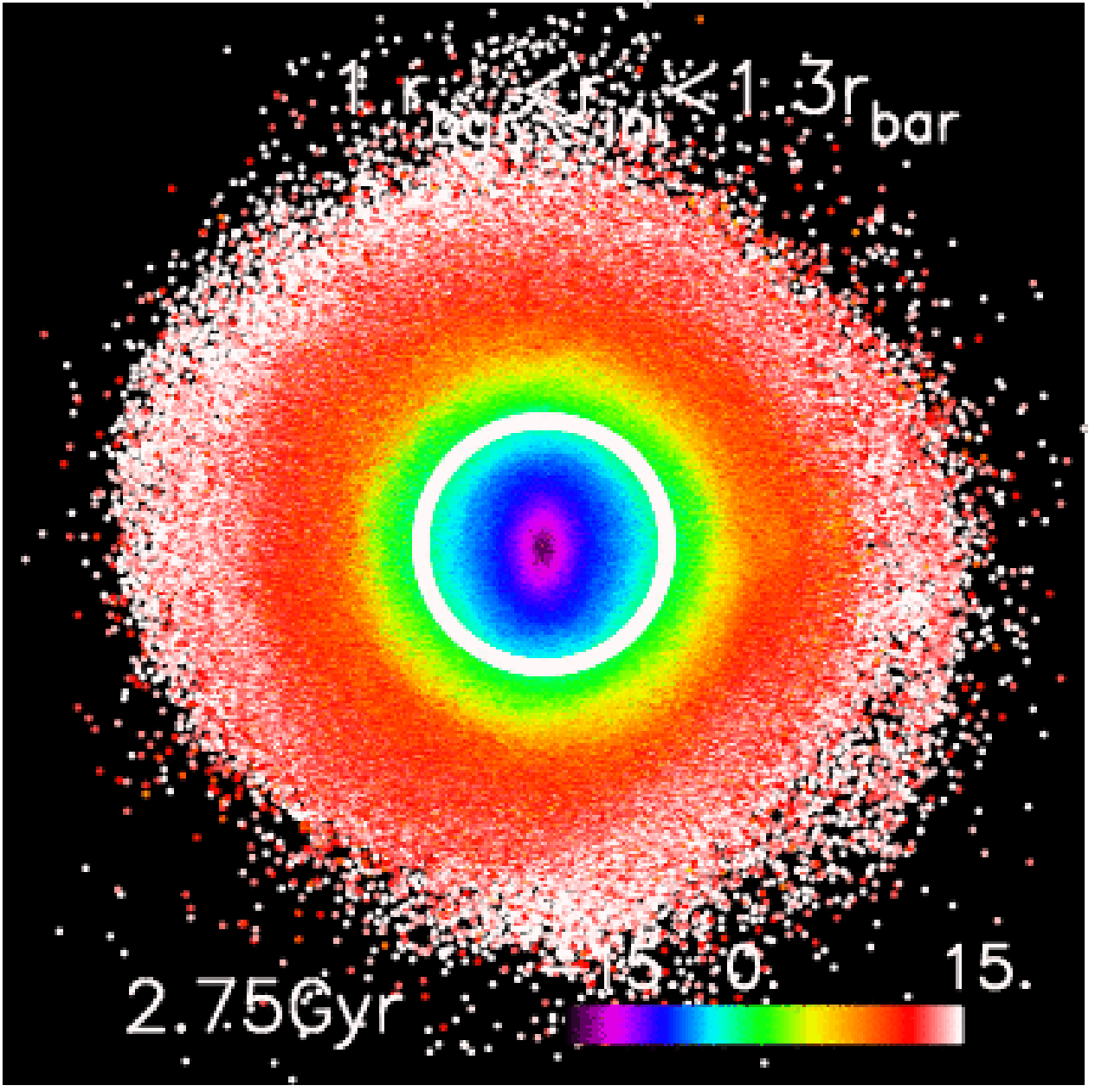}
\hspace{-0.2cm}
\includegraphics[width=3.cm,angle=270]{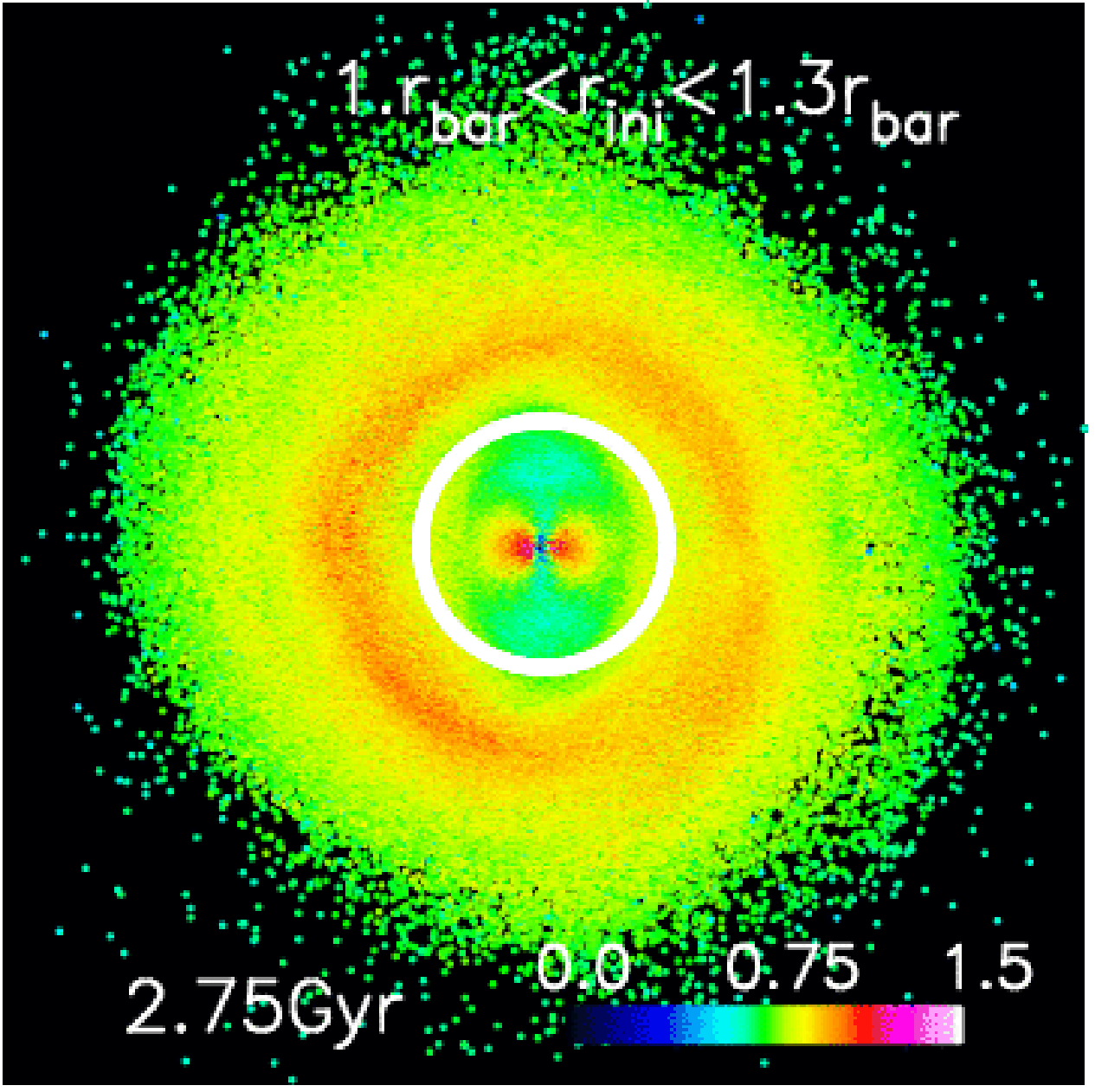}
\hspace{-0.2cm}
\includegraphics[width=3.cm,angle=270]{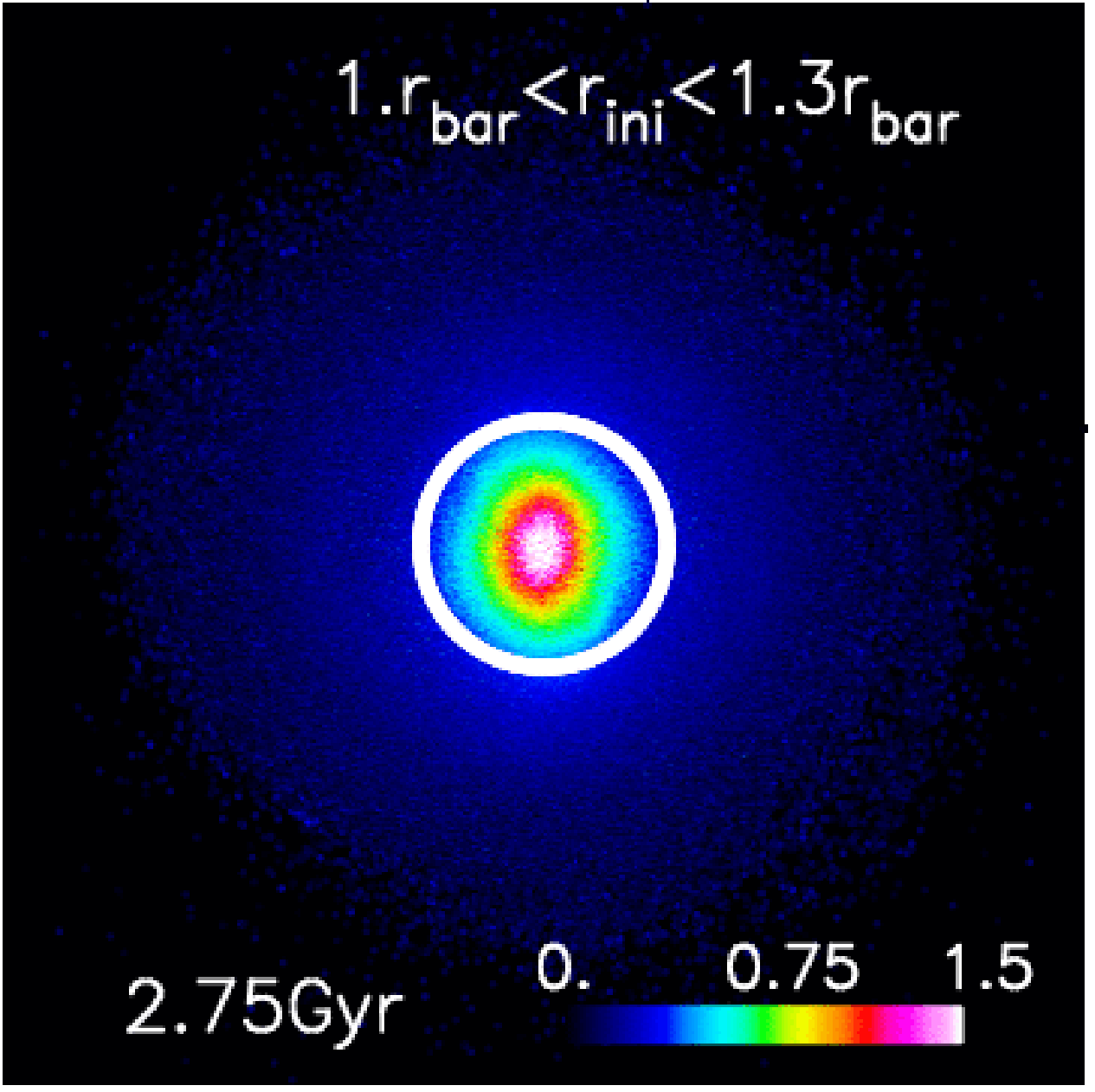}

\includegraphics[width=3.cm,angle=270]{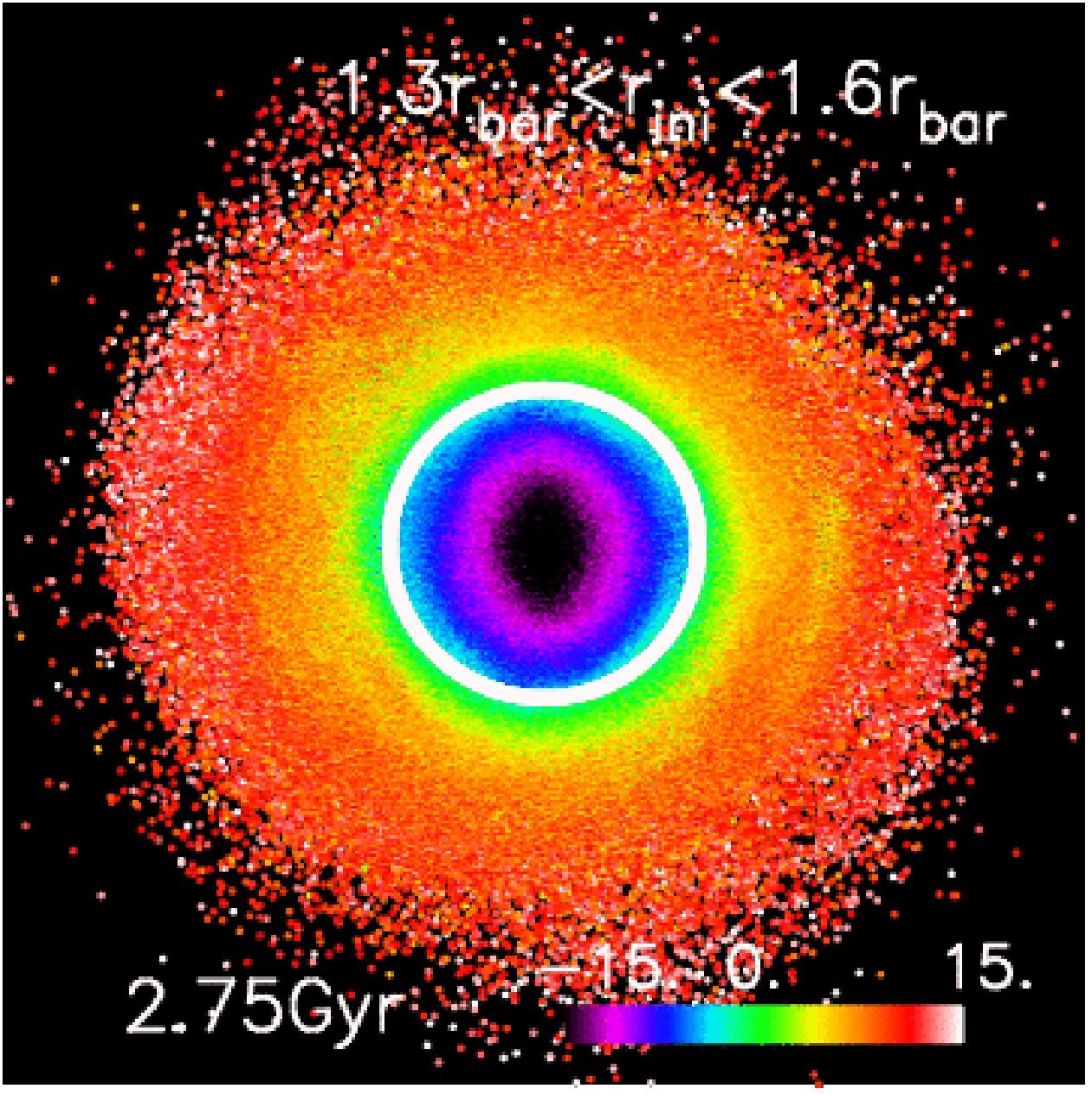}
\hspace{-0.2cm}
\includegraphics[width=3.cm,angle=270]{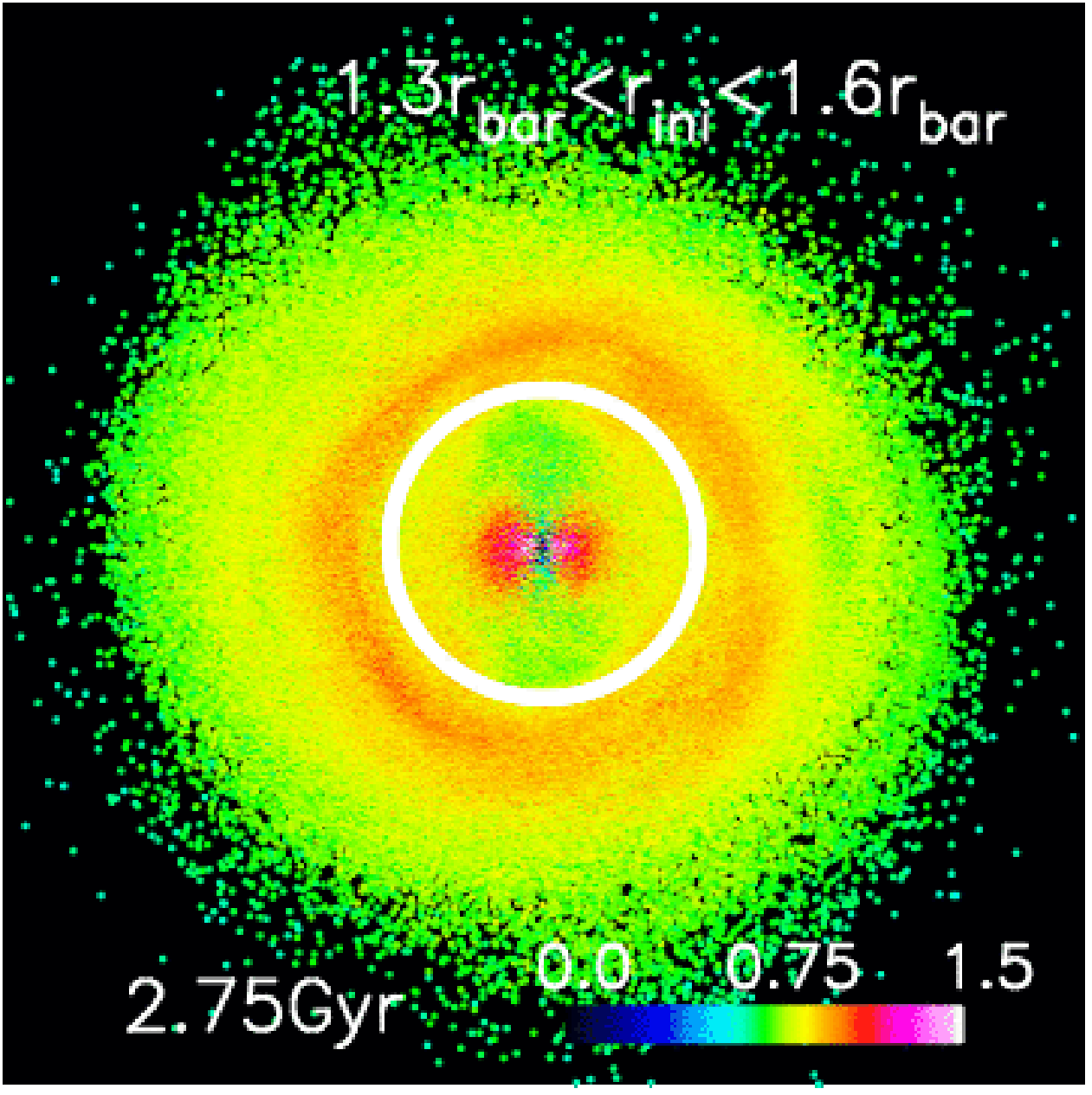}
\hspace{-0.2cm}
\includegraphics[width=3.cm,angle=270]{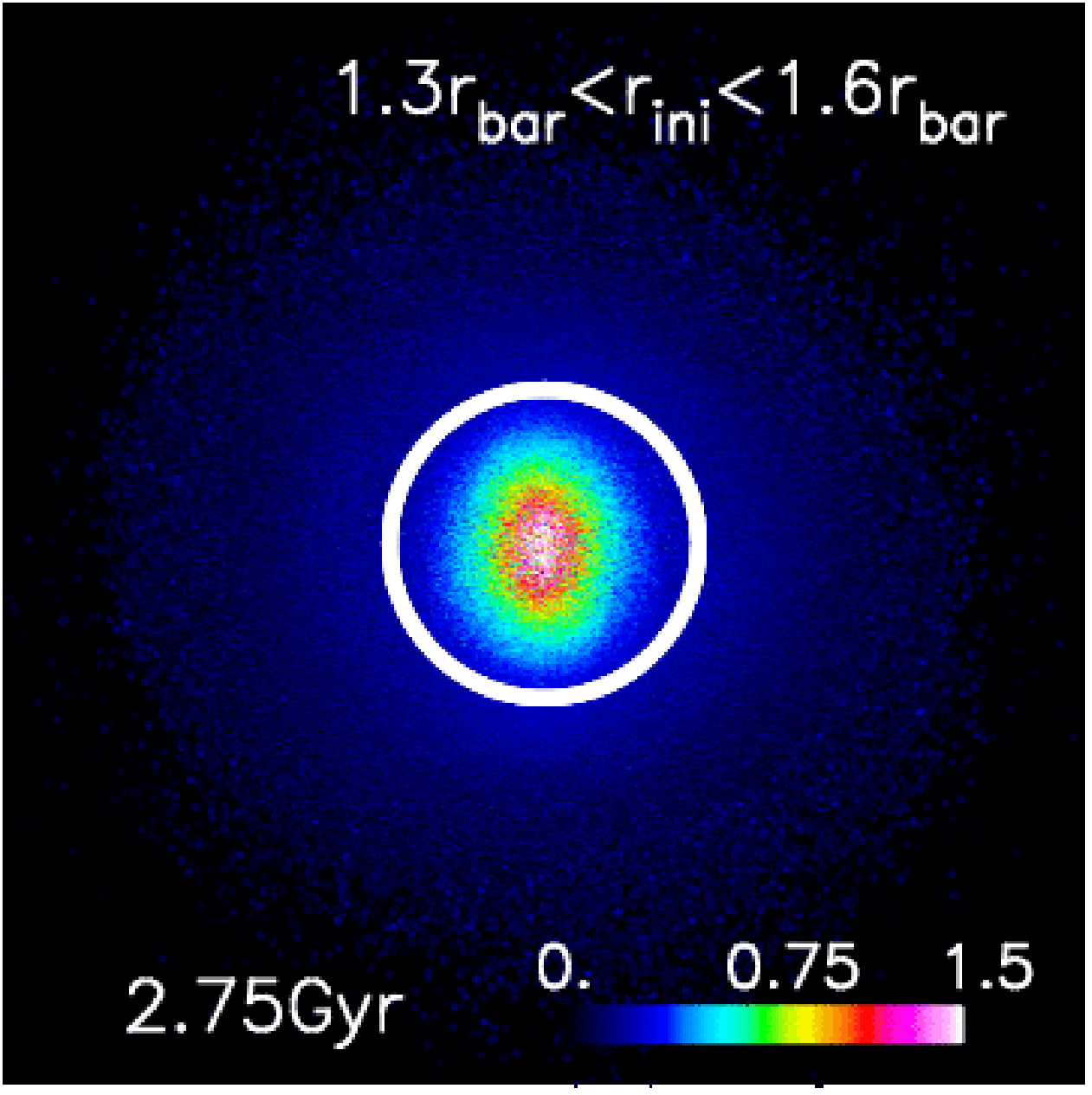}

\includegraphics[width=3.cm,angle=270]{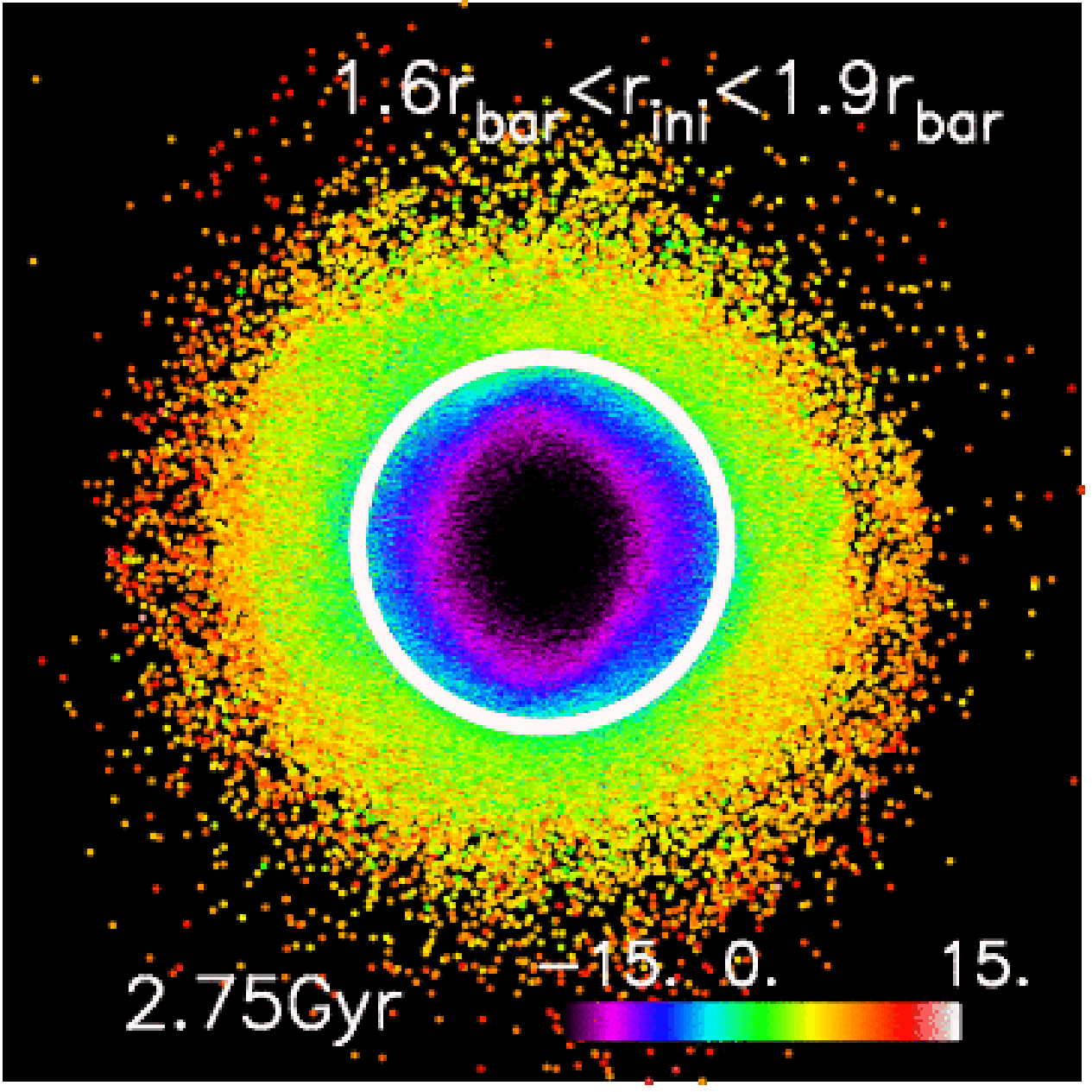}
\hspace{-0.2cm}
\includegraphics[width=3.cm,angle=270]{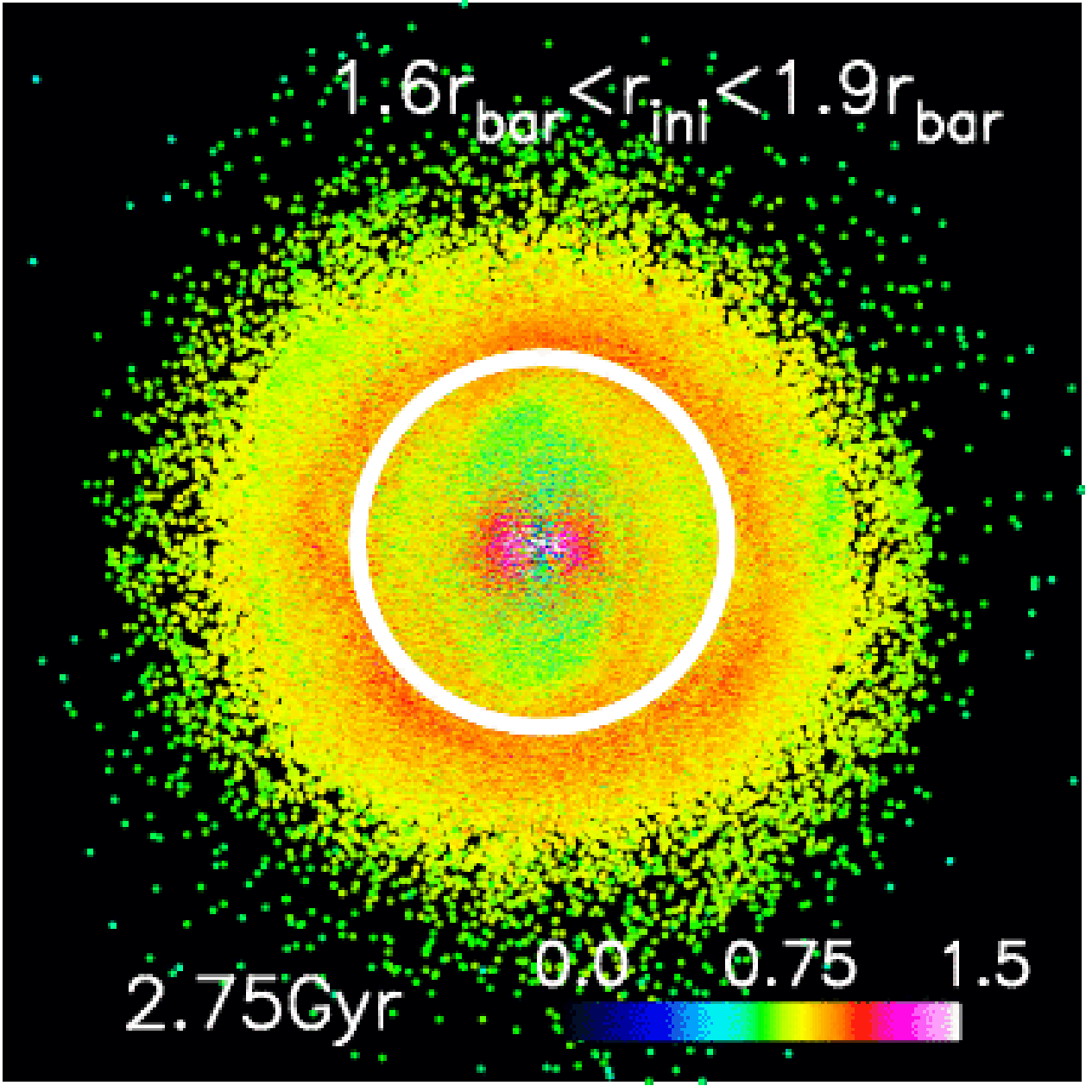}
\hspace{-0.2cm}
\includegraphics[width=3.cm,angle=270]{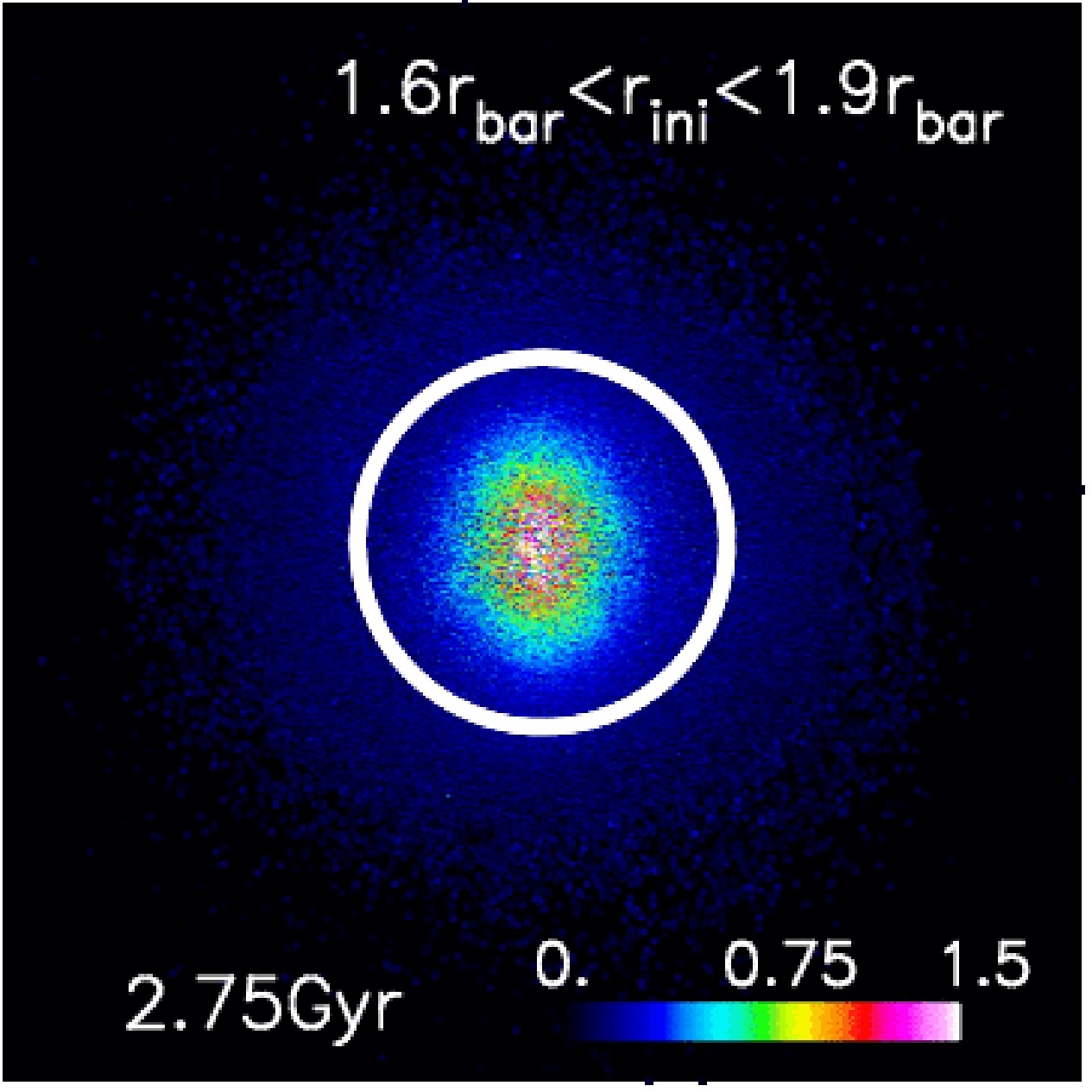}

\caption{\emph{(From left to right:)} Face-on maps of the angular momentum
variation $\Delta L$, of the rotational support $L_{norm}=L/L_{circ}$, and
of the vertical velocity dispersion $\sigma_z$  for stars with different
birth radii (from top to bottom: $\rm{r_{ini}} \le 0.4r_{bar}$;
$0.4r_{bar} \le \rm{r_{ini}} \le 0.7r_{bar}$; $0.7r_{bar} \le \rm{r_{ini}}
\le r_{bar}$; $r_{bar}\le \rm{r_{ini}} \le 1.3r_{bar}$; $1.3r_{bar}
\le \rm{r_{ini}} \le 1.6r_{bar}$; $1.6r_{bar} \le \rm{r_{ini}} \le
1.9r_{bar}$).  In this, and in all the following maps,
all the quantities have been calculated over pixels whose
size is 100pc$\times$100pc. The strong AM redistribution following bar
formation is evident from the $\Delta L$ maps. The rotational support
$L_{norm}$ shown in the middle column is the ratio between the average
$L$ at a given pixel, and the AM, $L_{circ}$, that a star in circular
orbit at the same location would have. In each panel, the stellar bar is
aligned with the $y-$axis and the average initial radius is indicated
by a white circle. All the plots correspond to the time $t=$2.75 Gyr,
as indicated. Angular momenta, and their variation, are in units of 100 km/s/kpc; velocities
in units of 100 km s$^{-1}$.}
\label{DL_reg}
\end{figure}

\subsection{Birth radii and radial migration: face-on view}\label{dissecting_faceon}

Our modeled bulgeless disk initially has an azimuthally symmetric
stellar distribution, with no sign of asymmetries for the first 0.8
Gyr of evolution, as shown by the Fourier analysis of the surface
density distribution of the face-on stellar disk (see Fig.~\ref{asym},
top panel). At about $t=0.8$~Gyr, a bar and spiral arms start to
develop, grow rapidly and stay strong for the following Gyr, until
about $t=2$~Gyr. At this time the stellar bar undergoes a vertical
instability, as a consequence, its strength diminishes and its scale
height increases considerably. This is the epoch when the bar changes
from a thin structure within the plane to a thick structure which may
appear either boxy or peanut-shaped depending on the angle between the
observer's line-of-sight and the bar major axis (see Fig.~\ref{asym},
bottom panels). In this simulation, the bar's resonances are located at
the corotation radius, $r_{CR}$=8--9 kpc, the inner Lindblad resonance,
$r_{ILR}$=2--3 kpc and outer Lindbland resonance, $r_{OLR}$=13 kpc for
simulation ages between 1 and 2~Gyr.  These resonances move outwards
slightly as the bar slows down and decreases in strength. Since the
OLR in these models coincides with the initial extent of the disk (see
Sect.~\ref{method}), the mapping that we will discuss in this section is
limited to the disk inside $r_{OLR}$. We will show in Appendix~\ref{app1}
that these results are also valid for simulated disks whose initial
extent and resonances location are different from those presented in
this Section, showing indeed that the disk up to the
OLR participates in the formation of the boxy-peanut shaped bulge
structure. During the whole simulation, the bar semi-major axis length,
$r_{bar}$,  is approximately 7-8 kpc, a factor of about two greater than
the length of the bar observed in the Milky Way. To present
our results in a more general way, independent on the bar size in the
following analysis all spatial scales are given in units of $r_{bar}$. In
this unit, between $t=1$ and $t= 2$~Gyr, $r_{ILR}=0.3-0.4 r_{bar},
r_{CR}=1.1-1.2r_{bar}, r_{OLR}=1.9r_{bar}$.

How does the formation and presence of asymmetries in the disk,
and in particular, the presence of the stellar bar -- the strongest
asymmetry in our simulations -- affect the spatial redistribution of
stars in the disk? To answer this question, similar to what was done in
\citet{dimatteo13},  we have selected stars according to their initial
radii\footnote{Hereafter, we will refer to this radius as the ``birth
radius''. By birth radius, we simply mean the distance from the galaxy
center a star has at the beginning of the simulation. This distance does
not change significantly for approximately the first Gyr of evolution
of the disk, until stellar asymmetries form, thus guaranteeing that the
definition is robust.}, defining six different regions in the disk,
as follows: 
$r_{ini} \le 0.4r_{bar}$, 
$0.4r_{bar} \le r_{ini} \le 0.7r_{bar}$, 
$0.7r_{bar} \le r_{ini} \le r_{bar}$,
$r_{bar}\le r_{ini} \le 1.3r_{bar}$, 
$1.3r_{bar} \le r_{ini} \le 1.6r_{bar}$, 
$1.6r_{bar} \le r_{ini} \le 1.9r_{bar}$, 
with $r_{ini}$
%$\rm{r_{ini}} \le 0.4r_{bar}$, $0.4r_{bar} \le \rm{r_{ini}} \le 0.7r_{bar}$, $0.7r_{bar} \le \rm{r_{ini}} \le r_{bar}$,
%$r_{bar}\le \rm{r_{ini}} \le 1.3r_{bar}$, $1.3r_{bar} \le \rm{r_{ini}} \le
%1.6r_{bar}$, $1.6r_{bar} \le \rm{r_{ini}} \le 1.9r_{bar}$}, with $r_{ini}$
being the distance, in the disk plane, of stars from the galaxy center.
This selection is shown in Fig.~\ref{redistrib}.

As a consequence of the angular momentum (hereafter AM) redistribution
initiated by the bar and spiral arms, stars tend to diffuse in the disk
as soon as stellar asymmetries start to develop. However, while stars
inside the  inner Lindblad resonance stay mostly confined in the inner
bar region \citep[see also][]{martinez13, pfenniger91}, outer disk stars,
in particular those at and beyond corotation, migrate both outward and
inward, reaching both the edges and the center of the disk. Within a few
rotational periods at the epoch of formation of the stellar asymmetries,
 outer disk stars are able to reach the inner disk, contributing to
populating the bar: their distribution shows a clear $m=2$ asymmetry
elongated with the bar major axis and tends to accumulate in two stellar
over-densities at the edges of this structure (see Fig.~\ref{redistrib}).
At the onset of the bar vertical instability, those stars that are close
to the vertical inner Lindblad resonance (VILR), which is at about $0.8
r_{bar}$ from the center, are scattered to greater heights becoming
part of the boxy/peanut-shaped structure. Since stars from a large range of initial birth radii are able to reach the bar
region before its vertical buckling, as a consequence, the resulting
bulge is populated by a mixture of populations, from stars born in situ
(i.e. in the inner disk) to stars coming from all of the outer radii, from
those just outside the bar to the outermost extent of the disk,
 at about 2$r_{bar}$.

It is interesting to note that the orbital characteristics of stars
that end up in the bar region depend on where they originated in
the disk. Figure~\ref{rmax}, for example, shows the distribution of
orbital pericenter ($r_{min}$) and apocenter ($r_{max}$) radii for
stars originating in five different disk regions: the median of both of
these radial distributions increases with increasing originating radii
of the stars.  This effect is such that outer disk stars which end up
in the bar/bulge region tend to orbit over a larger portion of the disk
than stars born in the inner disk. This memory of their initial
location in the disk translates directly into only limited variations (see
bottom panel in Fig.~\ref{rmax}) in their Jacoby energy \citep[see][for
a definition]{BT87}, as already pointed out by \citet{martinez13}. From
Fig.~\ref{rmax} and Table~\ref{table1} one can also see that while stars
born inside the VILR are trapped in the inner disk (for example, 90\%
of stars with  $0.4r_{bar}\le r_{ini}\le 0.7r_{bar}$ have apocenters
inside the bar), outer disk stars can have pericenters penetrating in the
bar region, and apocenters outside it; in other words not all the outer
disk stars penetrating the bar region are confined within the bar region
over their whole orbit. As an example, among the stars with $r_{bar}\le
r_{ini}\le 1.3r_{bar}$, 90\% of them have pericenters inside the bar
($r_{min} < r_{bar}$), while only 30\% have their whole orbit confined
inside the bar ($r_{max} < r_{bar}$). There is thus a high fraction --
about 60\% -- that have pericenters penetrating inside the bar region,
but apocenters outside which lie outside of it. It is also interesting
to note that, among stars with $r_{bar} \le r_{ini} \le 1.3r_{bar}$,
there is a significant fraction ($\sim$20\%) whose apocenters are inside
the bar region and are also smaller than their initial birth radii
(i.e. both the conditions $r_{max} < r_{bar}$ and $r_{max} < r_{ini}$
are satisfied, see Table~\ref{table1}):  those stars may be considered as
the true migrators, since their orbital radii have significantly changed
(this is guaranteed by the condition $r_{max} < r_{ini}$). However,
since all the stars whose orbit penetrates, at least partially, into
the bar contribute to its properties, in the following, we adopt a more
general definition: ``outside-in migrators'' are stars that enter the
bar region and spend part of their orbit at distances from the galaxy
center smaller than their initial birth radii.

 \begin{figure*}
\centering
\includegraphics[width=4.cm,angle=270]{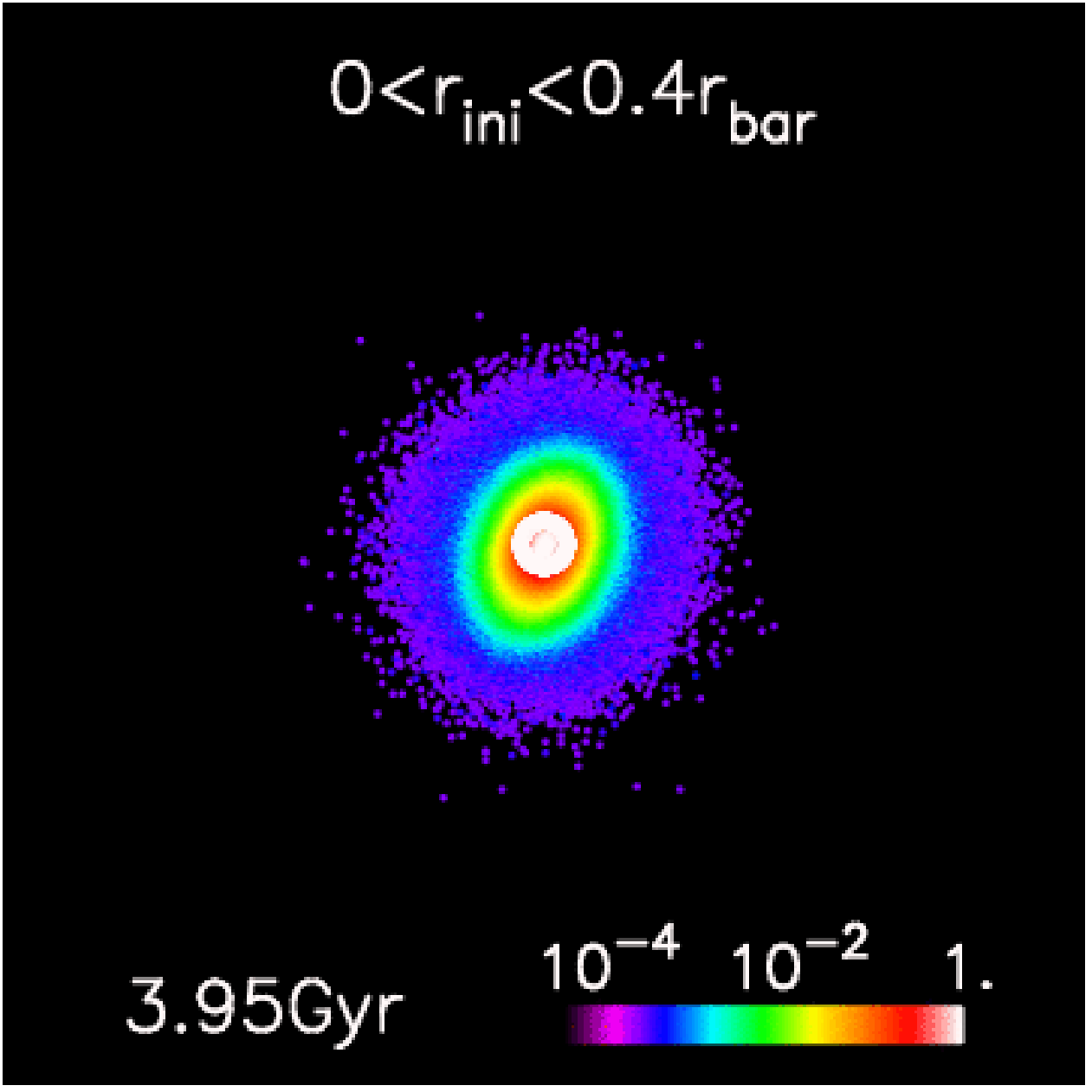}
\includegraphics[width=4.cm,angle=270]{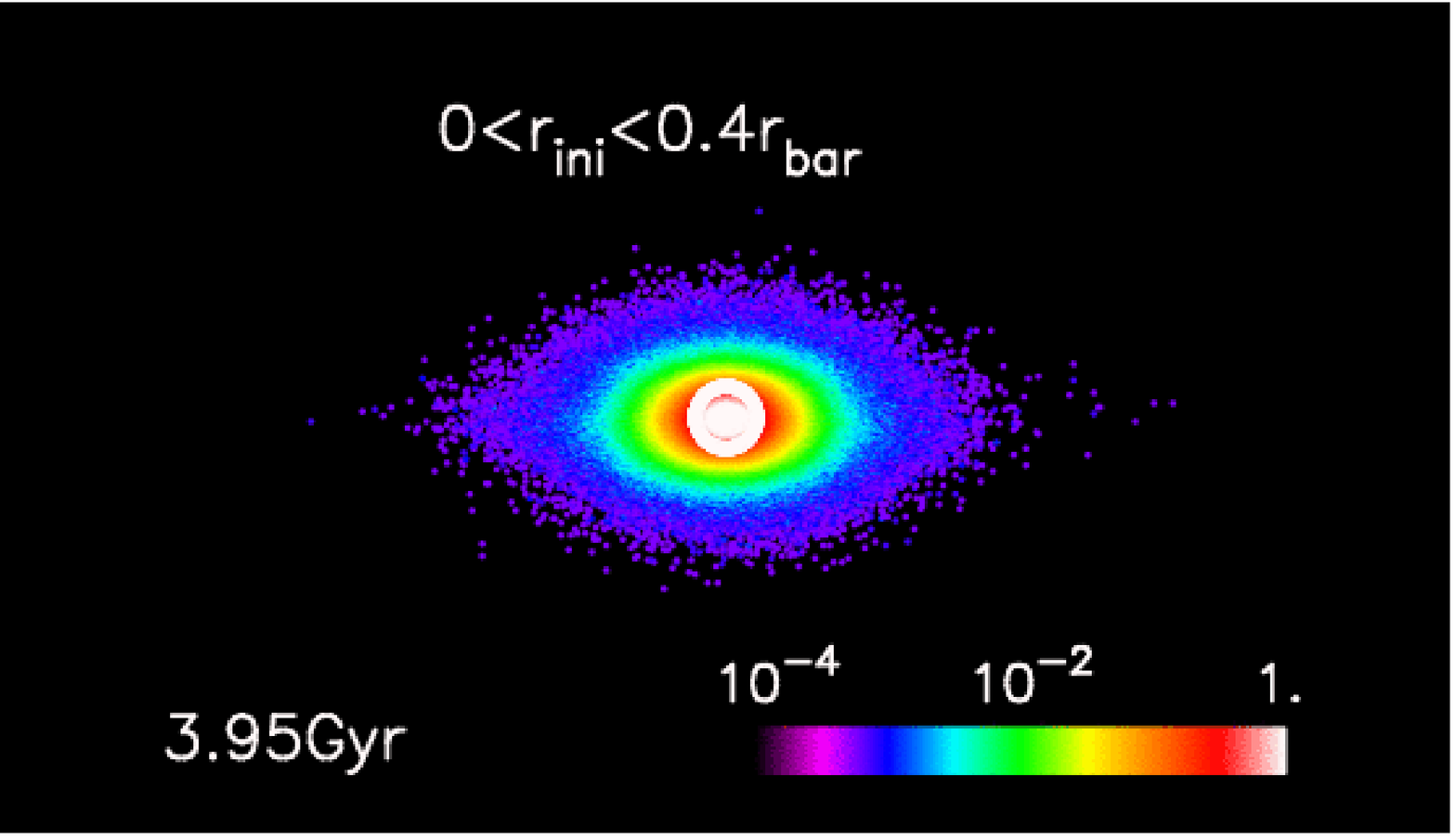}
\includegraphics[width=4.cm,angle=270]{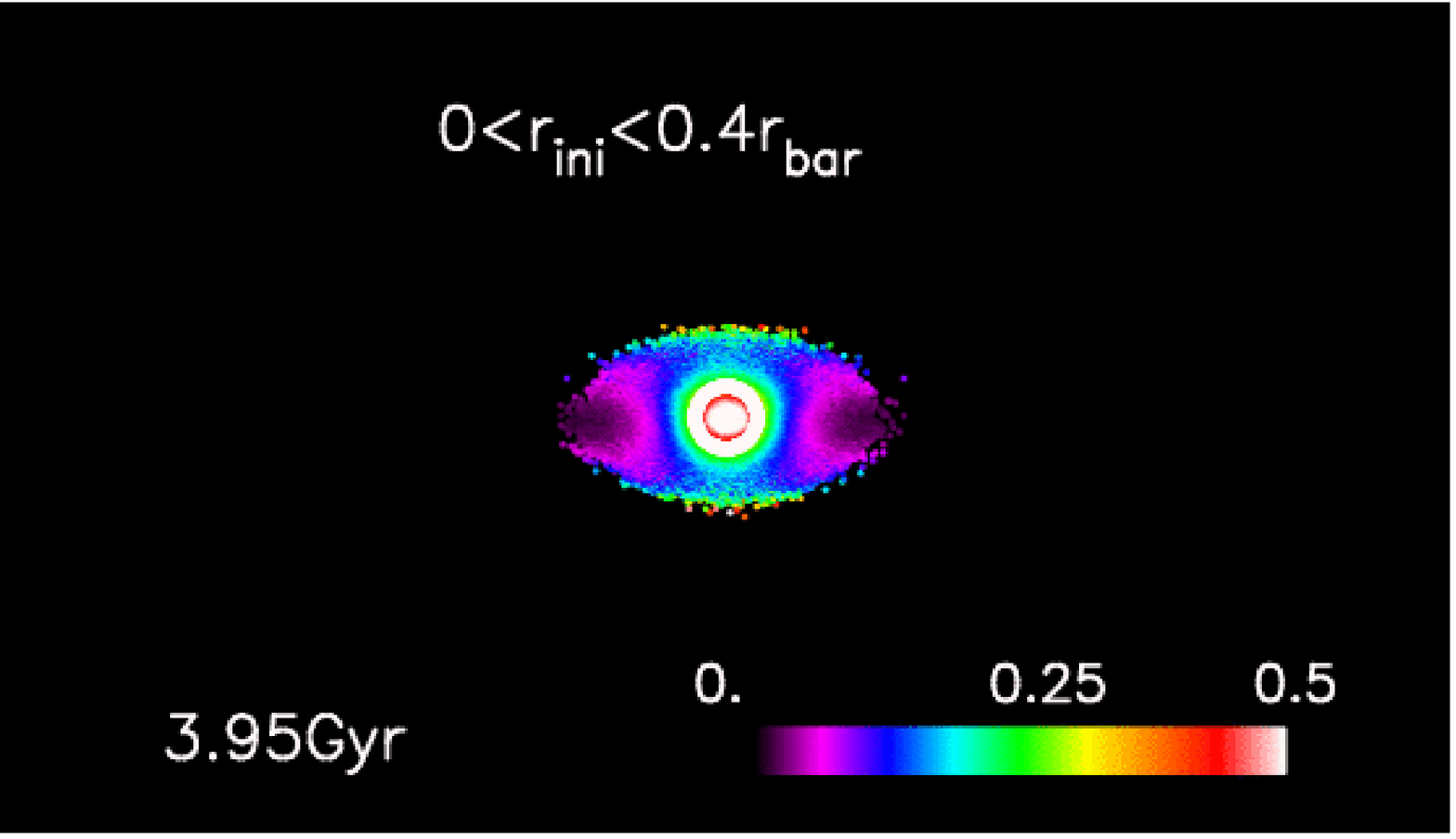}

\includegraphics[width=4.cm,angle=270]{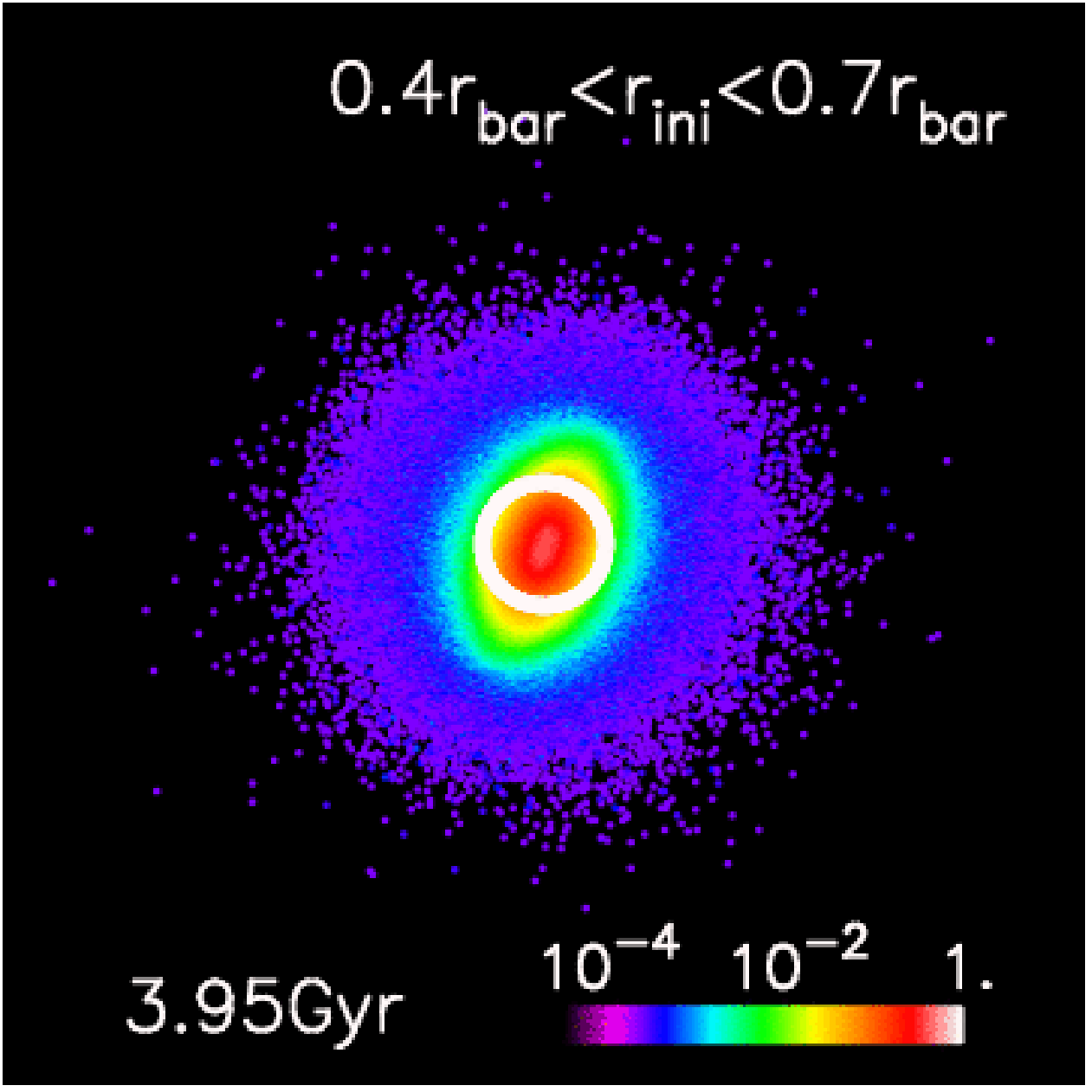}
\includegraphics[width=4.cm,angle=270]{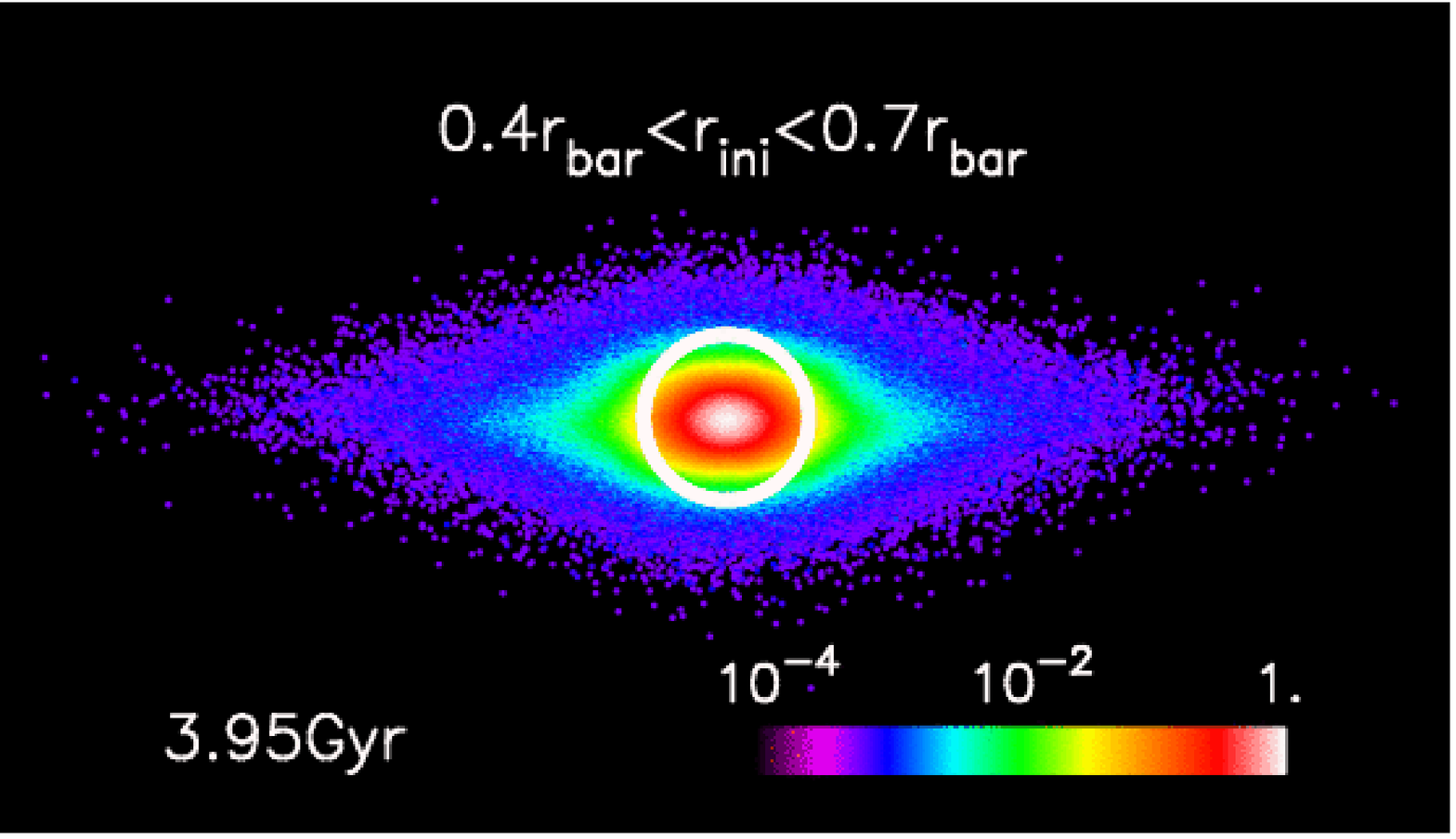}
\includegraphics[width=4.cm,angle=270]{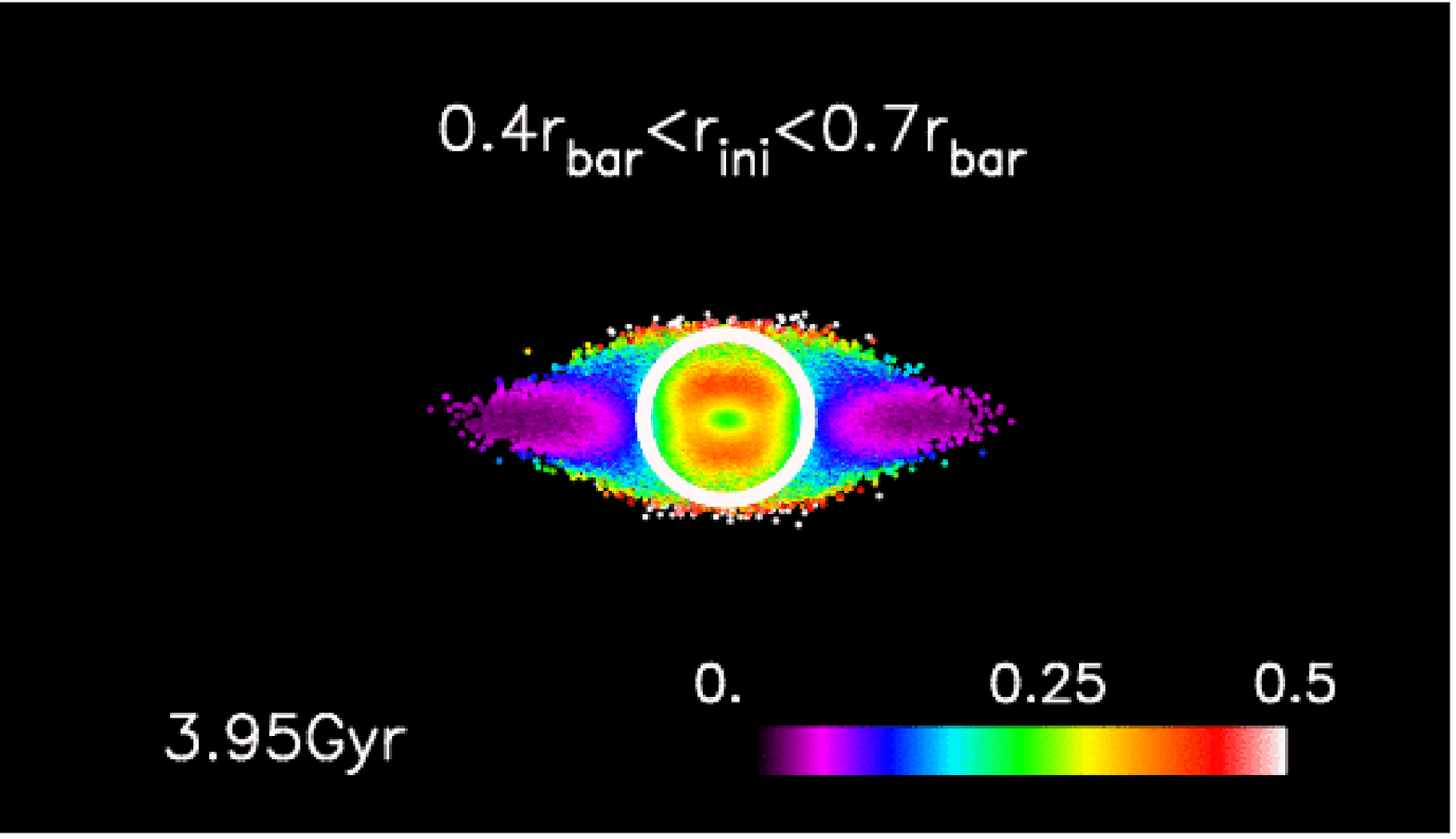}

\includegraphics[width=4.cm,angle=270]{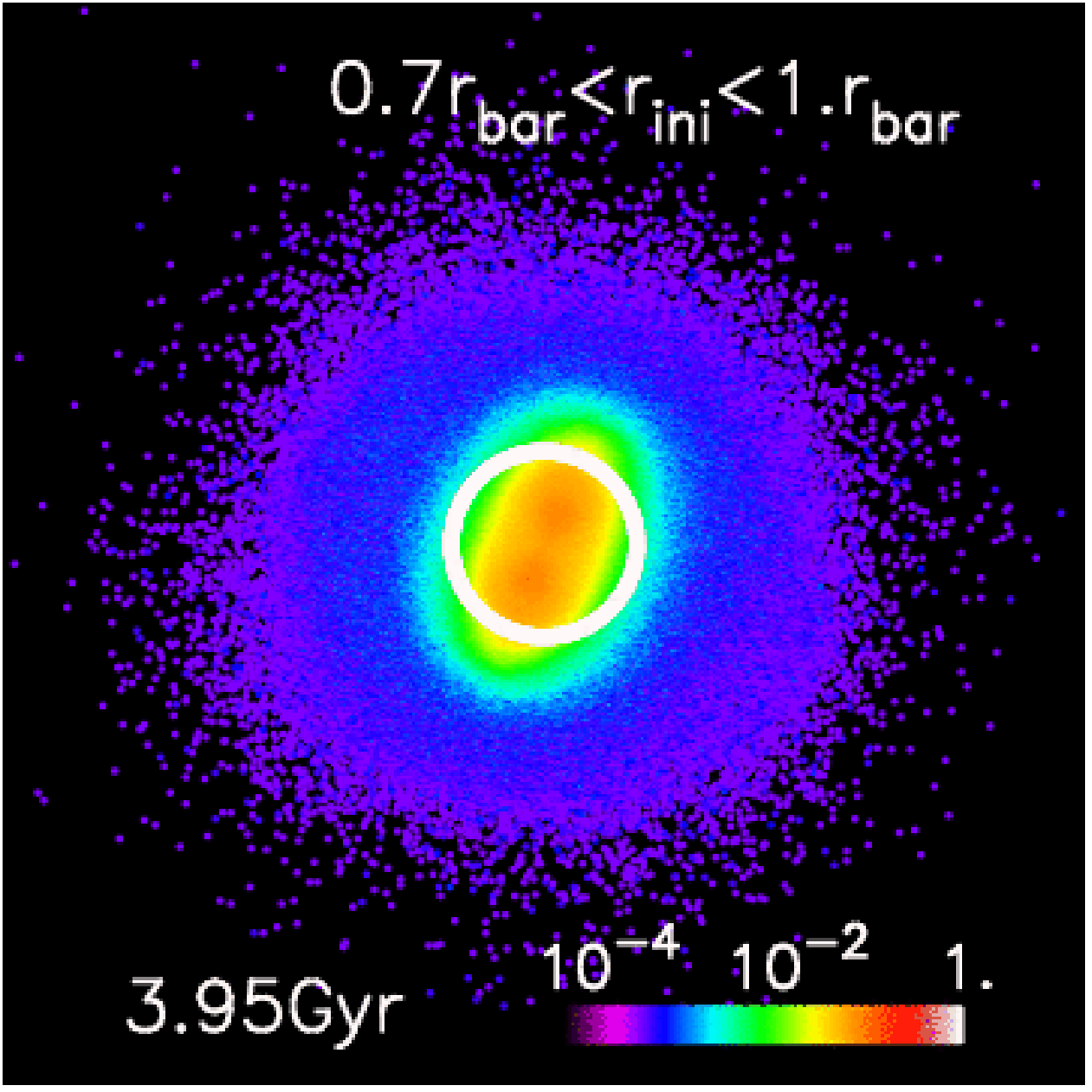}
\includegraphics[width=4.cm,angle=270]{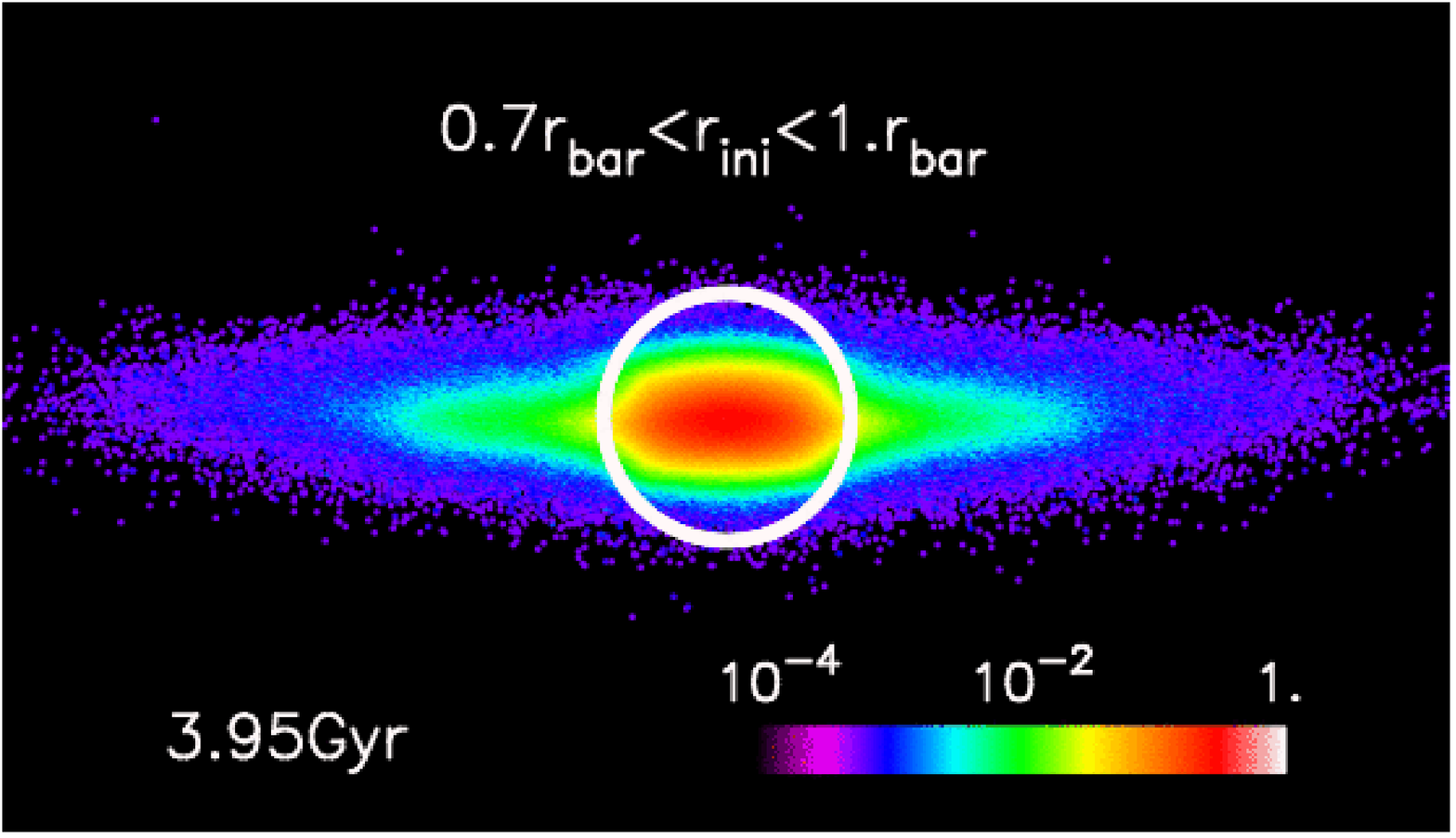}
\includegraphics[width=4.cm,angle=270]{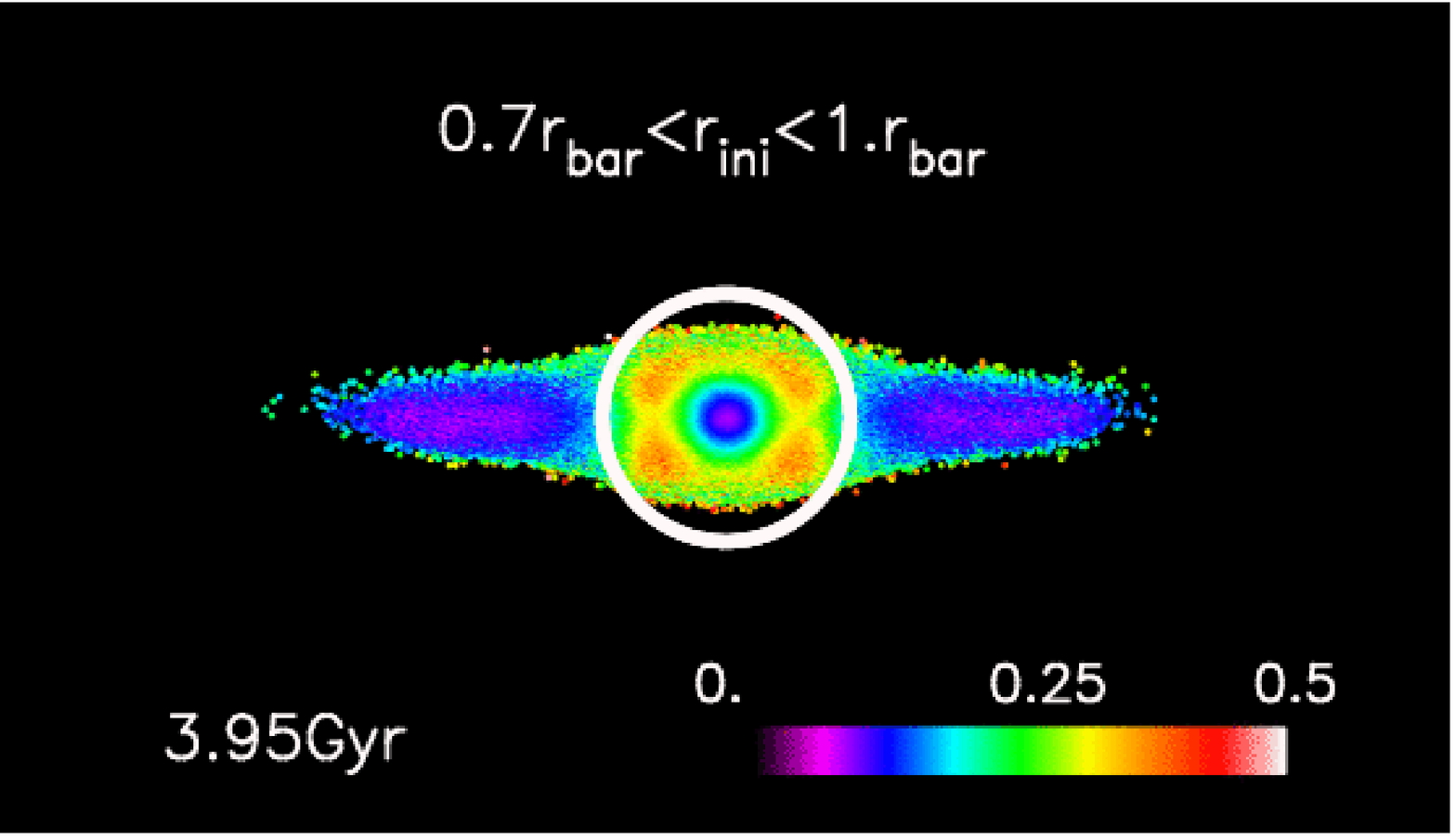}

\includegraphics[width=4.cm,angle=270]{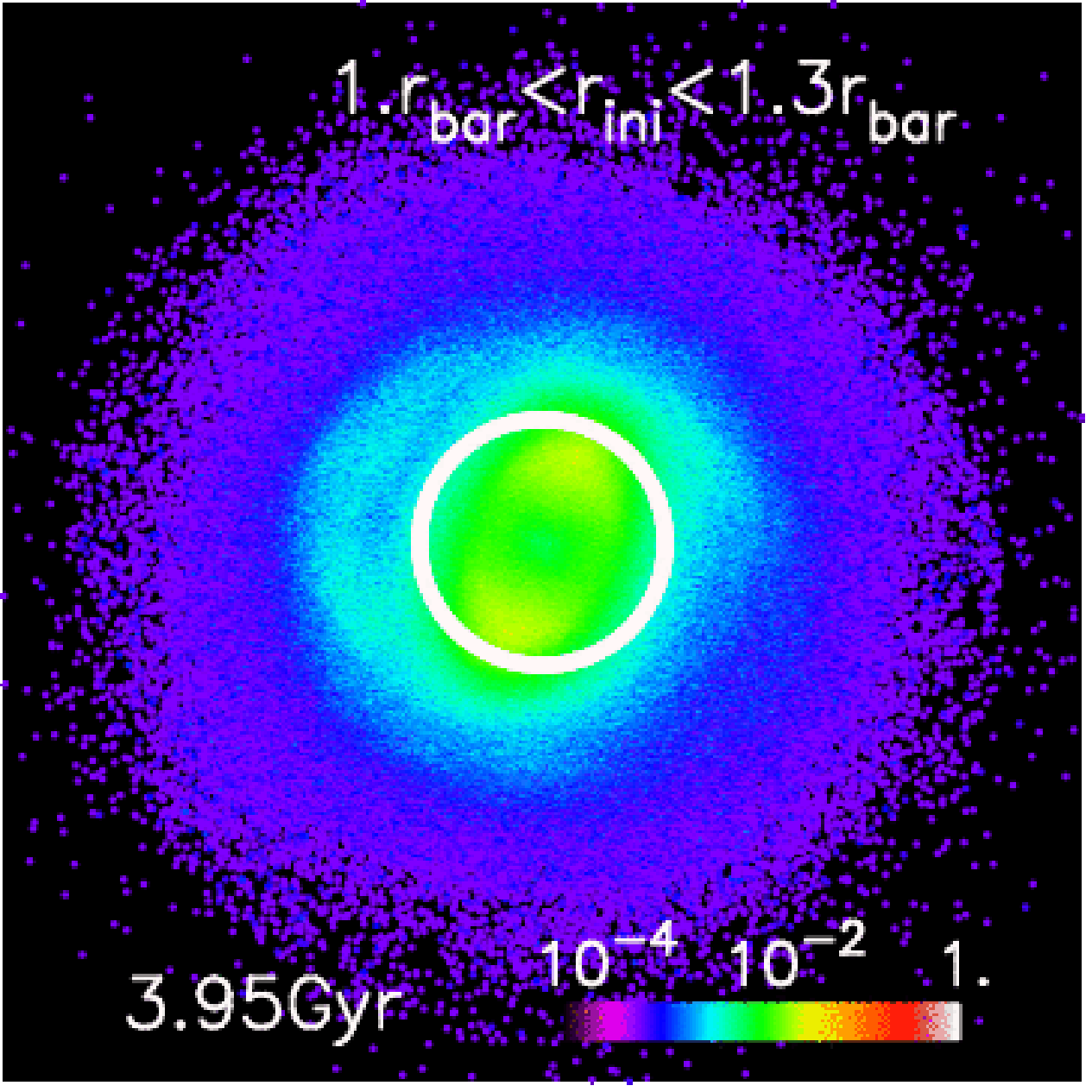}
\includegraphics[width=4.cm,angle=270]{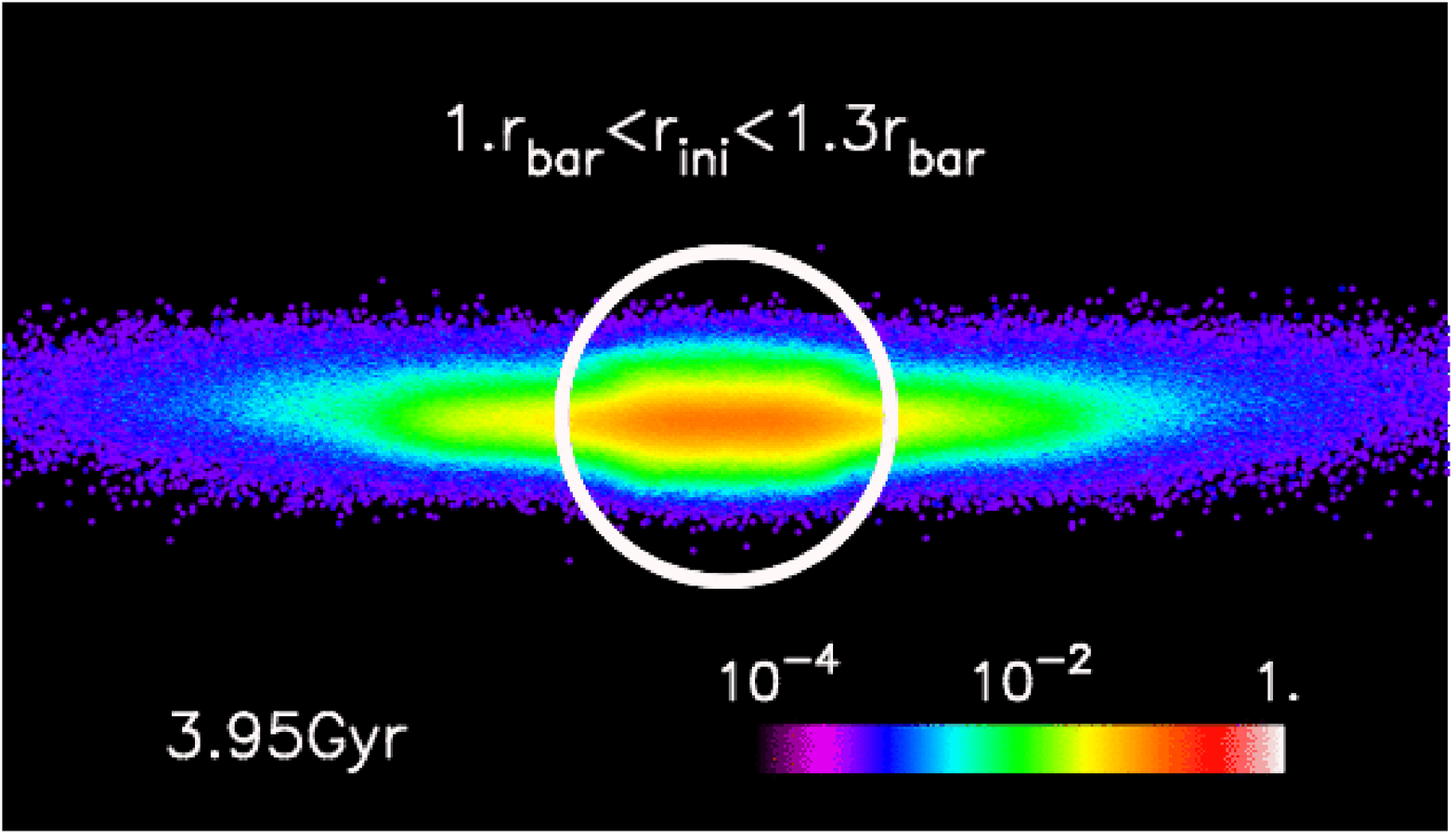}
\includegraphics[width=4.cm,angle=270]{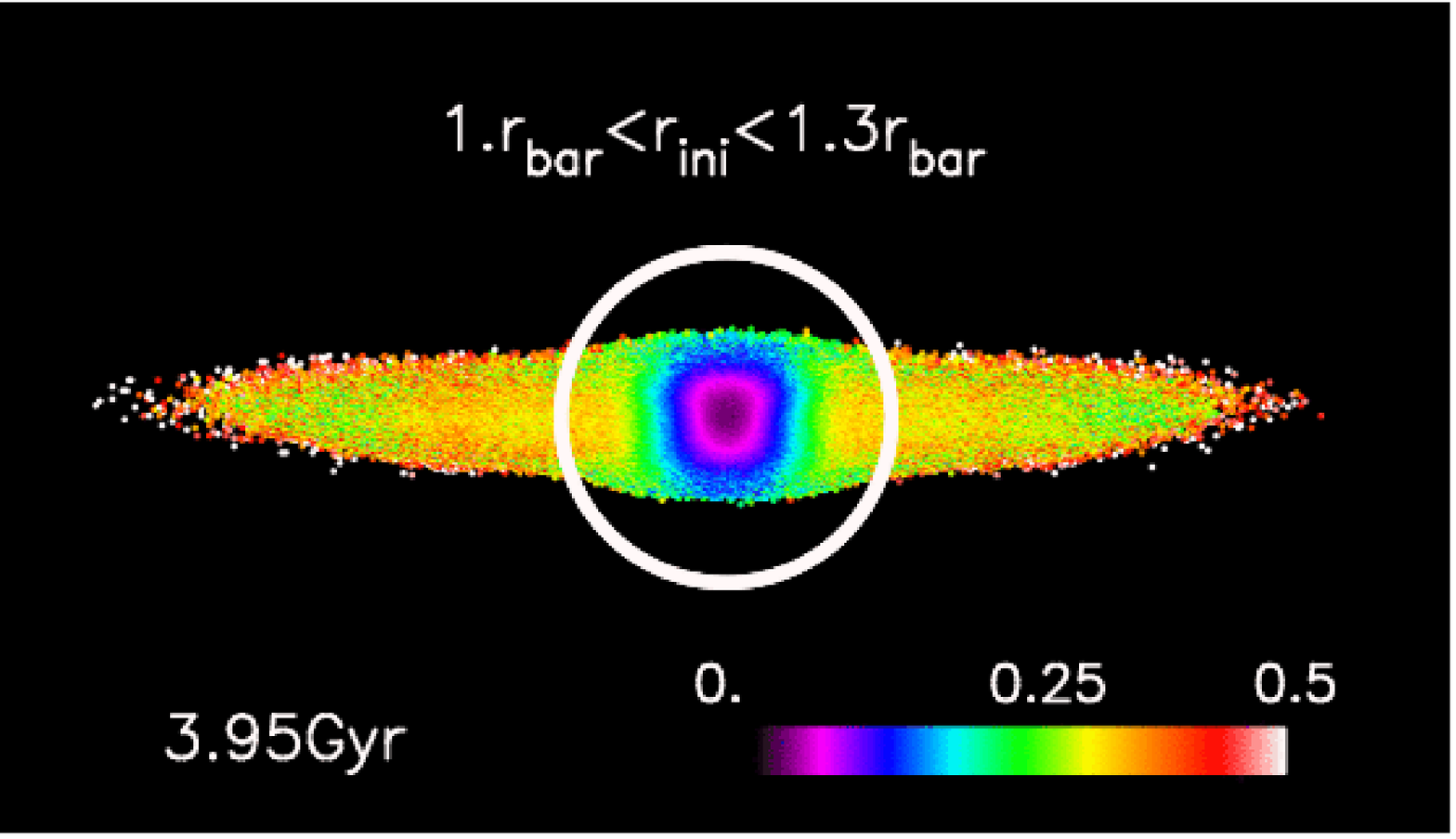}

\includegraphics[width=4.cm,angle=270]{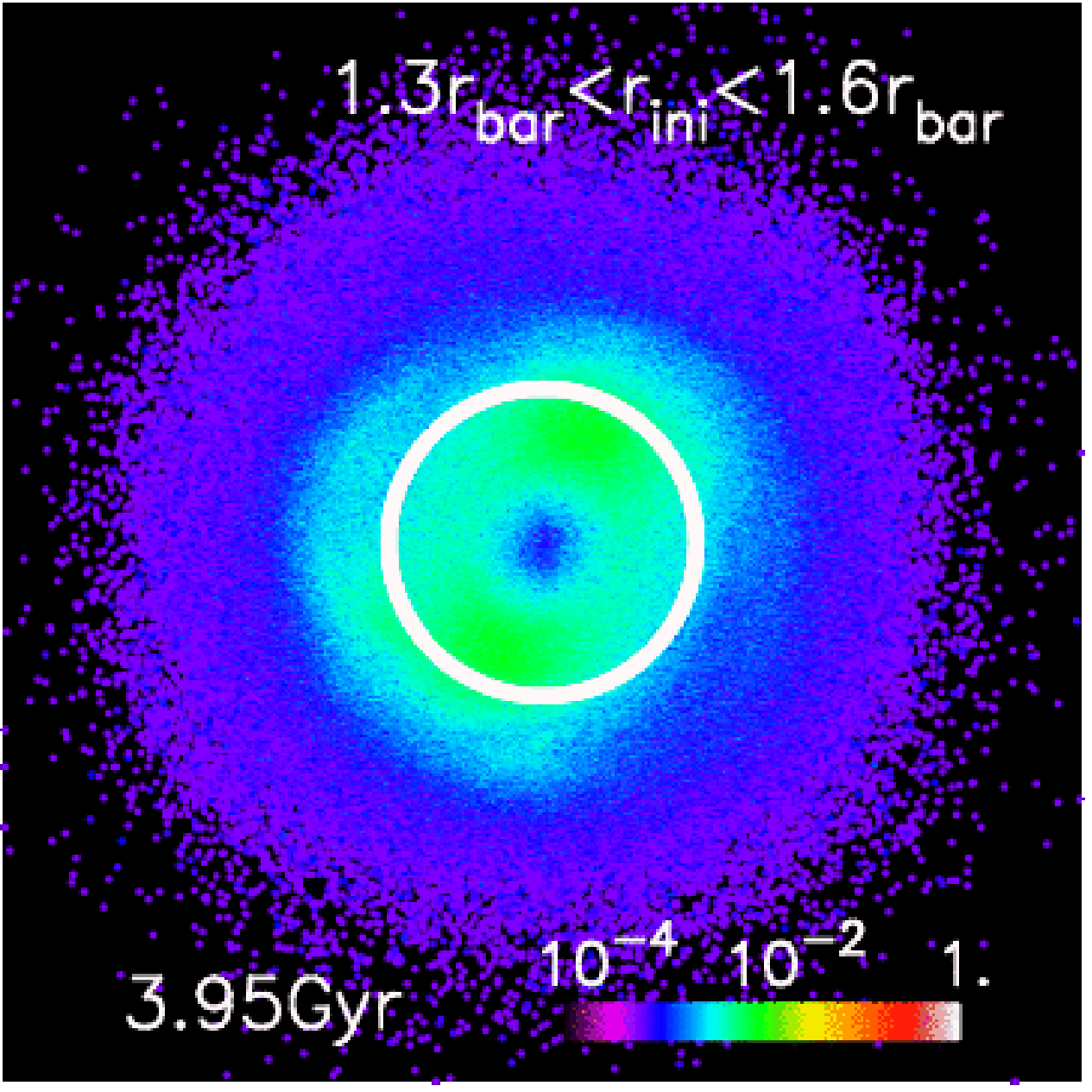}
\includegraphics[width=4.cm,angle=270]{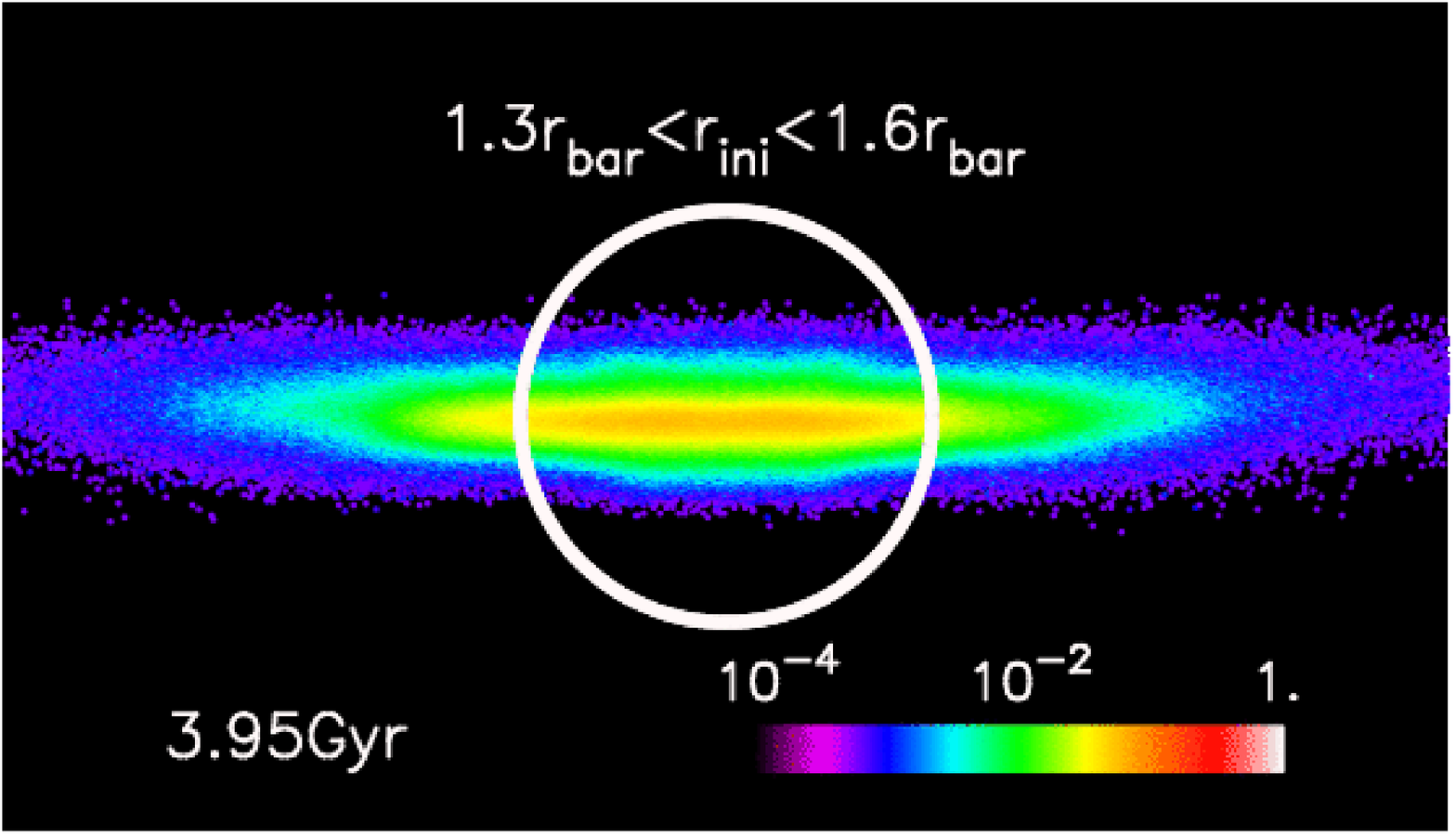}
\includegraphics[width=4.cm,angle=270]{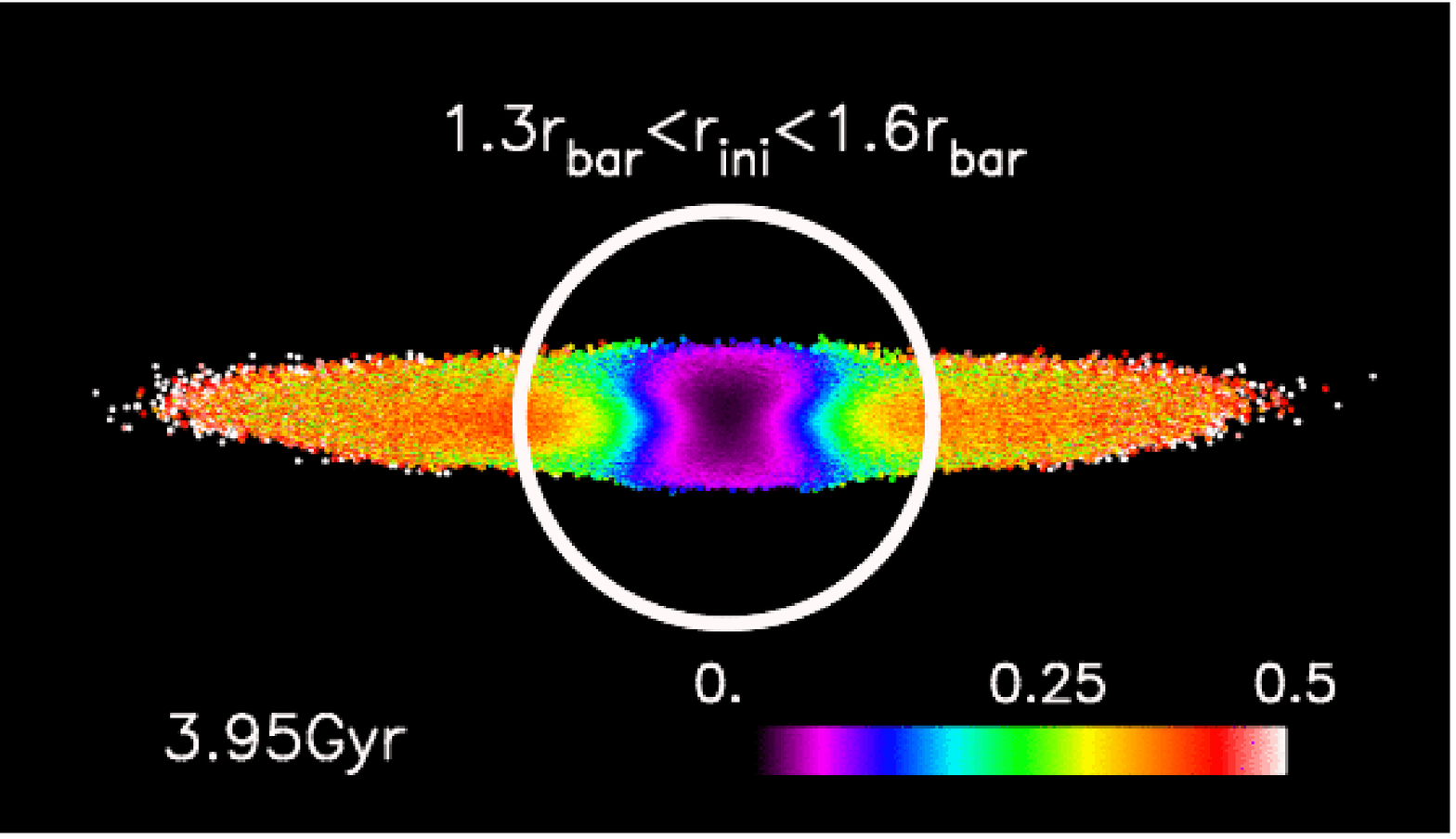}

\includegraphics[width=4.cm,angle=270]{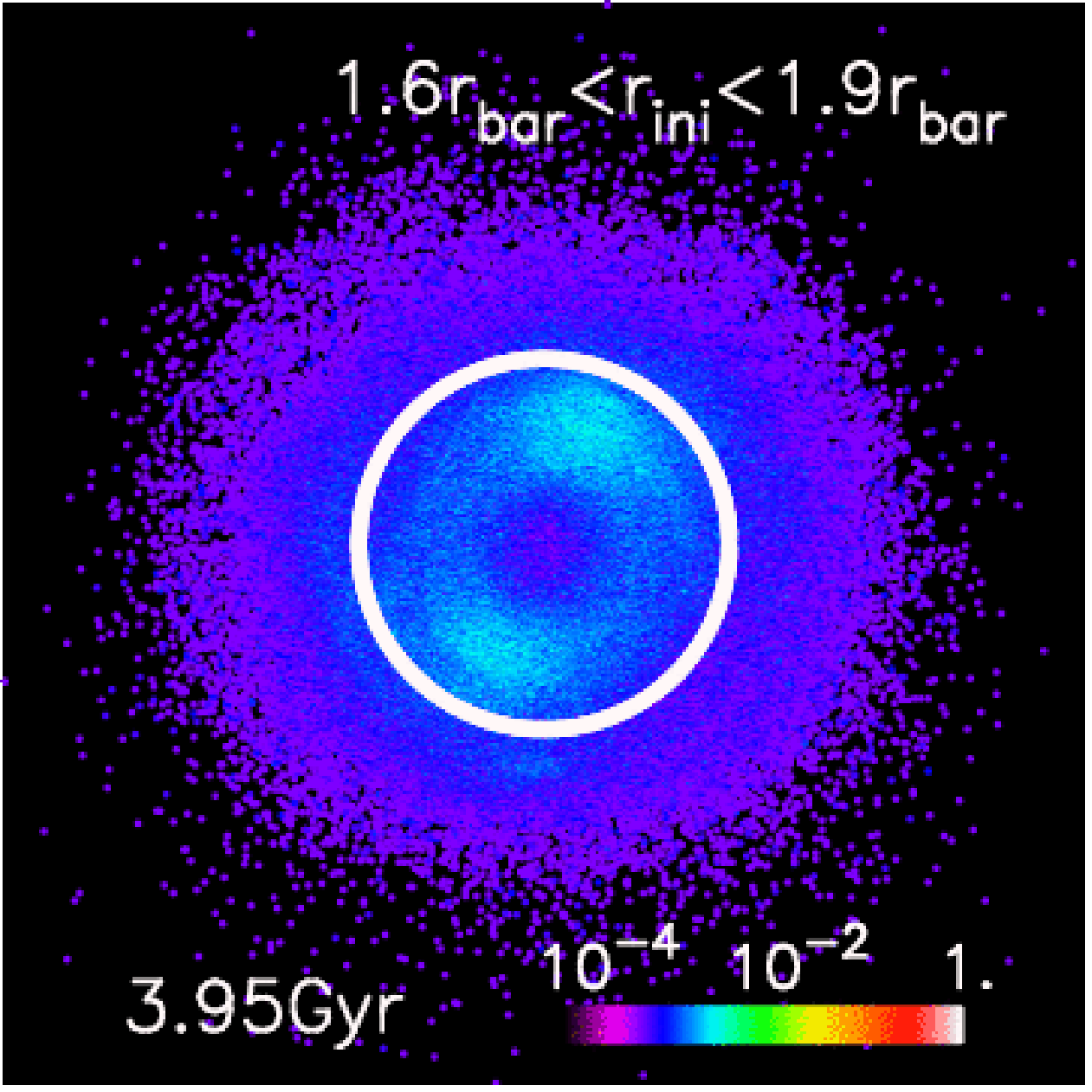}
\includegraphics[width=4.cm,angle=270]{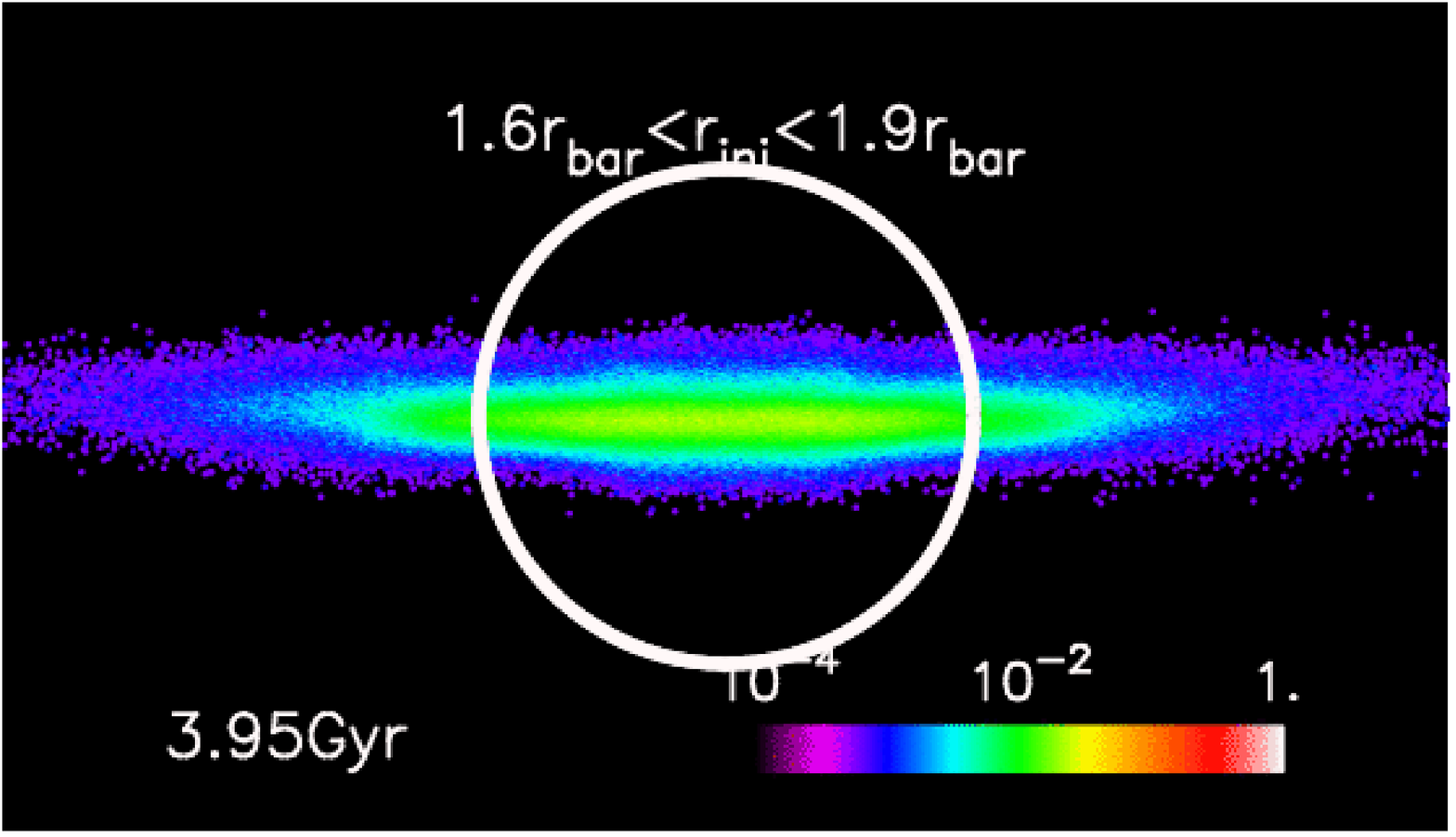}
\includegraphics[width=4.cm,angle=270]{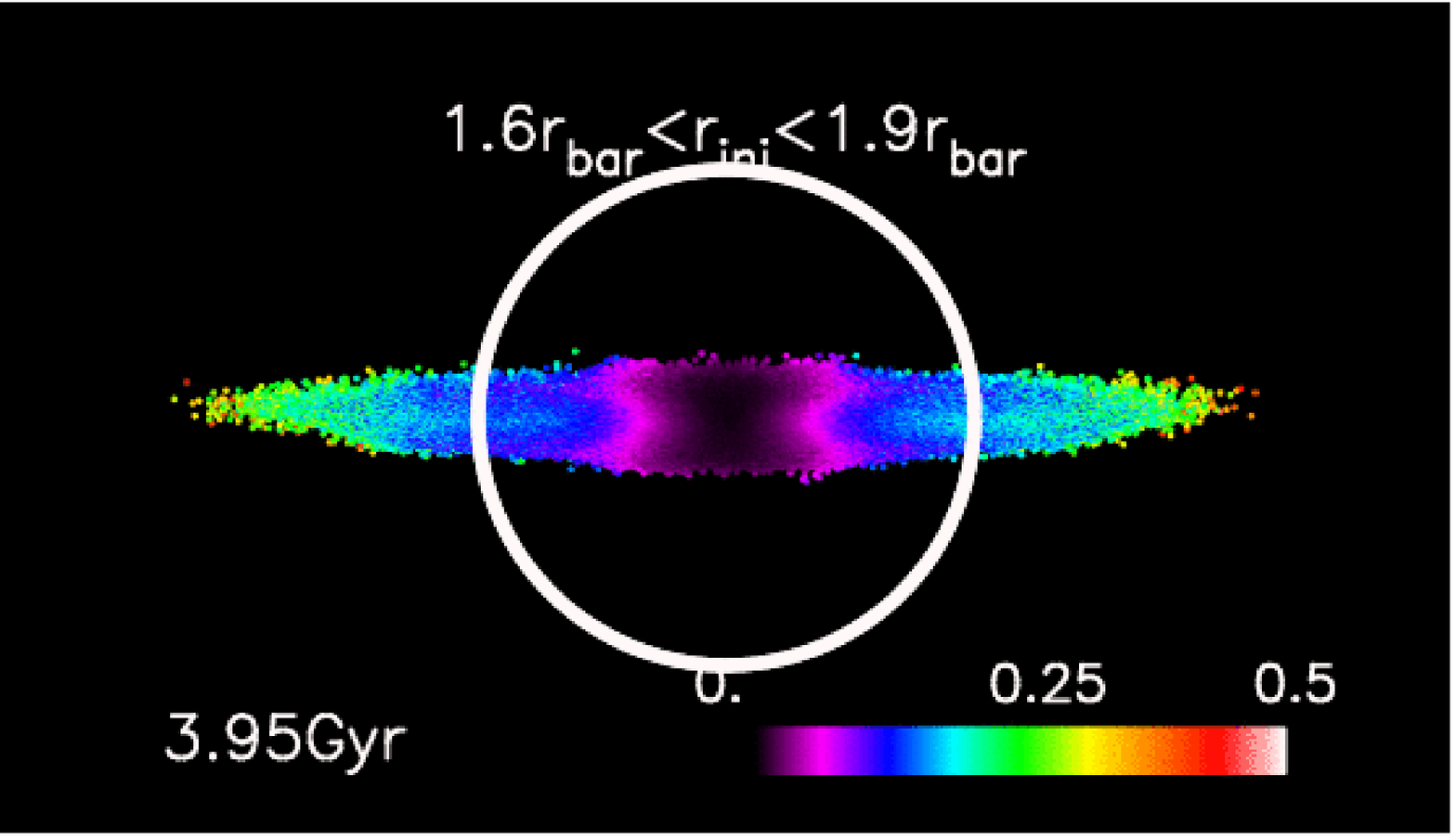}

\caption{(\emph{Left and central panels}): Face-on and edge-on density
distribution of stars born at different radii from top to bottom:
$\rm{r_{ini}} \le 0.4r_{bar}$; $0.4r_{bar} \le \rm{r_{ini}} \le
0.7r_{bar}$; $0.7r_{bar} \le \rm{r_{ini}} \le r_{bar}$; $r_{bar}\le
\rm{r_{ini}} \le 1.3r_{bar}$; $1.3r_{bar} \le \rm{r_{ini}} \le
1.6r_{bar}$; $1.6r_{bar} \le \rm{r_{ini}} \le 1.9r_{bar}$. The
color scale is in the same units in all left and middle panels.
(\emph{Right panel}): Fractional contribution of stars born at different
radii to the total local stellar density. All the plots correspond to
$t=$3.95 Gyr, as indicated. The average initial radius is indicated by
a white circle and the bar is inclined by 20 degrees with respect to
the observer's line-of-sight.}
\label{maps00_regrot}
\end{figure*}

\begin{figure*}
\centering
\vspace{1.5cm}
\includegraphics[trim = 0cm 0cm 2cm 0cm, clip=TRUE, width=4.5cm,angle=270]{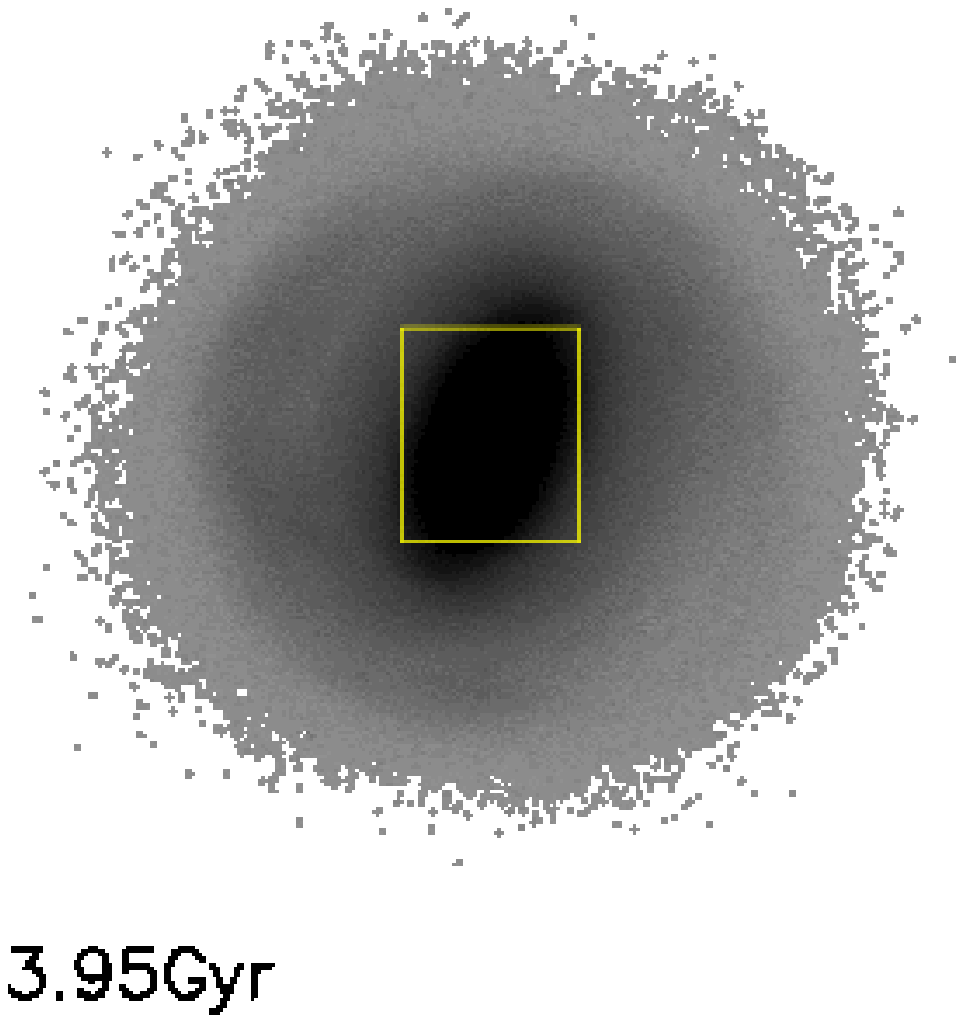}
\includegraphics[trim = 0cm 0cm 2cm 0cm, clip=TRUE, width=4.5cm,angle=270]{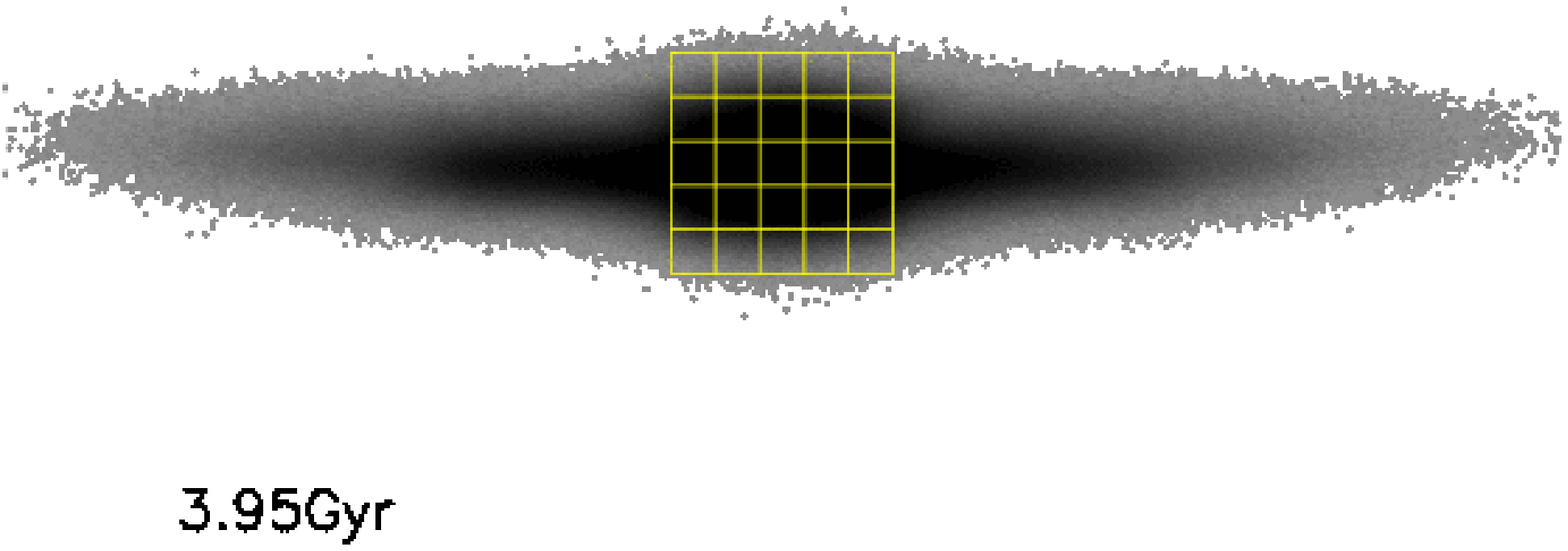}
\includegraphics[width=11.cm,angle=270]{phistorininoforegroundrot2_gS0_q1p8_BD0p00.ini.ps}
\caption{Histograms of the birth radii of stars which populate the boxy
bulge at $t=3.95$~Gyr, for a simulated galaxy with B/D=0. In each panel,
stars have been sorted according to their birth radius:  $\rm{r_{ini}} \le 0.4r_{bar}$ (red histogram); $0.4r_{bar} \le \rm{r_{ini}} \le
0.7r_{bar}$ (green histogram); $0.7r_{bar} \le \rm{r_{ini}} \le r_{bar}$ (blue
 histogram); $r_{bar}\le
\rm{r_{ini}} \le 1.3r_{bar}$  (cyan histogram); $1.3r_{bar} \le \rm{r_{ini}} \le
1.6r_{bar}$ (purple histogram); $1.6r_{bar} \le \rm{r_{ini}} \le 1.9r_{bar}$ (yellow histogram).
%$\rm{r_{ini}}  \le$ 3~kpc (red histogram); 3~kpc $\le \rm{r_{ini}}  \le$
%5~kpc  (green histogram); 5~kpc $\le \rm{r_{ini}}  \le$ 7~kpc  (blue
%histogram); 7~kpc $\le \rm{r_{ini}}  \le$ 9~kpc  (cyan histogram);
%9~kpc $\le \rm{r_{ini}}  \le$ 11~kpc  (purple histogram); 11~kpc $\le
%\rm{r_{ini}}  \le$ 13~kpc  (yellow histogram). 
Each panel in the bottom
figure corresponds to an element of the grid shown in the top-right
figure. For each of those elements, the contribution of stars of a
given provenance in the disk has been normalized to the total number
of stars which populate that element. Only stars in the bar region,
that is inside the yellow rectangle shown in the top-left figure, have
been included in this plot.}
\label{grid00rot}
\end{figure*}

\subsubsection{Angular momentum}

The spatial redistribution of stars in the disk initiated at the
epoch of bar formation is a consequence of AM redistribution, as
previously shown in several papers \citep[e.g.][, among others]{min10,
bru11, min11}. Figure~\ref{DL_reg} (left column) shows maps of face-on
projections in the variation of angular momenta, $\Delta L$, of stars
born at different radii. This variation is evaluated at times between
$t=0$ and $t=2.75$~Gyr. At this epoch, $t=2.75$~Gyr, the bar has already
acquired its boxy/peanut shaped morphology. As expected, these maps
clearly show that for outside-in migrators trapped in the bar region,
those with larger birth radii experience larger AM changes. However, it is
interesting to note that, even if outside-in migrators have experienced
the largest changes, their final AM still retains the memory of their
initial birth radii: Fig.~\ref{DL_reg}, middle column, shows projected
face-on maps of $L_{norm}=L/L_{circ}$, that is of the AM, at
time \emph{t}, normalized by the AM of the corresponding circular
orbit at that radius, at the same time \emph{t}. As usual,
stars have been selected with respect to their initial provenance in the
disk. From these plots, it is evident that, in the bar region, the AM
content depends on the birth radius, and that the AM content increases for
stars with larger birth radii.  For example, among the stars ultimately
found in the bar, those born inside 0.4$r_{bar}$ have, on average,
an AM content 6-7 lower than those of stars born beyond 1.6$r_{bar}$
that have subsequently migrated in the central kpcs. In other words,
in the bar region, \emph{the larger the initial birth radius of a star,
the greater its final AM}. As we will discuss in the following sections,
this finding has important consequences for the spatial redistribution
and for the kinematics of stars in the boxy/peanut-shaped structure. Note also that stars born
inside the VILR show also an interesting trend in the vertical velocity
dispersion, $\sigma_z$ (Fig.~\ref{DL_reg}): inside the bar region,
stars which have smaller birth radii also have the lower $\sigma_z$.

\subsection{Dissecting the structure of boxy/peanut bulges}\label{dissecting}

If the stellar bar is the result of the mixing of various stellar
populations with different AM contents and vertical velocity dispersions,
it is natural to investigate how these populations are mapped into the
vertical structure of the boxy/peanut-shaped bulge.

In Fig.~\ref{maps00_regrot}, we show face-on and edge-on maps of the final
evolved stat of our modeled bulgeless disk (i.e. $t=3.95$~Gyr). Stars
have been classified according to their birth radii. To facilitate the
comparison to the Milky Way bar, the final configuration has been rotated
so that the bar viewing angle is about 20 degrees \citep[][see
also Wegg \& Gerhard 2013 for a more recent estimate of the bar
inclination]{bissantz02, shen10}. Fig.~\ref{maps00_regrot} shows the
resulting face-on maps of stars which originate from six different
annuli (defined in Sect.~\ref{dissecting_faceon}).  Stars born in the
outer regions of the simulated disk that are close to the OLR,
 are found in the central kpcs of the galaxy and are part of the bar
structure. From the corresponding edge-on maps (Fig.~\ref{maps00_regrot})
one sees that stars with different birth radii do not redistribute
in the same way in the boxy/peanut shaped structure. Stars born in
the innermost disk regions tend to make up populations which exhibit
rounder isophotes in the projected maps than the population made up
of outside-in migrators.  The shape of the stellar distribution indeed
becomes increasingly boxy/peanut-like for stars which migrated inwards
from the outer disk or were born around the VILR. Even stars which
migrated from the edge of the disk participate in the formation of the
thick bar (most keenly visible in the edge-on view of their stellar
distribution which shows the characteristic X-shape profile). This is a
consequence of the AM redistribution initiated by the formation of the
bar: these stars migrate from the outer to the inner disk at the epoch
of bar growth (see Figs.~\ref{asym} and \ref{redistrib}), and are, at
the time of bar buckling, captured at the VILR and thus participate in
the orbital families supporting the boxy/peanut shaped structure.

The dependence of the isophotal shape on individual populations with
different birth radii implies that, at any given location in the bulge,
stars of different provenance have different relative contributions
to the local stellar density. This is clearly shown in the edge-on
maps of Fig.~\ref{maps00_regrot}, where the fractional contribution
to the local stellar density of stars of different provenance are
elucidated. Stars born in the inner disk ($r_{ini}\le 0.4r_{bar}$)
dominate the local stellar density only in the very inner regions of the
bulge, inside $\sim$0.4$r_{bar}$, and their contribution to the local
density decreases very quickly when moving horizontally or vertically
out from the center. The relative contributions of stars of increasing
birth radii reaches a maximum in progressively more distant regions
of the bulge. Also stars born outside the VILR are found in the bulge and their contribution
is most significant at its outer edges.

Another way of showing this inhomogeneous mapping of a stellar disk into
a boxy bulge is shown in Fig.~\ref{grid00rot}, where we have divided the
edge-on ($x-z$) projection of our modeled bulge into a 5$\times$5 grid,
with each panel in the grid having a size of 0.3$r_{bar}$ x 0.3$r_{bar}$,
divided in such a way as to cover the whole extent of the modeled bulge
(that is a $[-0.7r_{bar},0.7r_{bar}] \times [-0.7r_{bar},0.7r_{bar}]$
region; see the top-right panel for a x-z view of the grid). To avoid
contamination from foreground and background stars, in analogy with what
commonly is done in bulge surveys,  we have selected stars only inside the
bar region, which in our model corresponds to the region defined by $|y|
\le 0.8r_{bar}$ (see the top-left panel in Fig.~\ref{maps00_regrot}).
The resulting panels in Fig.~\ref{grid00rot} show the fractional
contribution to the local stellar density of stars born in different
regions of the disk.

From this analysis, we can deduce that:

\begin{itemize}

\item the fractional contribution to the boxy bulge of stars of
different provenance changes both with latitude (i.e. vertically)
and with longitude (i.e. horizontally);

\item  for increasing longitude at any given latitude, the contribution
of outside-in migrators increases, in agreement with the dependence of the
AM content on the star birth radius (Fig.~\ref{DL_reg}).  For example, for
the region, $-0.1r_{bar}<$x$<0.1r_{bar}$ and $-0.1r_{bar}<$z$<0.1r_{bar}$,
only 5$\%$ of stars have birth radii greater than $r_{bar}$ (i.e.,
external to the bar region itself). This fraction increases to more
than 25$\%$ at the edge of the bar, close to the galaxy midplane
($-0.7r_{bar}<$x$< -0.4r_{bar}$ and $-0.1r_{bar}<$z$< 0.1r_{bar}$;
$0.4r_{bar}<$x$<0.7r_{bar}$ and $-0.1r_{bar}<$z$<0.1r_{bar}$);

\item the redistribution depends also on latitude, as expected from the
trends found for $\sigma_z$ (Fig.~\ref{DL_reg}). Along the bar minor
axis, for example, at low latitudes about 50$\%$ of the stars have $r_{ini}
\le 0.4r_{bar}$, while, at high latitudes, half of the stars have $r_{ini}
\ge 0.7r_{bar}$, with more than 20$\%$ of the stars having an outer
disk origin ($r_{ini} >r_{bar}$);

\item at the edges of the X-shaped bulge structure, stars whose origin is
external to the bar (i.e. with $r_{ini} > r_{bar}$)  represent about
30$\%$ of the total local density.

\end{itemize}
  
We emphasize that it is stars from all of the disk, from the
center to the OLR, that contribute to the formation and structure of a
boxy bulge, and that stars formed outside the bar region and subsequently
by bar instabilities can represent a significant fraction of the stellar
density even at high latitudes. In the next section, we show how the
birth radius of a star reflects into its subsequent kinematics as a
contributor to the bulge morphology.

\subsection{Imprints of the stars birth radii on the bulge
kinematics}\label{kinematics}

\begin{figure*}
\centering

\includegraphics[width=4.cm,angle=270]{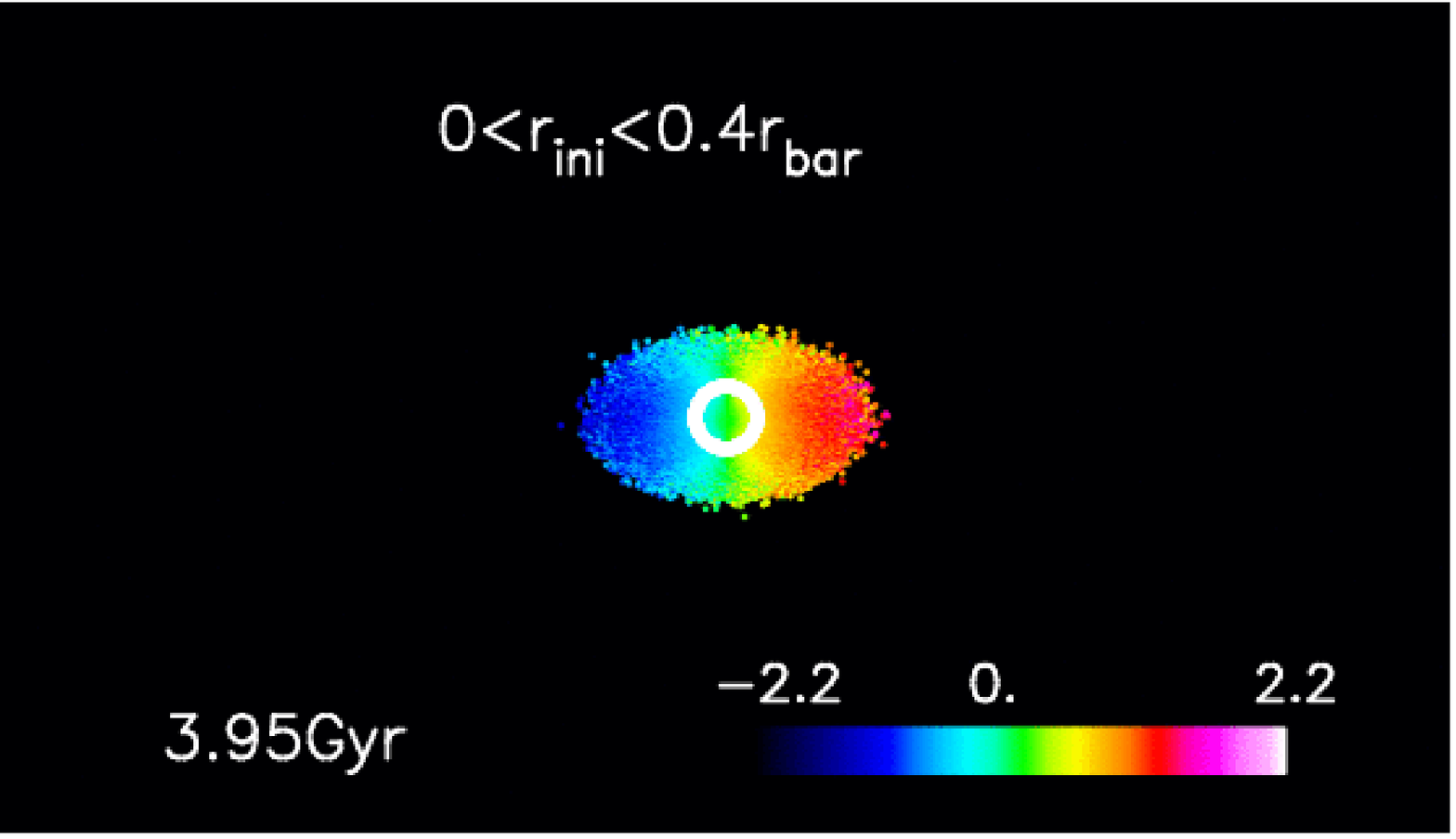}
\includegraphics[width=4.cm,angle=270]{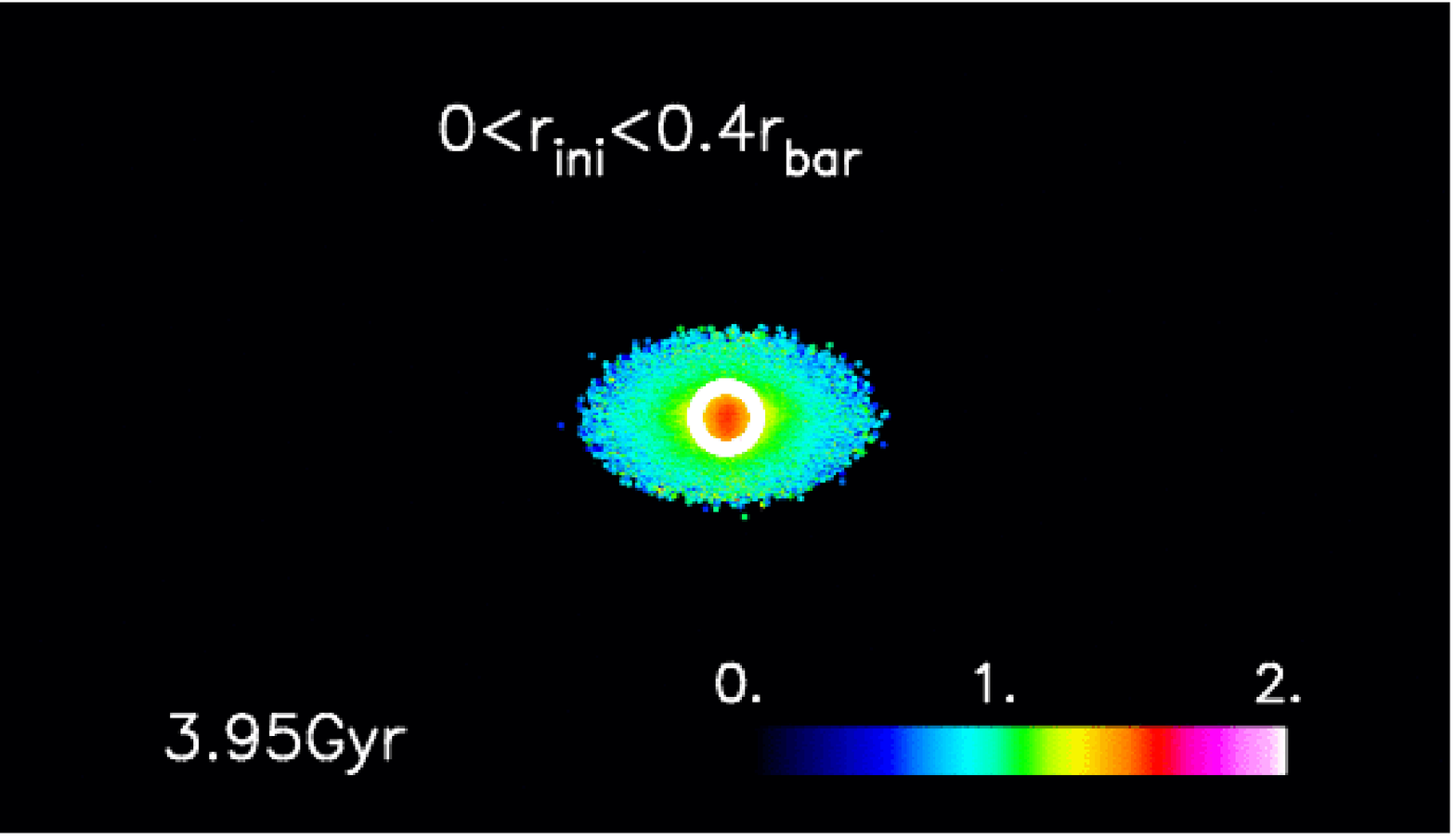}

\includegraphics[width=4.cm,angle=270]{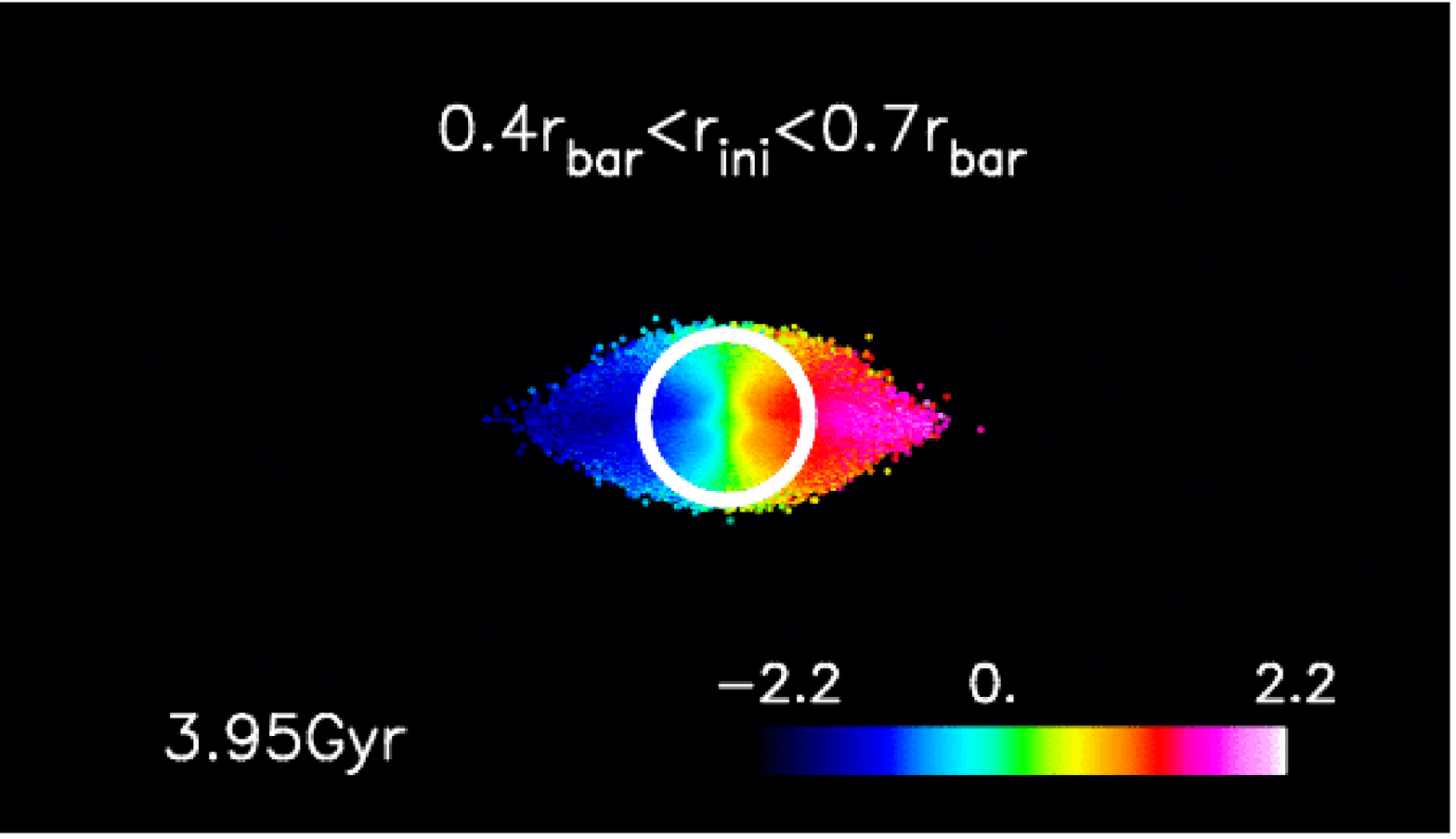}
\includegraphics[width=4.cm,angle=270]{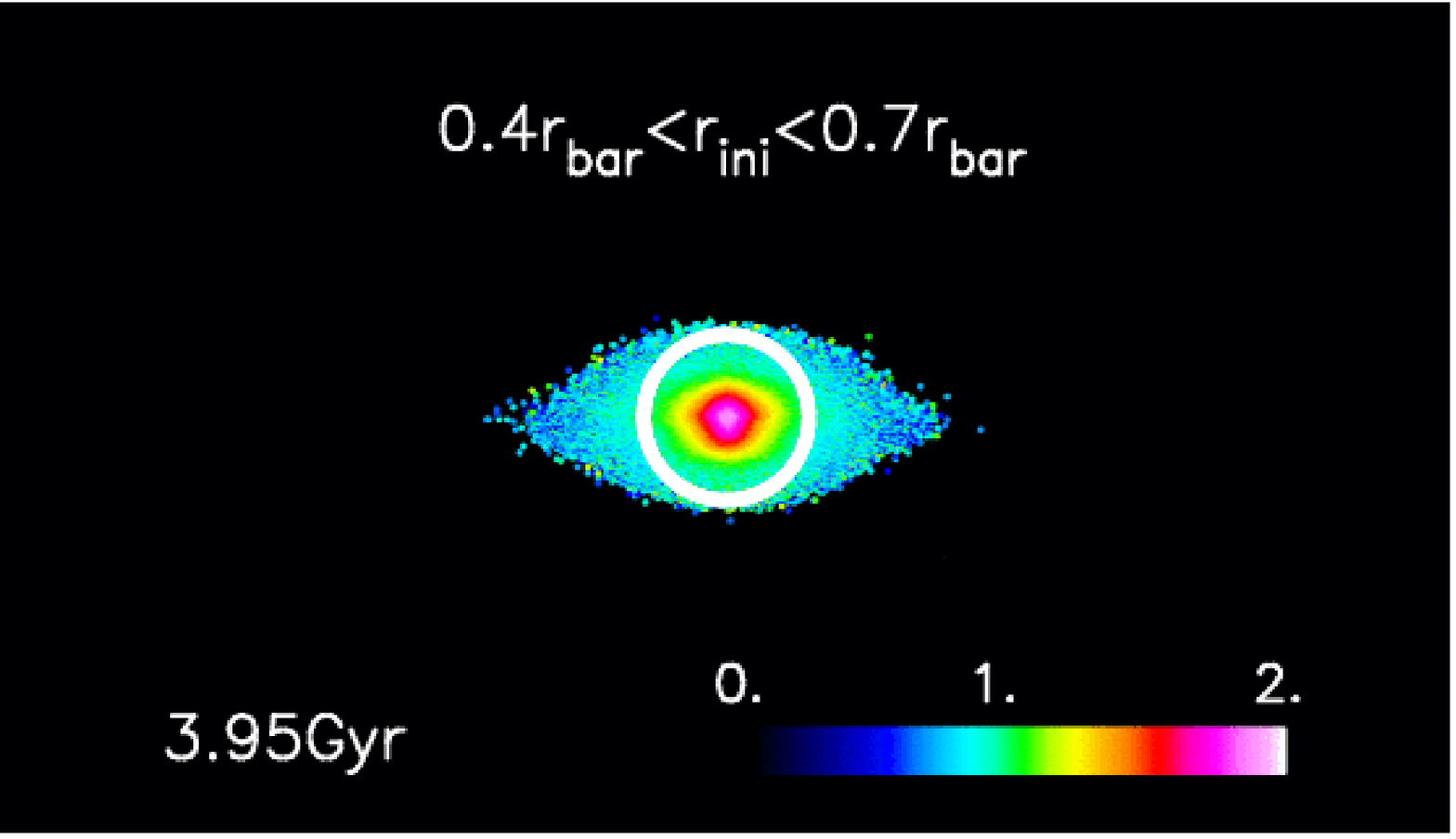}

\includegraphics[width=4.cm,angle=270]{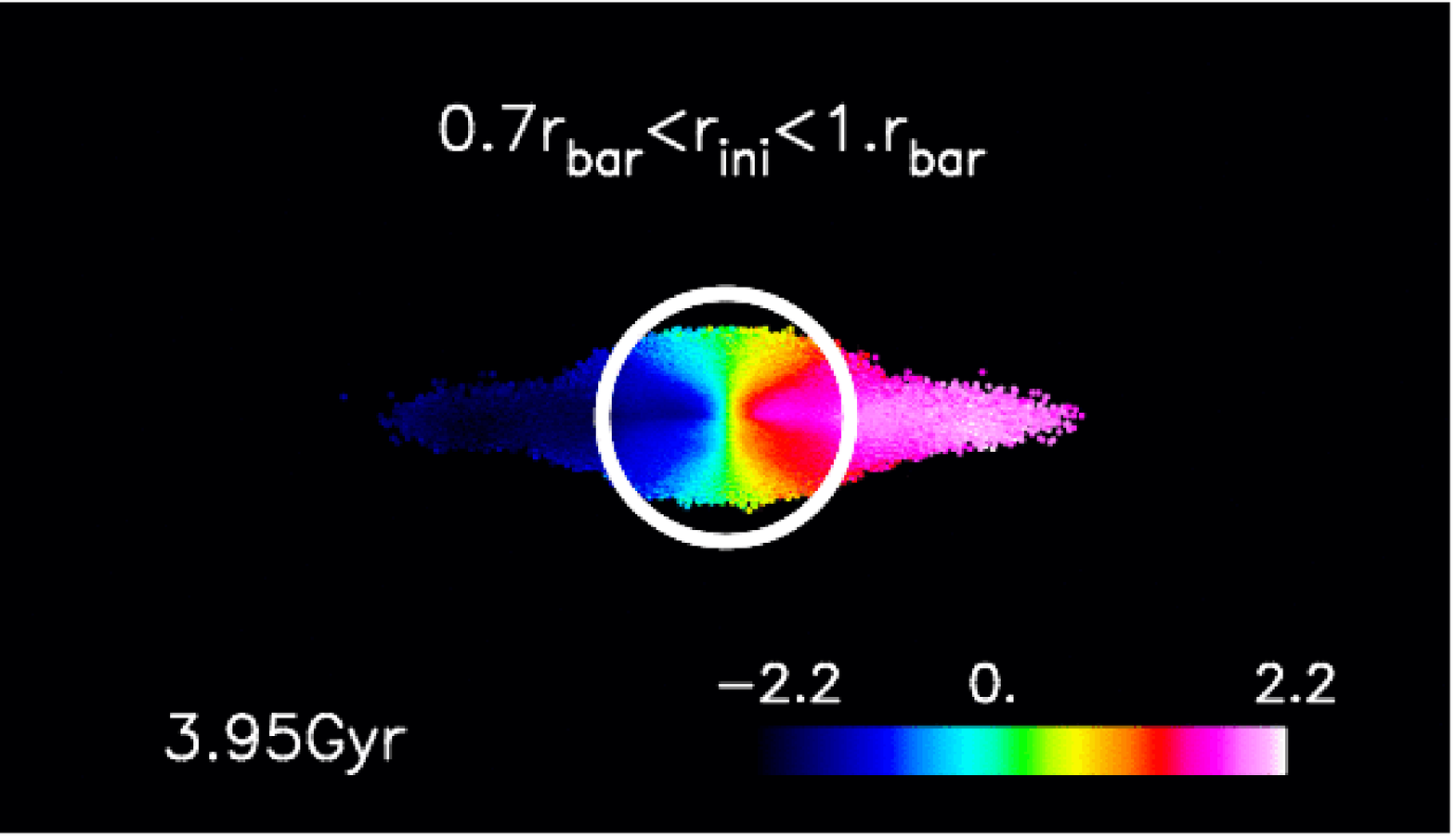}
\includegraphics[width=4.cm,angle=270]{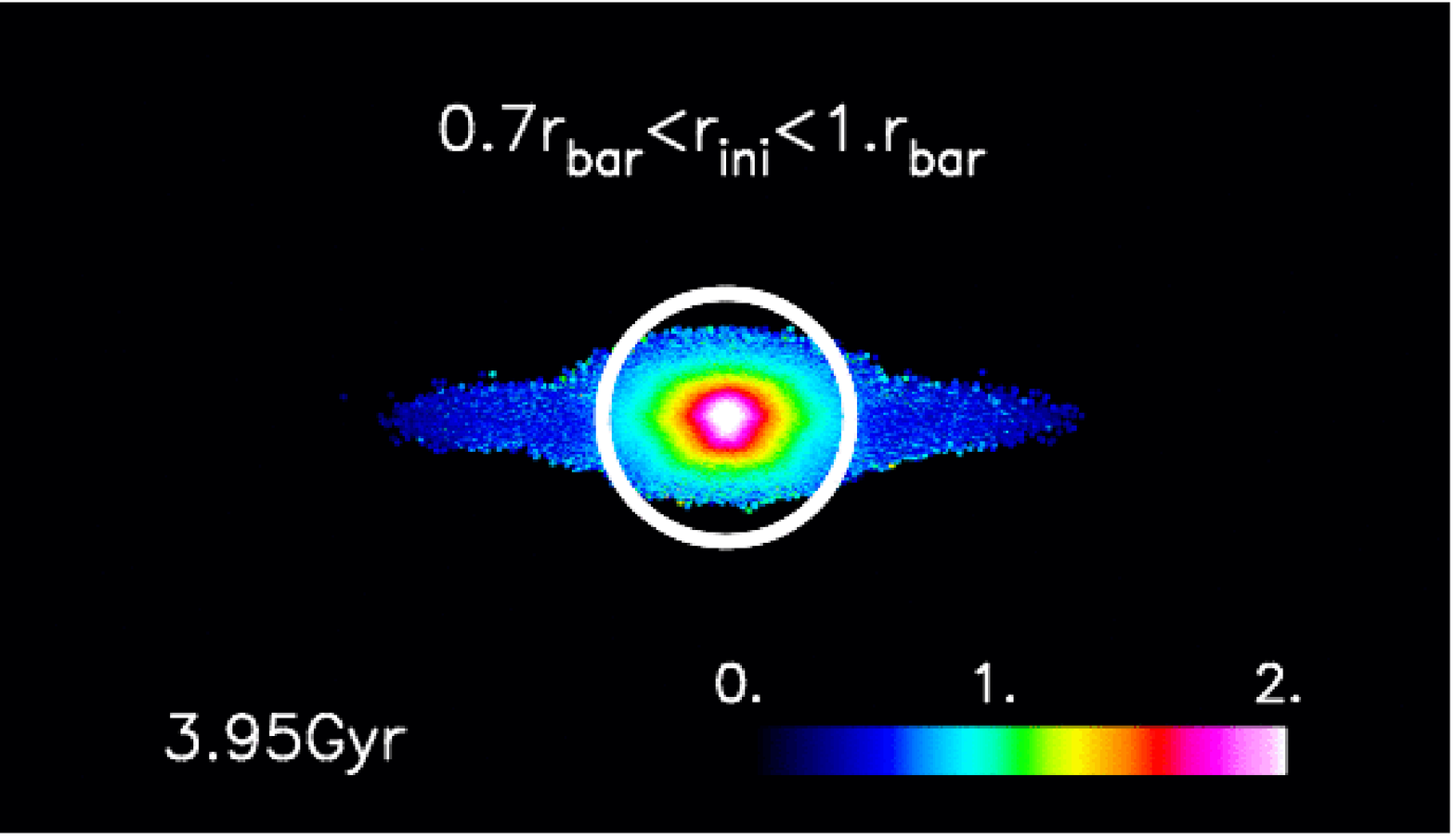}

\includegraphics[width=4.cm,angle=270]{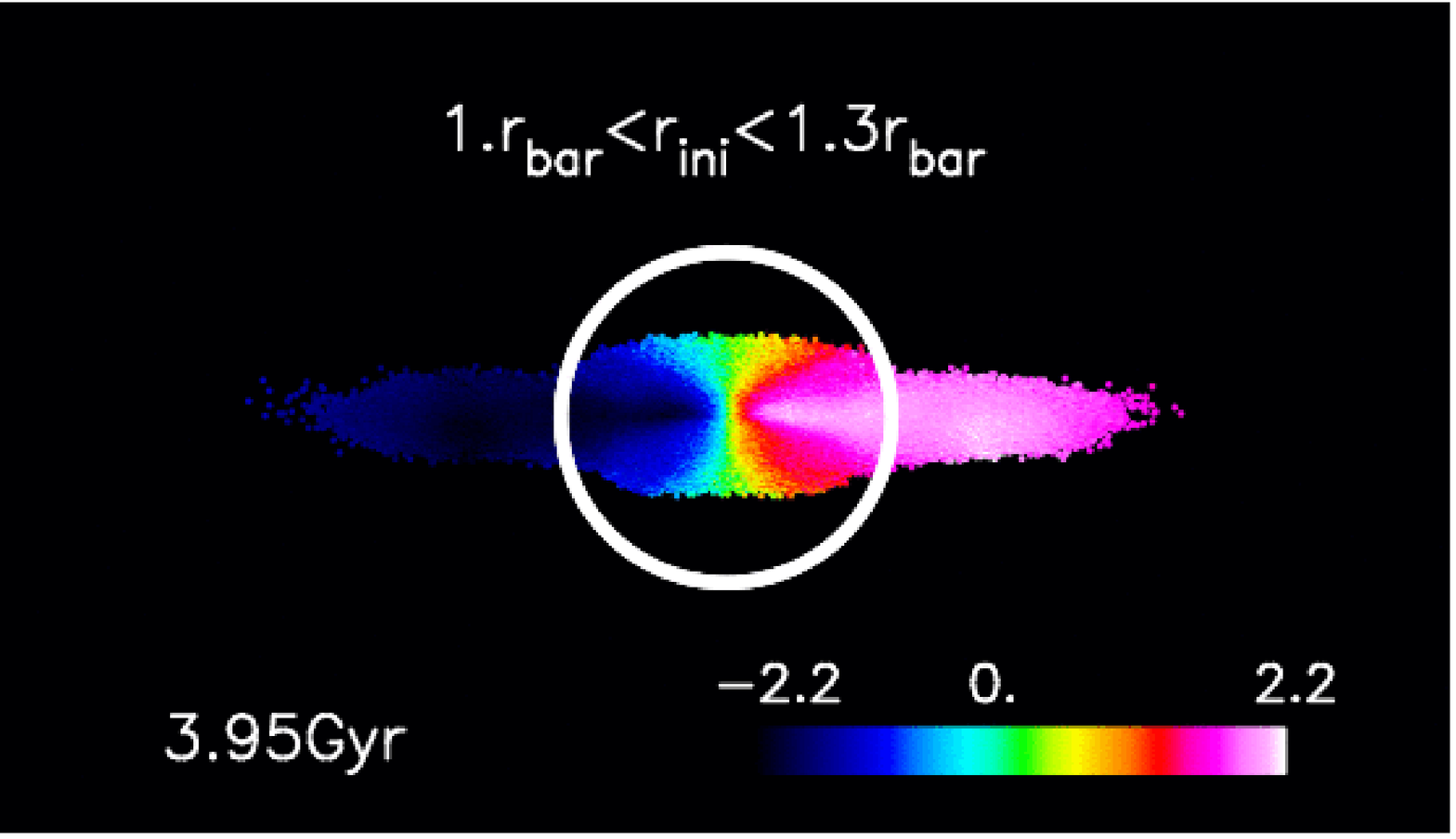}
\includegraphics[width=4.cm,angle=270]{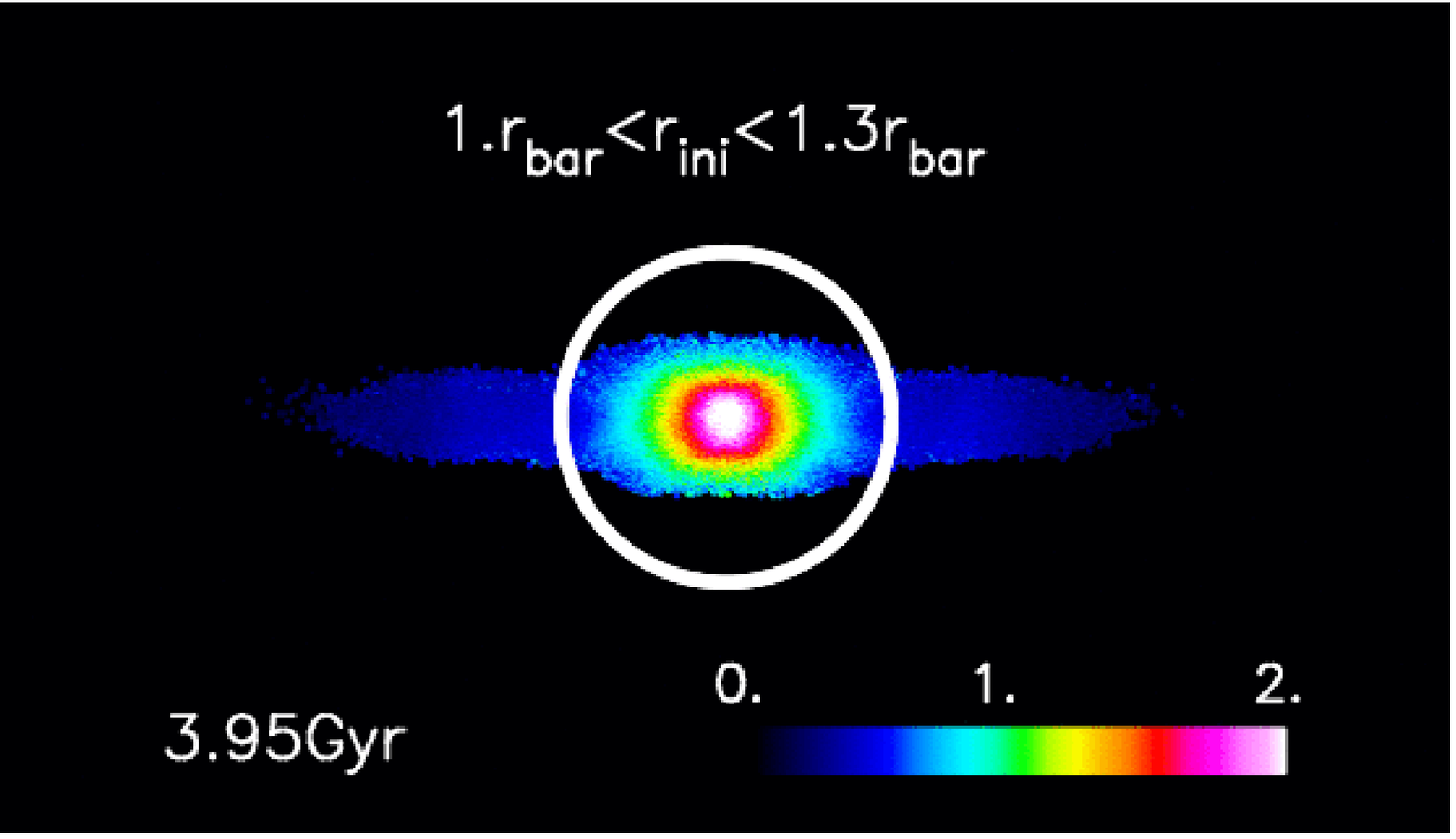}

\includegraphics[width=4.cm,angle=270]{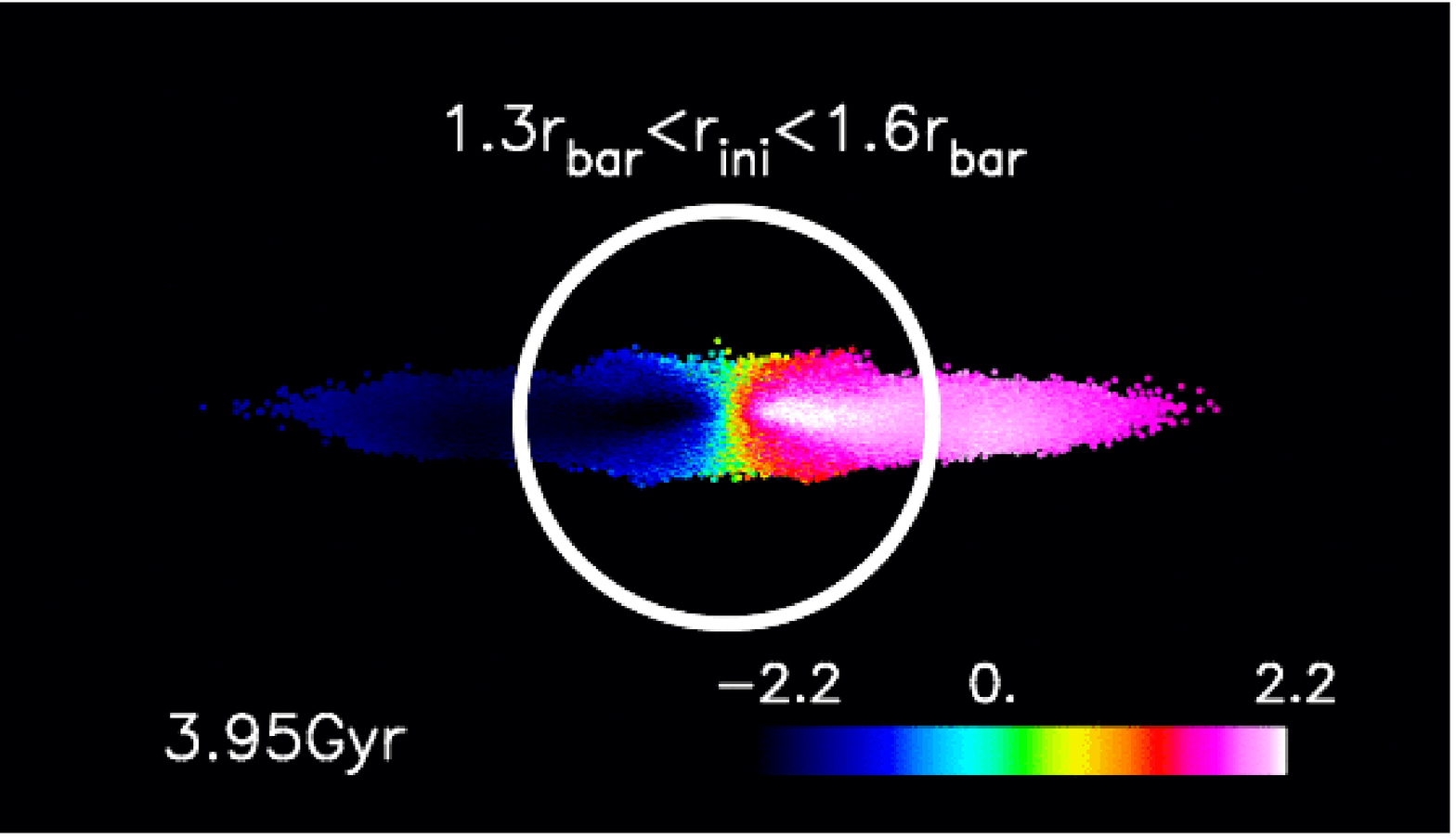}
\includegraphics[width=4.cm,angle=270]{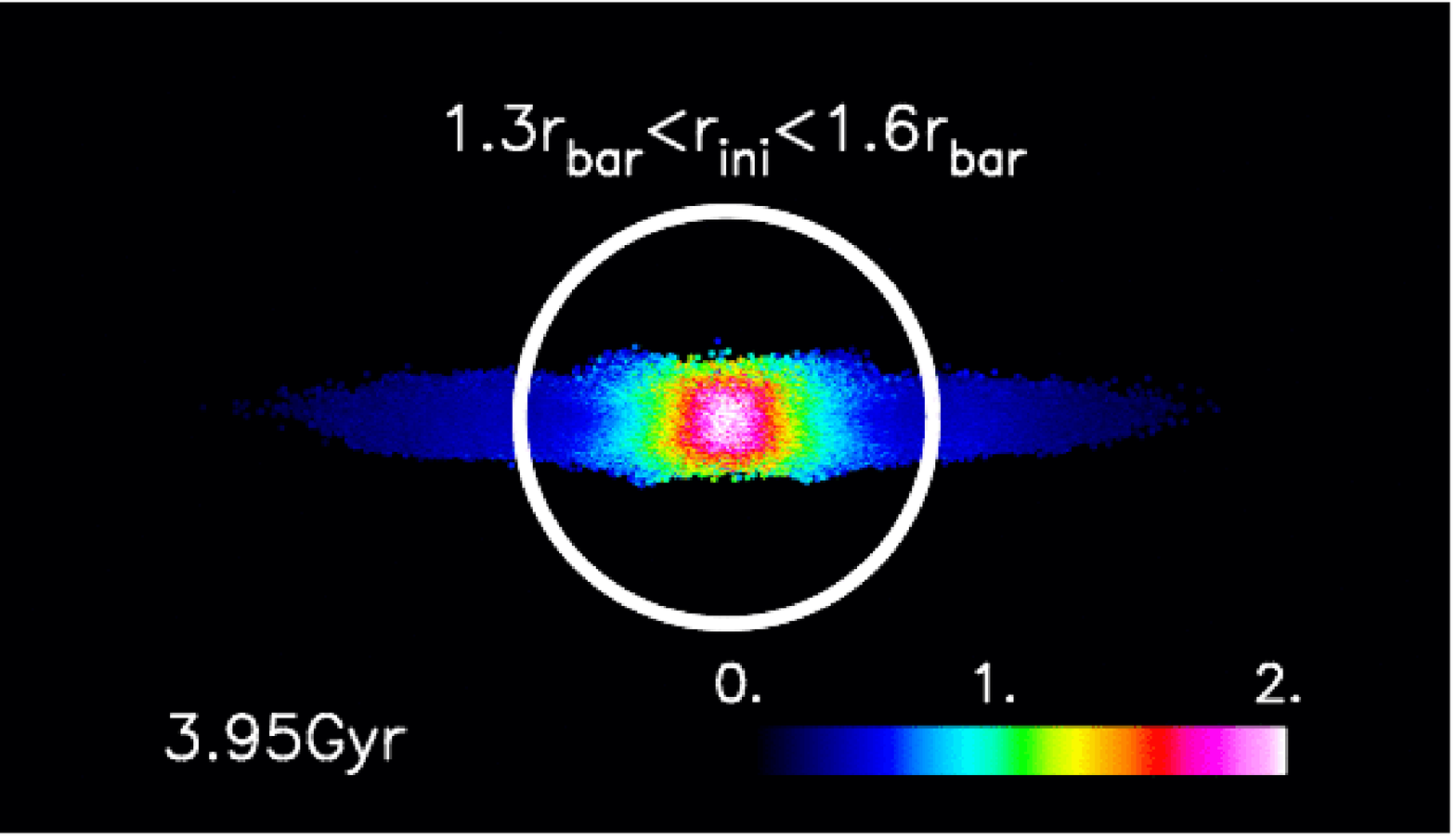}

\includegraphics[width=4.cm,angle=270]{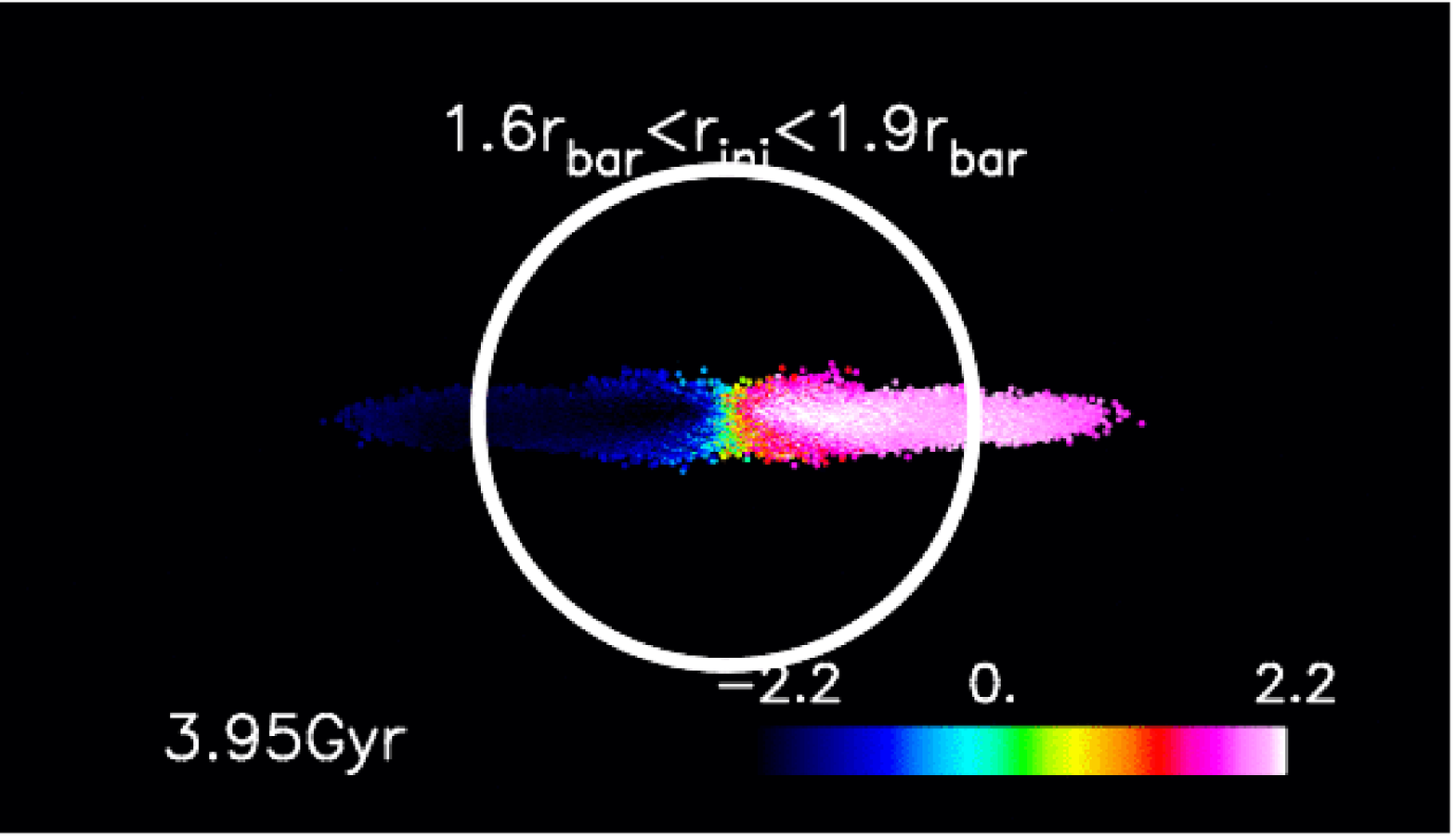}
\includegraphics[width=4.cm,angle=270]{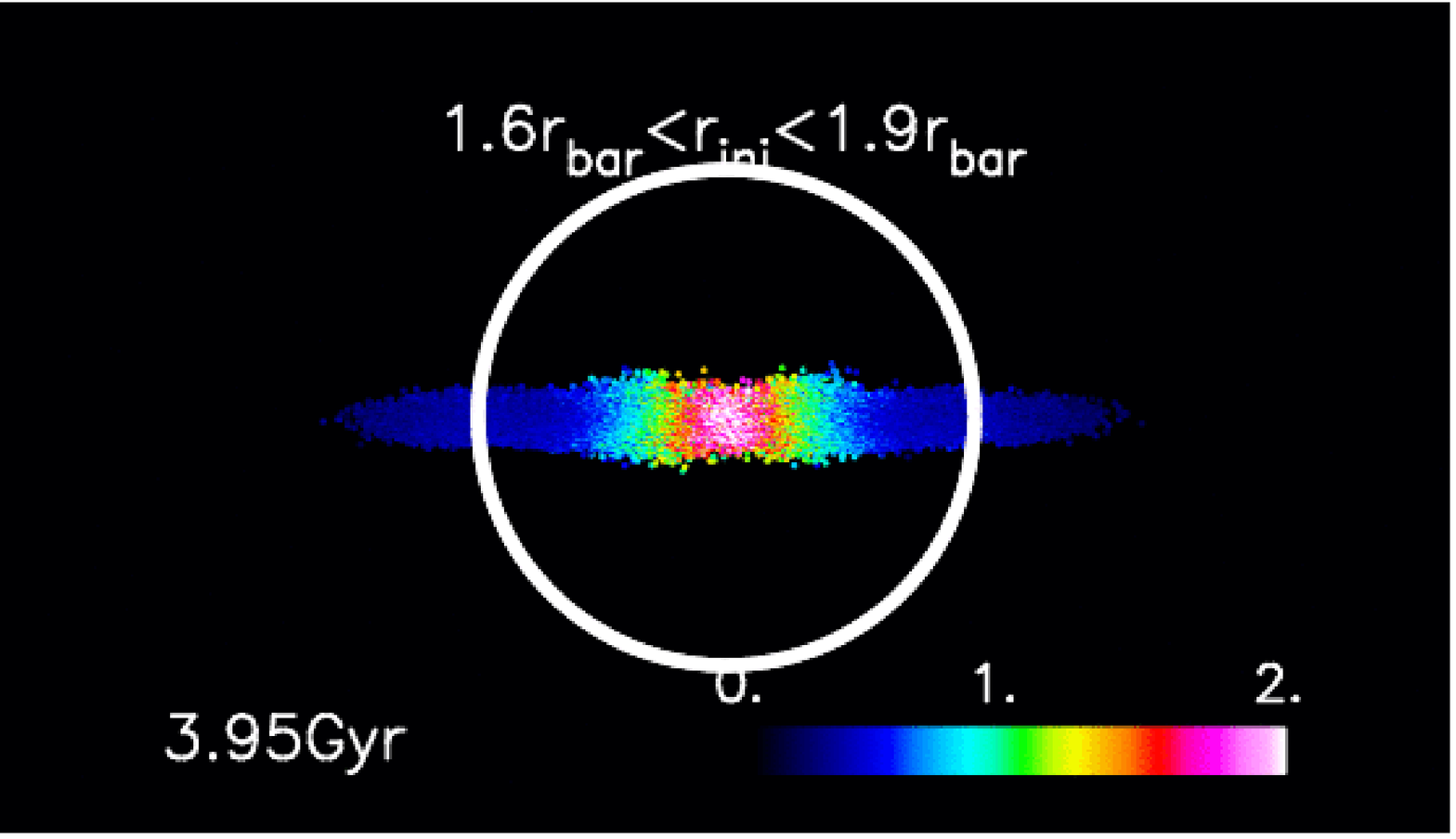}

\caption{Line-of-sight velocities and velocity dispersions maps for
the model with B/D=0. Each panel shows stars with different birth
radii (from top to bottom:  $\rm{r_{ini}} \le 0.4r_{bar}$;
$0.4r_{bar} \le \rm{r_{ini}} \le 0.7r_{bar}$; $0.7r_{bar} \le \rm{r_{ini}}
\le r_{bar}$; $r_{bar}\le \rm{r_{ini}} \le 1.3r_{bar}$; $1.3r_{bar}
\le \rm{r_{ini}} \le 1.6r_{bar}$; $1.6r_{bar} \le \rm{r_{ini}} \le
1.9r_{bar}$.) The bar is inclined by 20 degrees with respect to the
observer line-of-sight. Velocities are in units of 100km s$^{-1}$. In
each panel, the average initial radius is indicated by a  white circle.}
\label{vlos00_regrot}
\end{figure*}

In Fig.~\ref{vlos00_regrot}, we sort, as previously done, stars
according to their birth radius. The resulting line-of-sight velocity
and velocity dispersion maps for these different regions are shown in
Fig.~\ref{vlos00_regrot}. Note that in this plot, both quantities are
line-of-sight velocities, that is we are considering only the velocity
component parallel to the $y$-axis, that is perpendicular to the x-z
plane. As for Figs.~\ref{maps00_regrot} and ~\ref{grid00rot}, the bar is
inclined at an angle of 20 degrees with respect to the line-of-sight. We
will fix the inclination of the bar for all the following analysis to
facilitate direct comparison with the properties of the MW. From the
maps in Fig.~\ref{vlos00_regrot}, we can deduce the following trends:

\begin{itemize}

\item in the bulge region, the larger the birth radius of a star, the
higher its line-of-sight velocity. As shown in Fig.~\ref{vlos00_regrot}, 
moving from smaller to larger birth radii, the average absolute
line-of-sight velocity in the galaxy mid-plane increases from
$\sim$100 km s$^{-1}$ ($r_{ini}< r_{bar}$) to more then 220 km s$^{-1}$
($1.6r_{bar}<r_{ini}<1.9r_{bar}$). This result is clearly in agreement
with the results shown in Fig.~\ref{DL_reg}, that is the larger the birth
radius of a star, the higher its AM content. As a consequence of this
higher AM, while stars born in the inner-disk regions stay confined there,
outer disk stars which have reached the central kpcs, and which have a
higher AM, span a larger volume of the bar and disk than in situ, inner-disk stars (as already shown in Fig.~\ref{rmax} and Table~\ref{table1})

\item in the bulge region, the larger the birth radius of a star,
the greater its line-of-sight velocity dispersion. For example, in the
central regions of the bulge, the velocity dispersions increases from
about 150 km s$^{-1}$ (for stars with $r_{ini}<0.4r_{bar}$) to about 200 km s$^{-1}$ (for stars with $r_{ini}>r_{bar}$).   This trend is a consequence of the
previous point and of the orientation angle of the bar.  Indeed because
the AM content of outside-in migrators is higher than those of in situ
stars, these stars mostly contribute to populating the edges of the bar,
thus their orbit span a larger radial extent (i.e. a larger distance
from the galaxy center) than those of in-situ, inner-disk stars. The
bar being oriented at 20 degrees, the velocity component parallel to the
line-of-sight is determined mainly by motions along the bar major axis. As
a consequence, outer disk stars populating the bar span the larger extent
along the bar, and thus for an orientation of 20 degrees, show the highest
line-of-sight velocity dispersions. Note that this trend of increasing
velocity dispersion with increasing star birth radius is mostly evident
inside the VILR. Outside this radius, stars in the boxy bulge seem to
show a line-of-sight velocity dispersion which is independent on their
birth radius and similar to that of stars born near VILR.

\end{itemize}

\begin{figure*}
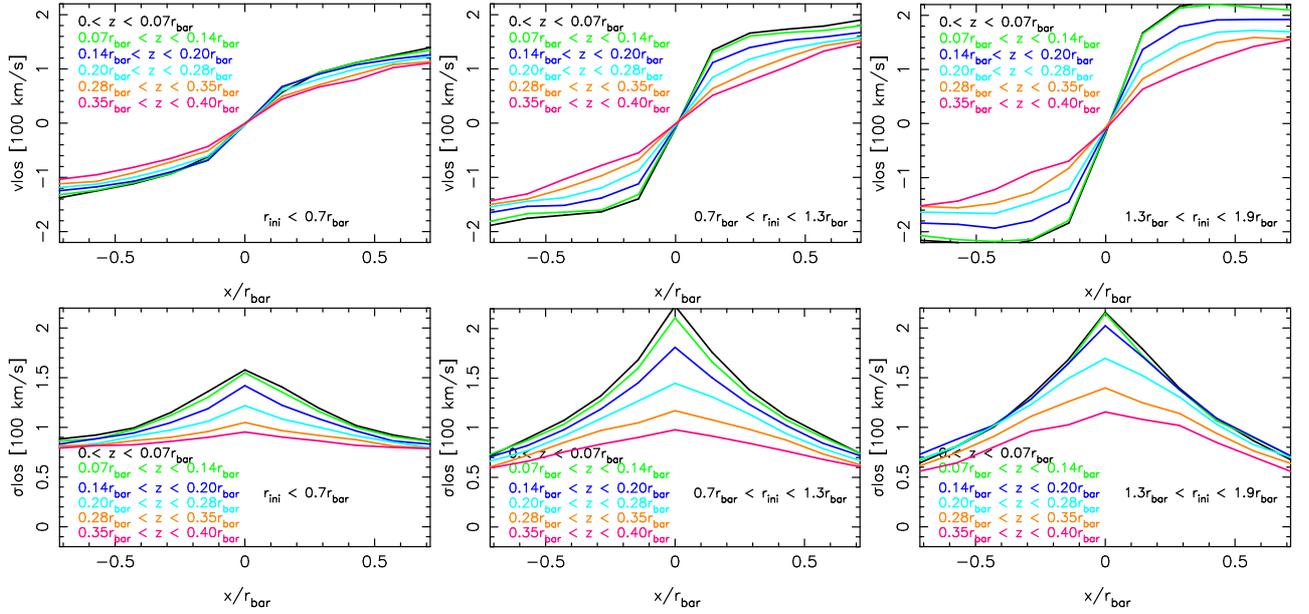

\centering

\includegraphics[width=4.cm,angle=270]{pvlos_horizxzreg1redrotnew_gS0_q1p8_BD0p00.ini.ps}
\includegraphics[width=4.cm,angle=270]{pvlos_horizxzreg2redrotnew_gS0_q1p8_BD0p00.ini.ps}
\includegraphics[width=4.cm,angle=270]{pvlos_horizxzreg3redrotnew_gS0_q1p8_BD0p00.ini.ps}

\includegraphics[width=4.cm,angle=270]{pslos_horizxzreg1redrotnew_gS0_q1p8_BD0p00.ini.ps}
\includegraphics[width=4.cm,angle=270]{pslos_horizxzreg2redrotnew_gS0_q1p8_BD0p00.ini.ps}
\includegraphics[width=4.cm,angle=270]{pslos_horizxzreg3redrotnew_gS0_q1p8_BD0p00.ini.ps}

\caption{Line of sight velocities (\emph{top panels}) and velocity
dispersions (\emph{bottom panels}) along the bulge major axis, subdivided
into six different projected vertical distances from the galaxy mid-plane:
$0 < z < 0.07r_{bar}$ (black curve); $0.07 r_{bar}< z < 0.14r_{bar}$
(green curve); $0.14r_{bar} < z < 0.20r_{bar}$ (blue curve); $0.20r_{bar}
< z < 0.28r_{bar}$ (cyan curve); $0.28r_{bar} < z < 0.35r_{bar}$
(orange curve); $0.35r_{bar}< z < 0.40r_{bar}$ (red curve). Stars have
been grouped accordingly to their initial birth radius: (\emph{from left
to right:}) $r_{ini} < 0.7r_{bar}$; $0.7r_{bar}< r_{ini} < 1.3r_{bar}$;
$1.3r_{bar}< r_{ini} < 1.9r_{bar}$.}
\label{kinhor_regrot}
\end{figure*}

\begin{figure}
\centering
\includegraphics[width=5.cm,angle=270]{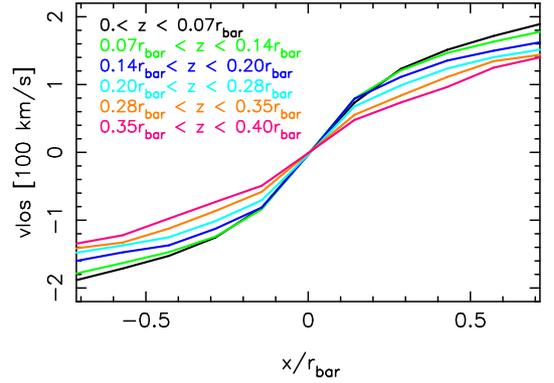}
\caption{Total line of sight velocity along the bulge major axis,
for six different projected vertical distances from the galaxy mid-plane: $0 < z < 0.07r_{bar}$ (black curve); $0.07
r_{bar}< z < 0.14r_{bar}$ (green curve); $0.14r_{bar} < z < 0.20r_{bar}$
(blue curve); $0.20r_{bar} < z < 0.28r_{bar}$ (cyan curve); $0.28r_{bar}
< z < 0.35r_{bar}$ (orange curve); $0.35r_{bar}< z < 0.40r_{bar}$
(red curve). All stars, independently on their birth radius, have been
considered for this plot.}
\label{kinhor_tot}
\end{figure}

To elucidate the dependence of the stellar kinematics on the birth
radius of the stars and on latitude and to facilitate comparison
with observations, we have plotted the line-of-sight velocity and the
velocity dispersion of stars in the bulge region, as a function of the
projected distance $x$ from the center, for different vertical distances
$z$ from the galaxy mid-plane. This is similar to the analysis done
for observations of stars in the Galactic bulge by \citet{ness13b},
except that, to avoid having to define the Sun position in our model --
which is always somewhat arbitrary -- we evaluate the line-of-sight
integrated quantities instead of radial velocities, i.e. instead
of velocities evaluated with respect to the Sun-star direction, as
done in their investigation. Since the relation between the two is
$v_{los}=v_rcos(l)$, where $l$ is the longitude of a star, it means that
at most, at the edge of the bulge \citep[$l \sim 15^\circ$, see for
example][]{freeman13}, we underestimate the radial velocities of stars
in our model by $\sim$4\%.  Figure~\ref{kinhor_regrot} shows the result
of this analysis. Line-of-sight velocities (upper panels) and velocity
dispersions (lower panels) are shown as a function of  $x$, for three
different regions of provenance of the stars ($r_{ini}\le 0.7r_{bar}$,
$0.7r_{bar}< r_{ini}\le 1.3r_{bar}$, and $r_{ini} >  1.3r_{bar}$) and
for six different vertical distances from the galaxy mid-plane. It is
clear that:

\begin{enumerate}

\item stars born in the inner disk show an almost cylindrical rotation,
while stars originating in the outer disk do not.  The resulting
total (i.e. taking into account all the stars, independent of their birth
radii) line-of-sight velocity curve (Fig.~\ref{kinhor_tot}) still shows
a cylindrical rotation,  with values similar to those observed in the
Galactic bulge. For example, comparing the vertical distances $0.14r_{bar}
\le z \le 0.35r_{bar}$ with the ARGOS fields between $b=-5^\circ$ and
$b=-10^\circ$: at $x/r_{bar}=\pm0.7$, which corresponds to approximately
$l=\pm20^\circ$, the differences in the velocity curves with latitude
are $\sim$20 km s$^{-1}$, similar to what observed.

\item at any given latitude, stars that have their origins further out in
the disk have higher line-of-sight velocities. For example (top
panels, Fig.~\ref{kinhor_regrot}), stars born between $0.7r_{bar}<
r_{ini}\le 1.3r_{bar}$ have a $v_{los}$ which is about 30$\%$ higher than
that of stars born inside $0.7r_{bar}$. This value increases up to about
40$\%$ if one compares the $v_{los}$ of stars born inside $0.7r_{bar}$
with that of stars born outside $1.3r_{bar}$. 
\item Whilst the total line-of-sight velocity curve is dominated by stars originating  between $0.7r_{bar}<
r_{ini}\le 1.3r_{bar}$ (cf. Fig.~\ref{kinhor_regrot} and  Fig.~\ref{kinhor_tot}),  innermost ($r_{ini} < 0.7r_{bar}$) and outermost ($r_{ini} >1.3 r_{bar}$)  stars represent, respectively, the low and high velocity tails of the line-of-sight velocity distribution at different latitudes and longitudes. This is a consequence of point 2.

\item the line-of-sight velocity dispersions show a trend with latitude.
This trend is in the sense that the velocity dispersion decreases with
increasing latitude (bottom panels, Fig.~\ref{kinhor_regrot})

\item the line-of-sight velocity trends are the same for outer and inner
disk stars, but their absolute values are larger for the former compared
to the latter (Fig.~\ref{kinhor_regrot}).

\end{enumerate}

The trends listed in points 3 and 4 are exactly those
found by \citet{ness13b} for populations A and B in the galactic
bulge. Moreover, population B rotates 20\% faster than A, as
would be the case if the average birth radius of the stars that make up this
component was larger on average than that of population A (point 2).
We will discuss on this important point in \S~\ref{discussion}.

\subsection{Classical bulges}\label{bulge}

\begin{figure*}
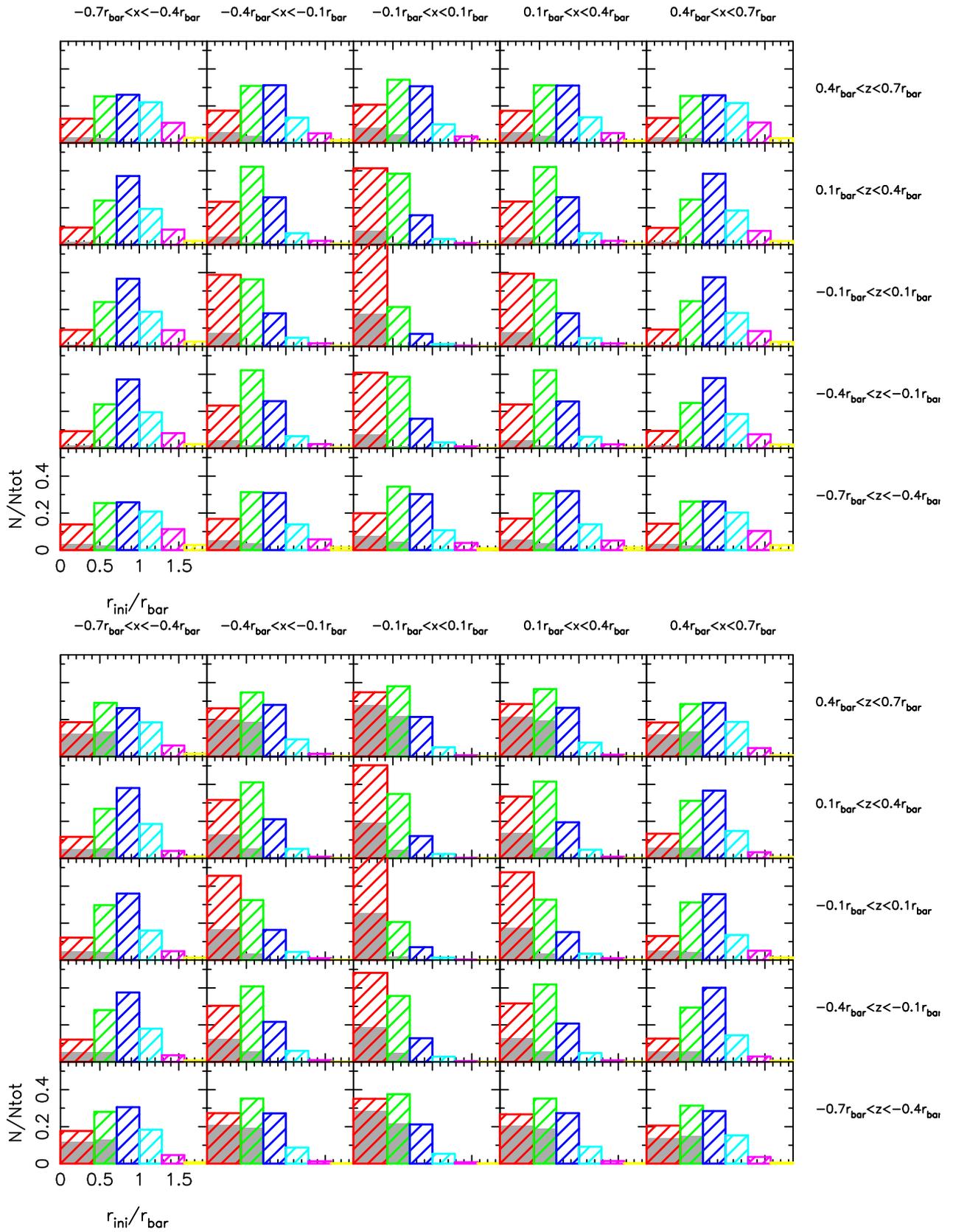

\centering
\includegraphics[width=11.cm,angle=270]{phistorininoforegroundrot2_gS0_q1p8_BD0p10.ini.ps}
\includegraphics[width=11.cm,angle=270]{phistorininoforegroundrot2_gS0_q1p8_BD0p25.ini.ps}
\caption{Same as Fig.~\ref{grid00rot} but for the case with B/D=0.1 (\emph{top panel}) and B/D=0.25 (\emph{bottom panel}). The contribution of the classical spheroid is indicated in grey.}
\label{grid00rotwithbulge}
\end{figure*}

\begin{figure*}
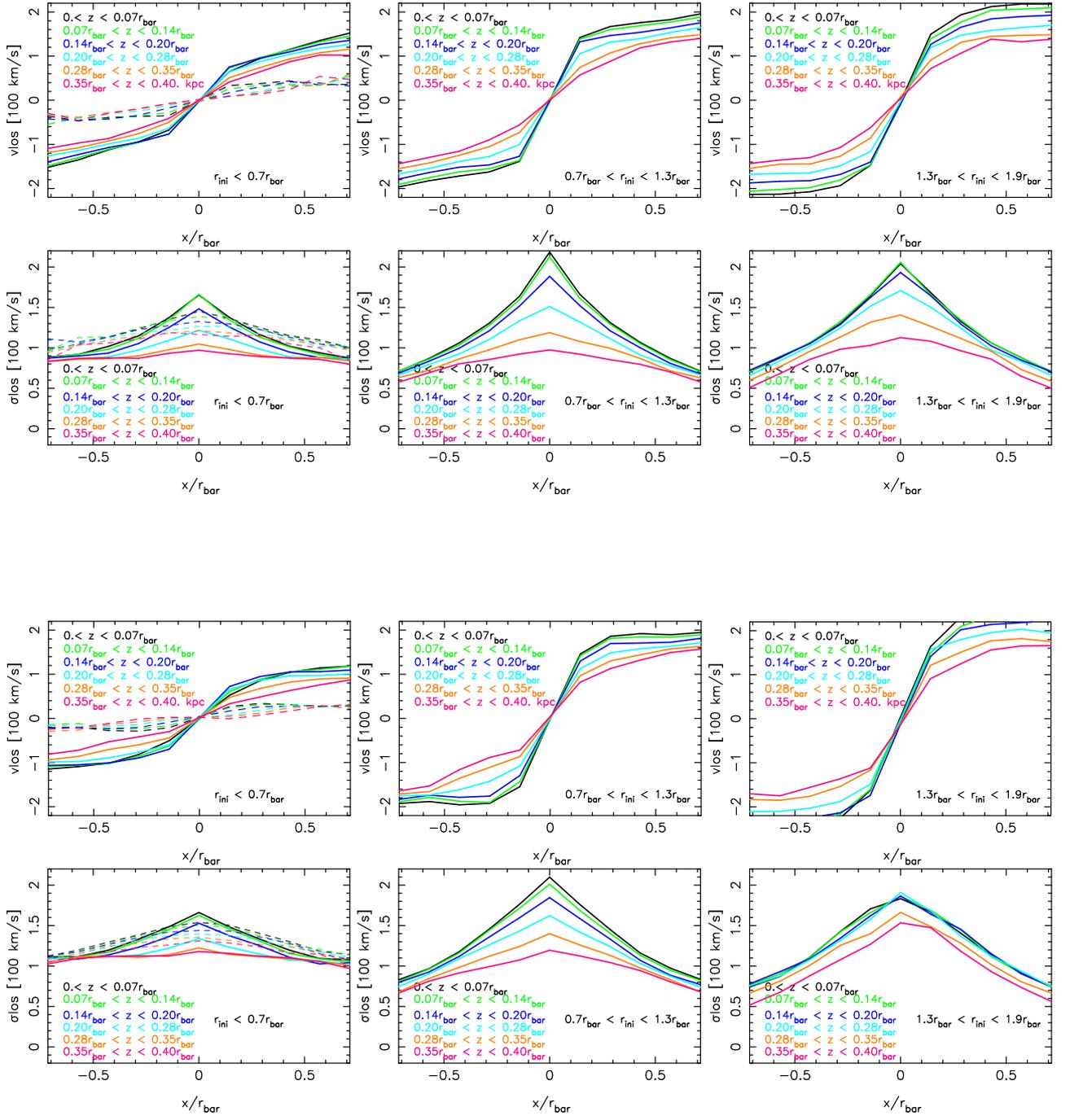

\centering

\includegraphics[width=4.cm,angle=270]{pvlos_horizxzreg1redrotnew_gS0_q1p8_BD0p10.ini.ps}
\includegraphics[width=4.cm,angle=270]{pvlos_horizxzreg2redrotnew_gS0_q1p8_BD0p10.ini.ps}
\includegraphics[width=4.cm,angle=270]{pvlos_horizxzreg3redrotnew_gS0_q1p8_BD0p10.ini.ps}

\includegraphics[width=4.cm,angle=270]{pslos_horizxzreg1redrotnew_gS0_q1p8_BD0p10.ini.ps}
\includegraphics[width=4.cm,angle=270]{pslos_horizxzreg2redrotnew_gS0_q1p8_BD0p10.ini.ps}
\includegraphics[width=4.cm,angle=270]{pslos_horizxzreg3redrotnew_gS0_q1p8_BD0p10.ini.ps}

\vspace{2cm}

\includegraphics[width=4.cm,angle=270]{pvlos_horizxzreg1redrotnew_gS0_q1p8_BD0p25.ini.ps}
\includegraphics[width=4.cm,angle=270]{pvlos_horizxzreg2redrotnew_gS0_q1p8_BD0p25.ini.ps}
\includegraphics[width=4.cm,angle=270]{pvlos_horizxzreg3redrotnew_gS0_q1p8_BD0p25.ini.ps}

\includegraphics[width=4.cm,angle=270]{pslos_horizxzreg1redrotnew_gS0_q1p8_BD0p25.ini.ps}
\includegraphics[width=4.cm,angle=270]{pslos_horizxzreg2redrotnew_gS0_q1p8_BD0p25.ini.ps}
\includegraphics[width=4.cm,angle=270]{pslos_horizxzreg3redrotnew_gS0_q1p8_BD0p25.ini.ps}

\caption{Same as Fig.~\ref{kinhor_regrot}, but for the model with B/D=0.1(\emph{top panels})  and B/D=0.25 (\emph{bottom panels}).  The contribution of the classical spheroid is indicated by dashed lines.}
\label{kinhor_regrotBD}
\end{figure*}

All our previous analysis has been restricted to the case of a boxy/peanut
shaped bulge formed in a pure stellar disk (B/D=0.). It is natural to
ask whether these results are sensitive to the inclusion of a classical
spheroid and what signature would be left by the presence of such a
component underlying a boxy/peanut-shaped structure. With the goal of
answering these questions, we have analyzed two simulations, with an
initial bulge-to-disk ratio respectively equal to B/D=0.1 and 0.25. In
this section we present the main results of this analysis.

To mimic the likely configuration of the bulge/bar of the MW,
we have rotated the stellar bar so that it has an inclination
of about 20 degrees with respect to our line-of-sight in the
analysis. After removing foreground and background stars as
we have done previously (Fig.~\ref{grid00rot}), we discuss the
mapping of the stellar disk into the boxy bulge for these two models
(Fig.~\ref{grid00rotwithbulge})\footnote{Note that we adopted the same
grid as in  Fig..~\ref{grid00rot}, because the bar spatial extent, in
both cases, is similar to that of the model with B/D=0.}.  Concerning the
birth radii of stars\footnote{In this analysis we consider both stars
initially belonging to the disk and stars originating in the classical
spheroid.} in the boxy bulge, we find the following results:

\begin{itemize}

\item as was the case for B/D=0, the fraction of inner stars ($r_{ini}
<0.4r_{bar}$) decreases with increasing longitude and/or latitude. For
the case with B/D=0.1, their fractional contribution changes from more
than 50$\%$, in the galaxy central regions, to about 20$\%$ when moving
vertically along the bar minor axis. For the case with B/D=0.25, at high
latitudes the contribution of inner disk stars is higher, about 35$\%$
of the total, because at these distances from the plane the contribution
of the classical spheroid increases (Fig.~\ref{grid00rotwithbulge}).

\item The relative contribution of inner stars to the local
density is very sensitive to changes in longitude: these stars
constitute the majority of the stellar mass in the inner bulge region
($-0.1r_{bar}<$x$<0.1r_{bar}$ and $-0.1r_{bar}<$z$<0.1r_{bar}$), while
their contribution is $\sim$10$\%$ in the outer bulge region, for a
B/D=0.1. For a B/D=0.25, their fractional contribution in the outer
regions of the bulge is higher (at most $\sim$20$\%$), and this is a
consequence of the presence of stars from the classical bulge at these
longitudes, as would be expected as the classical bulge contribution 
increases overall.

\item as was the case for B/D=0, outer disk stars ($r_{ini} >r_{bar}$, the
bar scale length in both cases) are part of the bulge structure. Within
the outer regions of the bar, indeed, they constitute between 20 and
30$\%$ of the local stellar content.

\end{itemize}

Because of the existence of a mass-size relation for bulges \citep[see,
for example,][]{gadotti09}, the fractional contribution of a pre-existing
spheroid to the local stellar density depends on the mass of the spheroid:

\begin{itemize}

\item a B/D=0.1 classical bulge is mostly confined to the inner bulge
regions (mostly inside $0.4r_{bar}$, for a bulge core radius equal to
$0.18r_{bar}$), and, in our models, its contribution to the total stellar
mass never exceeds 20$\%$, regardless of the longitude or latitude;

\item a B/D=0.25 classical bulge, which is more extended (its initial
core radius is equal to 2 kpc) and more massive, contributes stars all
along the boxy bulge structure. Moreover, its contribution increases
with latitude: as the thick bar stellar density diminishes, the classical
bulge stellar contribution increases (up to about 50$\%$ of the total, see
the behavior along the bulge minor axis in Fig.~\ref{grid00rotwithbulge}).

\end{itemize}

In our models, classical spheroids hidden in a boxy bulge thus
show trends which depend on their mass. To summarize:

\begin{itemize} 

\item while a B/D=0.1 bulge nowhere in the boxy structure contributes for more
than 20$\%$ to the local density, a B/D=0.25 bulge can contribute as
much mass as that of the bar at high latitudes (if no other structure,
like a thick disk, is present, of course).

\item While the fractional contribution of a B/D=0.1 bulge is constant
or slightly decreases with latitude, a B/D=0.25 bulge shows a trend
such that its contribution increases with latitude, with trends
similar to those observed for outside-in migrators. However, while
outside-in migrators are part of the boxy/peanut shaped structure (see
Fig.~\ref{maps00_regrot}), classical bulges are not, because of their
higher velocity dispersions. In other words, any stellar population
whose orbits support the boxy structure \citep[as populations A and B
in][]{ness13b} cannot be dominated by  a classical component. Thus even
if the metallicities of the stars are compatible with those expected
from a classical bulge, on the basis of a mass-metallicity relation
\citep{gallazzi05, thomas10}, the morphology and kinematics of these stars will not
be consistent with those expected for a classical bulge.  We discuss
the evidence for a classical bulge in the MW in \S~\ref{discussion5}.

\end{itemize}

A classical component leaves signatures in the kinematics of stars in
the bulge. As found for the case with B/D=0, we confirm that stars
that originate further out the disk have higher rotational support and
velocity dispersions. That is, outside-in migrators in the bulge are still
recognizable because of their higher AM and because they are a dynamically warmer replica
of populations born in the inner disk (Fig.~\ref{kinhor_regrotBD}). The
trends found in \S~\ref{kinematics} are thus robust and independent
of the mass of a possible (hidden) spheroid.  At the same time, stars
originally in the classical bulge:

\begin{itemize}

\item are characterized by a much lower rotational support than stars that
originate in the disk. Indeed even if stars that comprise the classical
bulge acquire some AM during the disk and bar formation and evolution
\citep[see also][and Fig.~\ref{kinhor_regrotBD}]{saha12,saha13}, these stars never reach rotational
velocities similar to those of disk stars, whatever their location in
the boxy structure.

\item have line-of-sight velocity dispersions which diminish with latitude,
but in such a way that the steepness of the relation is lower than that
observed for stars originating in the disk (moving vertically from low to
high latitudes, the central velocity dispersions decrease by $\sim$20$\%$)

\item have line-of-sight velocity dispersions which also diminish with
increasing longitude, but the gradient of the relation is flatter than
that measured for disk stars, regardless of their birth radii.

\end{itemize}

\section{Discussion: The Milky Way bulge}\label{discussion}

The models presented in the previous sections are intended to simulate
the formation of the bar and its subsequent buckling. We found that
the spatial redistribution the stars in the disk undergo during this
period generates a  structure that has properties similar
to that observed in the bulge of the MW.  However, these simulations
have their limitations.  These simulations were not intended to model
all the components present in a galaxy at the epoch of bar formation,
but modeled only those that are sufficiently kinematically cold to be
susceptible to having their orbital parameters changed significantly
by bar formation, evolution, and the instabilities and asymmetries that
this evolution generates.

The fossil chemical and dynamical record of stars in the solar
vicinity suggest that the thick disk of the MW formed during a period
where the gas out of which stars formed was well-mixed chemically and
highly turbulent \citep[][and references therein]{haywood13, snaith13}.
If the MW formed its bar at the end of its turbulent phase, about 8-9 Gyr
ago, more than 50$\%$ of its current stellar mass was probably already in place
\citep{snaith13}. This value is also in agreement with the extrapolation
of the MW thick and thin disk masses \citep[e.g.][]{fuhrmann12} and
with the estimates of the stellar mass growth of MW-type galaxies as
a function of redshift \citep{vandokkum13, leitner12}.  
The oldest part of this stellar disk  \citep[old thick disk in the nomenclature
of ][]{haywood13} had probably too high vertical velocity dispersions to be involved in the bar instability process Ð current values of the vertical velocity dispersions of old thick disk stars at the solar neighborhood are around 40 km/s, see Haywood et al. (2013). 
However, the youngest disk \citep[young thick disk in the nomenclature
of ][]{haywood13}  had much lower velocity dispersions -- current values at the solar neighborhood are $\sim$ 25- 30 km/s, see \citet{haywood13}, similar to the values characterizing the stellar disks modeled in this paper.
In other words, the youngest and thinnest component
of the MW thick disk, which formed between 8 and 10 Gyr ago, must have
been much more susceptible to the influences of bar instabilities than
the oldest component of the disk, the old thick disk, which formed more than 10 Gyr ago \citep{haywood13}.

With these considerations in mind, we now investigate how much of the
general scheme presented in the previous section is applicable to our
Galaxy and its stellar populations.  We will mainly compare our results
to the set of ARGOS papers, which provide a detailed mapping of the MW
bulge, and a ''dissection'' of its stellar populations that allows us
to make a direct comparison to the results of our simulations.

\subsection{The disk origin of the Milky Way bulge}

Briefly recapitulating the Introduction, the current picture of the MW
bulge suggests that it is made of several distinct stellar populations,
which have different chemical and kinematic characteristics, and
whose fractional contribution to the bulge stellar density depends on
latitude \citep{ness12, ness13a, ness13b}. Interestingly, as shown by
\citet{ness13a}, the three main components found in the bulge -- A, B
and C -- form a sequence in chemical characteristics, from population A
with a high mean abundance and a low $\alpha$ enrichment (mean [Fe/H] and
[$\alpha$/Fe]$\sim$ 0.1 dex), to population B with a intermediate mean
metal abundance and moderate $\alpha$ enhancement (mean [Fe/H]$\sim$
-0.25 dex and [$\alpha$/Fe]$\sim$ 0.2 dex), to population C, the most
metal-poor (mean [Fe/H]$\sim$ -0.7 dex) and $\alpha-$enriched ($\sim$
0.3 dex) of the 3 populations.  This sequence is reminiscent of that
found at the solar vicinity \citep[see, e.g.][]{gonzalez11}. This
is also confirmed by the findings of \citet{bensby13}: the chemical
characteristics of the bulge almost perfectly tracks that of the
local disk.\footnote{Over the last decade, the chemical differences
originally found between the bulge and the thick disk stellar sequences
\citep{zoccali06, lecureur07, fulbright07} have been rather consistently
decreasing as we learn more \citep{gonzalez11, bensby10, bensby13}.
In spite of the very impressive similarities that are now seen between
stars in the solar vicinity and bulge, the results of \citet{bensby13}
suggest a possible (small) shift between the two samples.  However, we
caution that this can still be due to some residual systematic differences
in the selection, analysis, and data quality between the two samples,
at least until these results are otherwise confirmed with some other
independent high quality data set.  Note also that, when comparing the
disk and the bulge populations \citep[e.g. Fig.~27 of][]{bensby13},
the metal-poor tail of the thin disk ([Fe/H]$<$$-$0.2 dex), should not
be considered for such a comparison: these are outer (thin) disk stars
and are almost non-existent in the inner disk and bulge (lower blue
sequence at [Fe/H] $<$$-$0.2 dex in the Fig. 27 of \citet{bensby13}).}
The similarity in the elemental abundance distributions
may imply, for stars in the bulge, the existence of an age sequence
similar to that found within the solar vicinity \citep{haywood13}. In
such a scheme, stars comprising component B would be associated with
the stars that were originally part of the young thick disk, and thus
would be kinematically colder than its more metal-poor counterpart in
the old thick disk which may be the stars that comprise component C.

Therefore, the chemical properties of the Galactic bulge suggest a
possible disk origin for all its main components. In the following, we
will discuss how such an evolutionary path is supported by our modeling,
by showing in particular that: \emph{(a)} the kinematic characteristics of
stars that comprise components A and B are indicative of their different
provenance in the disk, with stars that makeup component B formed at
larger distances from the galactic center on average than stars that
comprise component A; \emph{(b)}  a significant classical spheroid can
be excluded in the MW, thus leaving a disk origin as the only possibility
for most of the stellar mass present in the MW bulge.

\subsubsection{Components A and B}

Of the three main components, A and B are the only ones that
participate in supporting the boxy peanut-shaped morphology of
the bulge \citep{ness12}. This indicates that they must have been
sufficiently kinematically cold originally to be influenced by bar
instabilities. Components A and B are detected throughout the bulge
region, but in different relative proportions: B becomes dominant at
intermediate ($b = -7.5$~deg) and high latitudes ($b= -10$~deg), while
its relative contribution is comparable to that of A at lower heights
($b = -5$~deg) above the plane \citep{ness13a}. Component B rotates 20$\%$
faster than A, and has radial velocity dispersions that, at all latitudes
and longitudes, are a dynamically warmer replica of those of component
A. Our models suggest that all these characteristics can be explained
if the stars that makeup component B formed, on average, further out in
galactic disk than the stars of component A.\footnote{This is
in fact expected if stars in B originated as part of the young thick
disk because the surface density distribution of the young thick disk
extends further from the galactic center compared to the more metal-rich
 thin disk. Both of these components have similar local densities and
scale lengths, but the young thick disk scale height is approximately
two times larger than that of the metal-rich thin disk. Hence we
expect to find more stars of the young thick disk further from the
galactic center compared to those of the metal-rich thin disk.} We have seen,
indeed, in \S~\ref{kinematics} that that stars with larger birth radii
that end up in the boxy bulge have higher final rotational support and
radial velocity dispersions (this last trend is valid for the small bar
viewing angle of the MW bar). Moreover, the fractional contribution to
the peanut-shaped density distribution of component B tends to increase
with latitude. This is another argument in favor of the scenario where
the stars that comprise component B formed, on average, further away
from the Galaxy center than the stars that makeup component A.

When did these two components become part of the boxy structure? A
first possibility is that, components A and B were both present in
the MW disk at the time of bar formation. Because of the chemical
characteristics of these two components, and because of the fact that
A must have been, on average, more centrally concentrated than B, this
would imply the existence of a negative [Fe/H] gradient and of a positive
[$\alpha$/Fe] gradient in the disk  at the onset of the bar formation and
evolution phase of the Galaxy. Note that, with a difference $\Delta_{\rm{
[Fe/H]}}=<\rm{[Fe/H]}>_A - <\rm{[Fe/H]}>_B \sim$ $-$0.4 dex over most
of the extent of the disk, the resulting radial gradient would be
significantly flatter than the $-$0.4 dex/kpc needed in the work of
\citet{martinez13} to reproduce the vertical bulge abundance gradient
observed in the MW.  In our opinion, a radial gradient as steep as that
proposed by  \citet{martinez13} is unrealistic: it would imply that the
typical metallicity of the MW disk at 4--5 kpc from the Galaxy center was
between $-$1 and $-$1.4 dex. Such low metallicities are never reached
in the MW thin disk \citep[see, for example][]{fuhrmann99, bovy12,
haywood13}, but  are typical of the
metal-poor tail of the  thick disk and of the MW halo \citep[see, among
others, ][]{beers95, reddy08, nissen10}.

The second possibility is that,  when the bar formed, the young thick
disk was already in place, while the thin disk had not yet formed. This
second scenario is compatible with a formation epoch for the MW bar around
$z\sim1$, the epoch of bar growth in external galaxies \citep{sheth08,
melvin13}. This age also corresponds to the transition epoch between the
thick and thin disk formation at the solar vicinity and a significant
dip in the star formation rate \citep[][]{haywood13, snaith13}. In this scenario,
the young thick disk was captured first in the bar instability, forming
component B of the bulge. Subsequent star formation formed the thin disk,
which was then also involved in the formation of the peanut structure:
stars close to the vertical resonance were scattered to great heights from
the plane, contributing to the thick part of the peanut (in agreement
with stars of component A being detected at high latitudes as well),
while the stars from the inner disk contributed to the formation of
population A close to the mid-plane. To differentiate between these
scenarios will require N-body models which include realistic treatments
of gas dynamics, star formation and chemical enrichment. We note, however, that stars younger than 8-9 Gyr seem to be present in the Galactic bulge \citep{bensby13}. If confirmed, this result will favor this scenario of continuous enrichment of the Galactic bulge by younger disk populations.

\subsubsection{Component C}

Of the three main components, C is the only one in the bulge that does not
participate in contributing to the boxy/peanut structure. Its kinematics
is indeed compatible with a dynamically hotter population, characterized
by radial velocity dispersions nearly constant with longitude and latitude
\citep{ness13b}. As we will discuss in the following section, it
is difficult to reconcile the chemical and kinematic properties of
component C with a classical bulge. The other possibility is that stars
that comprise component C are stars which are part of the old thick disk
-- a population that was not significantly influenced by the buckling of
the bar because of its high velocity dispersions. Consistent with this
picture is the fact that its relative contribution to the bulge increases
with latitude as would be expected if it originated from a component
with a larger scale height than components A and B (see the
discussion in \citet{ness13b}).  Even if the models discussed
in this paper do not include this component, such a scenario seems
compatible with the N-body simulations which do include a thick disk and
which show that indeed such a kinematically hot component is much less
sensitive to being influenced by bar instabilities \citep{bekki11}. While
further studies taking into account both thick and thin disk populations
are needed to quantify the evolution of these thicker components in the
presence of a stellar bar, and thus to validate the suggested scenario,
we emphasize that the chemical properties of population C (mean [Fe/H] and
[$\alpha$/Fe]) are also characteristics of old thick disk stars within
the solar vicinity \citep{haywood13}. Similarly, the high rotational
support of component C, which is similar to that of component A,
also supports a ({\it in situ}) disk origin for component C.

\subsection{Is there any classical bulge in the MW?}\label{discussion5}

The possibility that the MW hosts a classical bulge, hidden within the
overall boxy/peanut-shaped structure, has been debated for at least
the last two decades \citep{zhao94, soto07, zoccali08,  babusiaux10}.
Some general arguments should be considered in this debate.  First,
if a classical spheroid is underlying the MW bulge, how massive can
it be? For a MW-type galaxy, the average bulge-to-total light ratio,
B/T, is $\sim$0.1 (Laurikainen et al 2007, Weinzirl et al. 2009), with
typical values between 0 and 0.2 (Weinzirl et al. 2009).  This implies
bulge-to-disk ratios between 0. and 0.25, with an average value of
about 0.11.  However, the bulge-to-disk ratio depends on the environment
as well. \citet{kormendy10}, for example, remarked that out of the 17
giant disk galaxies found within 8~Mpc from the MW, 15 are compatible
with being bulgeless or having very small bulges.  Extending the sample
to 11~Mpc, \citet{fisher11} confirm these results, with 80\% of disk
galaxies being either bulgeless or hosting a pseudobulge.  This suggests
that, for the MW, a B/D greater than 0.1 is perhaps unlikely.

In any event, if the Galaxy hosts a classical bulge, the expected
metallicity of this component can be estimated from the results
of bulges in other galaxies.  Assuming a MW stellar mass of about
$5\times10^{10}M_{\odot}$ \citep{mcmillan11},  the mass-metallicity
relation suggests typical [Fe/H] values around $-$0.3--0.4
dex, for a spheroid $\sim 10^{10}M_{\odot}$ \citep{gallazzi05,
thomas10}. Specifically, from \citet{gallazzi05}, one sees that for
a $\sim 10^{10}M_{\odot}$ spheroid, a metallicity below $-$0.7 --
-0.8 dex is unlikely, with about only 15$\%$ of galaxies with those
masses having such a low metal content.  Another independent approach
is to use the $i-$band absolute magnitude versus metallicity relation
of classical bulges given by \citet{zhao12}. For an absolute $B$-band
of the MW bulge $M_B= -17.65$ \citep{kormendy01}, and thus an $i-$
band absolute magnitude necessarily lower than this value,  one obtains
[Fe/H] values generally not below $-$0.35 -- $-$0.4 dex.  Both arguments
suggest that the classical bulge, if present, likely would have
metallicities compatible with the component B of Ness et al. According
to our models, at B/D=0.25, a spheroid this massive  would imply that
a substantial fraction of stars (between 30\% and 50\%)
at high latitudes should have kinematic characteristics typical of a
spheroid\footnote{This is also
consistent with the mass-size relation, which predicts that a classical
bulge with mass $\sim$few $\times$ 10$^{10}$ M$_{\odot}$ (that is $\sim$ 25\%
of the disk mass) would have an effective radius $\sim$ 1 kpc. In this
case, in the kinematic fields at least up to $|b|\le 7.5^\circ$, one
would expect to see a non-negligible number of classical bulge stars.
} (Fig.~\ref{grid00rotwithbulge})  but the kinematics of component B excludes this possibility.

Could the classical spheroidal population be masquerading as component
C or hidden among its stars?  For a number of good reasons, this is an
appealing idea. For example, component C shows the trends expected for a
rather massive bulge: its contribution increases with latitude and it does
not contribute to the boxy/peanut-shaped morphology of the bulge. However,
the mass-metallicity relation of bulges does not support this
hypothesis:  a bulge metallicity around $-$0.7 dex is typical of
$10^8 M_{\odot}$ spheroids \citep{lee08, mcconnachie12, kirby13}.
Such a mass would be an absolutely negligible contribution to the bulge
for a galaxy with mass and size of the MW, or conversely, in order to
maintain the necessary B/D ratio of the MW, the MW would have to have an
unrealistically low mass of only about a few $\times$ 10$^9$ M$_{\odot}$.

From these considerations, we favor a scenario where a classical
bulge, if present,  is hidden in component B, and it is small
(B/D$\la$0.1). The results presented in \S~\ref{bulge} suggest that a
possible signature of this classical component may be left in the radial
velocity distribution of the stars. We have shown, indeed, that even
if the classical component acquires some rotation during the secular
evolution, its rotational support is always lower than that of the disk
counterparts. Moreover, the difference in the rotational support of these
two components increases with longitude. As a result, the presence of
a low velocity tail in the radial velocity distribution whose strength
increases with longitude, may be indicative of the presence of a small
classical bulge component.

But is a classical spheroid really needed to explain the characteristics
of the MW bulge? The works cited at the beginning of this section
\citep{zoccali08, babusiaux10, hill11} all recognized that the complex
kinematics and chemistry found in the MW bulge is at odds with the
presence of a single stellar population with bar-like kinematics. They
suggested that the observed trends may be explained by a double-component
bulge, with an inner boxy-peanut structure formed by vertical
instabilities, and an outer classical bulge.  However, is this component
really a classical spheroid? \citet{babusiaux10} noted that, at high
latitudes,  the total radial velocity dispersion of stars along the bulge
minor axis is dominated by metal-poor stars that are dynamically hot.
In other words, it is the metal poor component that dominates the observed
trend of the radial velocity dispersion at high vertical distances from
the Galaxy plane \citep[e.g. Fig.~2 in][]{babusiaux10}.  Similar trends
have been found in models where the classical bulge becomes the dominant
component at high latitudes and determines the global trend of the
velocity dispersion \citep{fux99}. However, in the Fux model, the classical
bulge is very massive -- in fact, too massive for a MW-type galaxy --
the bulge-to-disk ratio being of the order of 0.5. With a lower bulge
mass,  around 10\% of that of the disk, consistent with that of the MW,
it would not have been possible to explain the kinematic trends observed
in the Galactic bulge: in the absence of a thick disk component, indeed,
at all latitudes, disk stars with bar-like kinematics would have been the
dominant component to the local stellar mass content, rather than stars
associated with a classical spheroid (Fig.~\ref{grid00rotwithbulge}).

This implies that the metal-poor, $\alpha-$enhanced, dynamically hot
population dominating the MW bulge at high latitudes must have a different
origin. The arguments presented at the beginning of \S~\ref{discussion}
point indeed to a different scenario.  There is a massive stellar
component (that could be about half of the present-day MW stellar mass)
already in place by z$\sim$1. Even assuming that an old (i.e. ages $>$
8 Gyr) classical spheroid about one tenth of the current MW disk mass
exists in the Galaxy, this would imply that we are still left with about
80$\%$ of the stellar mass formed at $z\ge$1 \footnote{This comes from
the following argument: a B/D=0.1 implies that $M_{bulge}=0.1\times
M_{disk}=0.1\times(M_{thin}+M_{thick})=0.2 M_{thick}$.  That is,
of the total stellar mass formed above z=1, only 20$\%$ can be in a
classical old spheroid, if $M_{thin}=M_{thick}$.} that is not in a
classical spheroid. As a consequence, it is this massive, disk-like,
metal-poor, $\alpha-$enriched component -- the thick disk -- the one
that is impossible to exclude from consideration when interpreting the
observed trends in bulge populations with vertical distances within the
MW bulge.  We know the thick disk exists and it must be contributing to
the characteristics of the bulge.

\section{Conclusions}\label{conclusions}

By means of idealized, dissipationless N-body simulations that follow
dynamical influence of the formation and subsequent buckling of a stellar
bar on disk stars, we have studied the formation and characteristics of
boxy/peanut-shaped bulges and compared them with the properties of the
stellar populations of the Milky Way's bulge.  The main general results of
our modeling, valid for the general family of boxy/peanut shaped bulges,
are the following:

\begin{enumerate}

\item Because of the radial migration initiated at the time of the bar
formation, boxy bulges are populated by stars born both in the inner
disk and in the outermost regions of the disk, up to the OLR. That is, it is essentially
the entirety of stellar disk that is mapped into the stellar populations
that comprise the boxy bulge.

\item Stars formed outside the bar radius can constitute a non-negligible
fraction of the stars that currently comprise the boxy bulge, in our
modeling up to 30\% of the stellar mass at high latitudes.

\item The fraction of outside-in migrators in a boxy bulge increases
both with galactic latitude and longitude.

\item The contribution of stars to the local bulge density depends on
their birth radius: stars born in the inner disk tend to stay confined
in the innermost regions of the boxy bulge, while stars born close to
or beyond the vertical inner Lindblad resonance tend to populate the
more extended regions of the boxy/peanut-shaped structure.

\item The stellar birth radii are imprinted on the stellar kinematics of
the bulge stars: bulge stars with larger birth radii have higher levels
of rotational support and line-of-sight velocity dispersions compared
with stars with smaller birth radii (but note that the trends in the
line-of-sight velocity dispersion depend on the angle at which the bar
is viewed).

\item If a classical spheroid is hidden among the stellar components of
a boxy bulge, then our modeling and empirical relationships between
spheroidal mass and size and metallicity indicate the following:

\begin{itemize}

\item  Because of the existence of a mass-size relation for spheroids,
its contribution to the local density of the boxy bulge depends on
its mass: the larger the classical bulge-to-disk ratio, the greater is
its fractional contribution to the stellar densities at high vertical
distances from the galaxy mid-plane.

\item A boxy bulge which contains a small classical spheroid (B/D=0.1)
is dominated everywhere by stars originating in the disk. In our
models, the classical spheroid indeed contributes, at most, about
20\% of the local mass density regardless of galactic longitude or
latitude.

\item For a more massive classical spheroid (B/D=0.25), its contribution
to the boxy bulge mass density stays constant or increases
with latitude. In the absence of a thick disk component, our models
predict that the ratio between the classical spheroid and disk stars
mass can become equal to 1 at high latitudes, that is the classical
spheroid becomes a non-negligible fraction of the stellar content of
the boxy structure far from the mid-plane.

\item Even if classical spheroids acquires some rotational
angular momentum during the secular evolution of the bar \citep[see
also][]{saha12, saha13}, their rotational support is generally
smaller than that of disk stars at the same location: classical spheroids
may be thus be revealed as a low velocity tail in the line-of-sight
velocity distribution of stars in the boxy bulge, whose strength increases with longitude.

\end{itemize}

\end{enumerate}

Comparing these results with the properties of the stellar populations
in the Milky Way bulge obtained by the ARGOS survey, we conclude that:

\begin{enumerate}[I]

\item The two most metal-rich components of the MW bulge have a disk
origin, with component B formed, on average, at larger radial distances
than component A.

\item Because of their chemical and kinematic characteristics, we suggest
that component A and B are dominated by the Galactic inner
thin disk and by the young \citep[ages $\le$ 10 Gyr; see][]{haywood13}
thick disk, respectively.

\item Because of the existence of a mass-metallicity relation for
spheroids, and because of the properties of extra-galactic classical
bulges, it is difficult to associate component C with a classical
spheroid. Its high level of rotational support, as deduced by radial velocity measurements, suggests a disk origin for
this component as well. On the basis of its chemical characteristics,
we suggest that it is indeed associated to the old galactic thick disk
\citep[ages in the range 8 --10 Gyr; see][]{haywood13}.

\item The presence of a massive classical spheroid, with B/D$\sim$0.2
can be excluded for the MW. If present, such a massive component would
indeed be significant at high latitudes (30--50\% of the local stellar
density) and should have metallicities comparable to those of population
B in the bulge. The bar-like kinematic and morphology of population
B excludes this possibility. As a result, if a classical bulge is
hidden in the populations of the boxy-peanut structure, it cannot be
massive (B/D$\le$0.1). This result is in agreement with those found
by \citet{shen10, kunder12}, on the basis of kinematic arguments, thus
supporting the scenario that most of the mass of the MW's bulge has a
disk origin.

\end{enumerate}

Many observational arguments discussed in this paper support the
presence of a massive thick disk 
in the inner Galactic regions. Such a component cannot be neglected when
tallying the mass budget of the MW bulge and inner disk. Modeling its
interplay with and contribution to the bar and the thin disk will be
the focus of future studies.

\section*{Acknowledgments}
All the simulations have been performed on CURIE at CCRT, CEA, within
the framework of the ``Projet Grand Challenge'' and with 
support provided by the GENCI grant x2012040507. The authors acknowledge
the support of the French Agence Nationale de la Recherche (ANR) under
contract ANR-10-BLAN-0508 (GalHis project). PDM warmly thanks M. Ness,
O. Gerhard, and L. Origlia for comments, suggestions and criticisms on
the content of this paper, and C. Babusiaux for remarks on a first version of this manuscript. We are grateful to the anonymous referee for
a careful report which helped us to improve the presentation and clarity
of our results.

\begin{appendix}
\section{Exploring different galaxy models}\label{app1}

The galaxy simulations presented in \S~\ref{method} and
\ref{results} develop a bar whose length ($r_{bar}=$7 kpc) is about
half of the initial disk size (13 kpc). In these simulations, the
corotation radius and the OLR are located at $7-8$ kpc, and 13 kpc,
respectively. Since the initial disk does not extend further, it is
natural to investigate what is the response and redistribution of stars
to bar formation and evolution that are initially at even larger radii
than in our original simulations. In this Appendix, we have analyzed
two additional simulations which are part of the sample of simulations
described in \citet{halle13}. These simulations include gas physics,
implementing a cooling prescription which allows the gas to cool low
temperatures and the formation of a molecular hydrogen component, together
with star formation and several different stellar feedback efficiencies
\citep[see ][ for details]{halle13}. For the purpose of the comparison
to the models presented in this paper, we have chosen two simulations
from this suite without molecular hydrogen and alternatively, with and
without including stellar feedback. We have chosen these two simulations
because they develop stellar bars with different characteristics, thus
allowing us to investigate the impact of those characteristics, and
specifically, the impact of the location of the associated resonances,
on the overall disk evolution. We summarize the main features of the
simulated galaxies for these two simulations from \citet{halle13}: they
consist of a gSb-like galaxy, composed of a stellar disk whose mass is
$M_d=4.5\times10^{10}M_{\odot}$, and whose scale length is $a_d=$5 kpc;
a classical bulge whose mass is 0.25$M_d$ and characteristic radius
$a_b$=1 kpc; and a gaseous disk whose mass is 0.2$M_d$, and whose scale
length is 11.8 kpc. The gaseous and stellar disk components follow a
Miyamoto-Nagai density profile while the stellar bulge and dark halo
components have Plummer density profiles \citep[see][for a complete
description of the models]{halle13}.  In particular, in these models
the initial stellar disk extends to 36 kpc from the galaxy center, thus
allowing us to trace the evolution of stars up to distances of $\sim$8
disk scale lengths. The total number of particles employed is $1200000$,
equally distributed in number as gas, stars and dark matter particles. 

\begin{figure*}
\centering
\includegraphics[width=14.cm,angle=0]{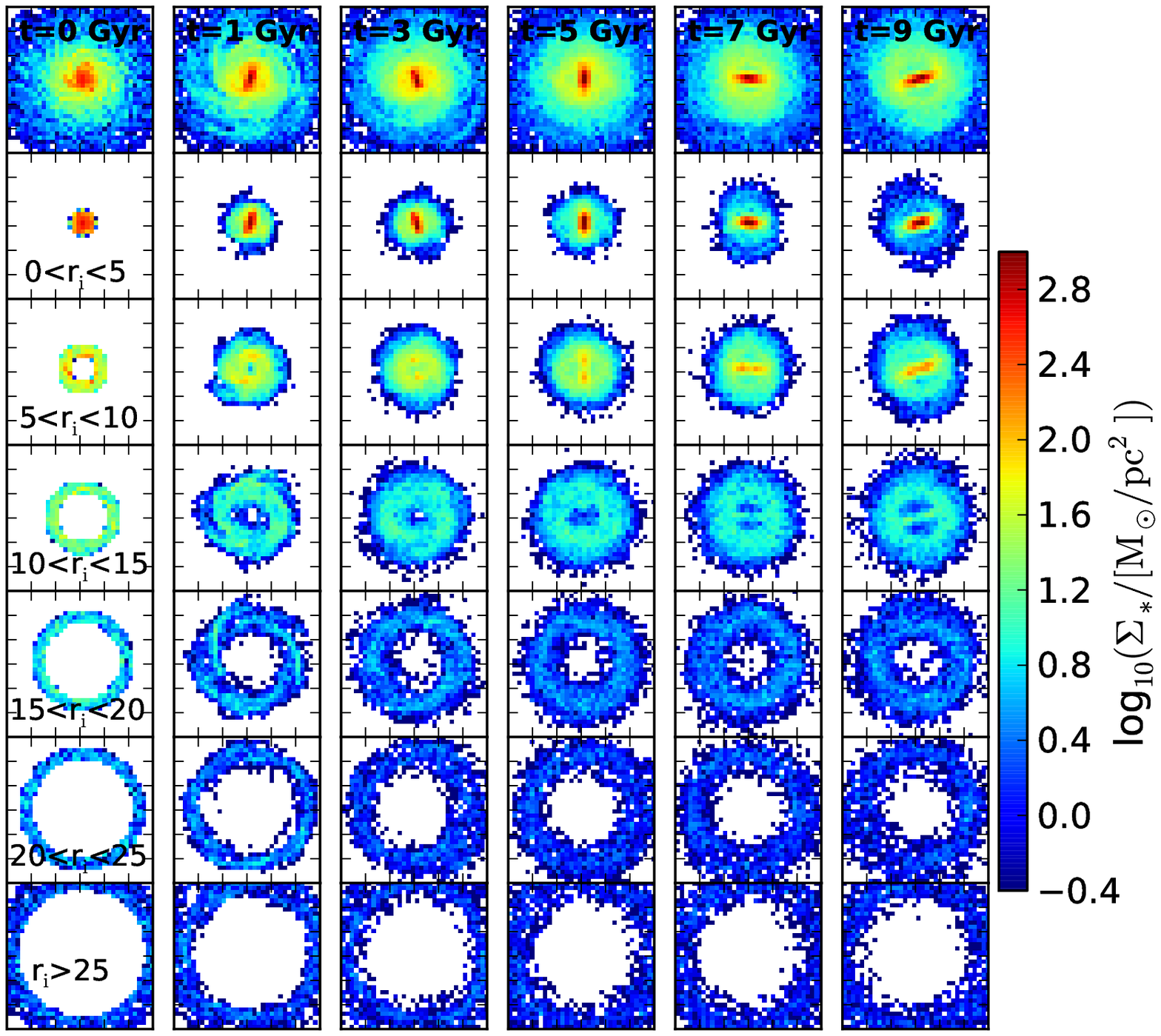}
\caption{\emph{(From top to bottom:)} Face-on density distribution
of stars with different birth radii: $r_{ini} \le$ 5 kpc; 5 kpc$\le
r_{ini}\le$ 10 kpc; 10 kpc $\le r_{ini} \le$15 kpc; 15 kpc $\le r_{ini}
\le$20 kpc;  25 kpc $\le r_{ini}$. Different columns correspond to
different times, as indicated. The total stellar density distribution
is given in the top row, at different times. }
\label{app1_fig1}
\end{figure*}

\begin{figure*}
\centering
\includegraphics[width=14.cm,angle=0]{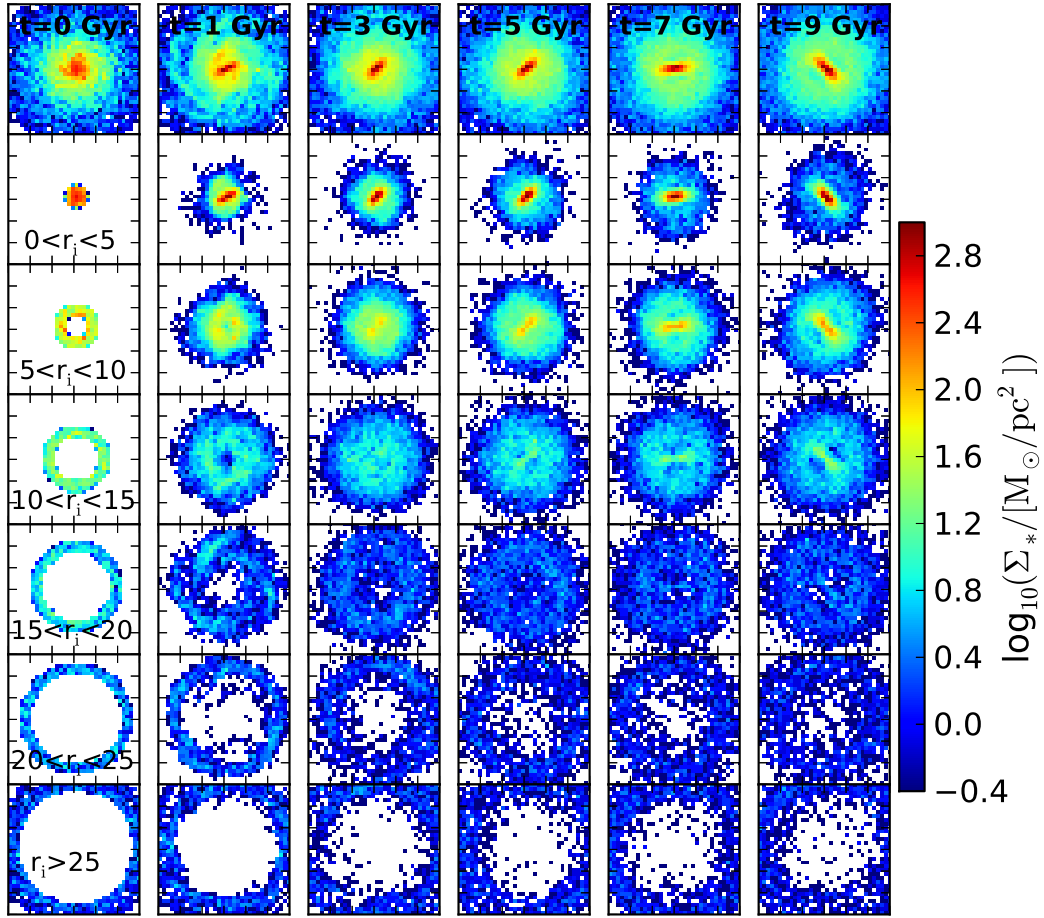}
\caption{Same as Fig.~\ref{app1_fig1}, but for the model without
stellar feedback.}
\label{app1_fig2}
\end{figure*}

Both models develop prominent stellar bars, whose length is
about 7~kpc, at 1 Gyr after the start of the simulation. At about t=5~Gyr, the corotation
and OLR are located at r=10~kpc  and r=17~kpc respectively for the
simulation with stellar feedback (Fig.~\ref{app1_fig1}), while they are
located slightly further out in the simulation without stellar feedback
(at r=12~kpc and r=20~kpc; Fig.~\ref{app1_fig2}). This difference is a
result of the fact that the simulation without stellar feedback has a
lower bar pattern speed than the simulation with stellar feedback. As was
the case in our original simulations (\S~\ref{results}), the formation
of the bar is accompanied by a significant redistribution of the stars
in the disk, with stars initially as far as 15-20~kpc, reaching the inner
regions in less than 2 Gyr. These outside-in migrators become part of the
stellar bar, as is evident in their asymmetric distribution in the inner
5--10~kpc  which tracks the orientation of the bar. However, it is also
interesting to note that the outermost regions of the simulated disks,
those at distances greater than 15~kpc (Fig.~\ref{app1_fig1}) and 20~kpc
(Fig.~\ref{app1_fig2}) in the two simulated galaxies, stay mostly confined
to the outer disk, with only less than 1/1000 of those stars reaching the bar region, after 9~Gyr in
the simulations. The transition radius where the different responses of
the disk to the spatial redistribution seems to occur is approximately
the radius of the OLR. In particular, comparing Figs.~\ref{app1_fig1}
and \ref{app1_fig2}, one see that the greater the radius of the OLR, the
larger the disk region involved in contributing stars to the outside-in
migration. As a corollary to this, it also means that the regions where
stars do not  migrate inwards also occurs beyond a larger radius. While
the study of this behavior will be the subject of subsequent studies, the
point we wish to emphasize here is that in the MW all kinematically cold populations inside the solar circle, 
where the MW bar OLR is probably located \citep[see,
for example,][]{gerhard10}, may have contributed to make up the Galactic boxy/peanut-shaped bulge.

\end{appendix}

\end{document}